\documentclass[fleqn,usenatbib]{mnras}

\usepackage{newtxtext,newtxmath}

\usepackage[T1]{fontenc}
\usepackage{ae,aecompl}

\usepackage{hyperref}
\usepackage{graphicx}	
\usepackage{blindtext}  
\usepackage{natbib}
\usepackage{rotating,times,graphicx,latexsym}
\usepackage{color}
\usepackage{longtable}
\usepackage{lscape}
\usepackage{lipsum} 
\usepackage{array}
\usepackage{flafter}

\usepackage{orcidlink}

\usepackage{soul} 
\usepackage{xargs} 

\graphicspath{ {images_pdf/} }
\newcommand{\hii}{H{\sc ii}~}
\newcommand{\vampset}{{$\hat{A}^o_{\{V \}}$}}

\title[Spot properties on YSOs]{A survey for variable young stars with small telescopes: IX - Evolution of Spot Properties on YSOs in IC\,5070}

\author[Carys Herbert et al.]{Carys Herbert\orcidlink{0000-0003-4217-8811}$^{1}$\thanks{E-mail: cbah2@kent.ac.uk}, 
Dirk Froebrich\orcidlink{0000-0003-4734-3345}$^{1}$, 
Siegfried Vanaverbeke$^{2,3,4}$, 
Aleks Scholz$^{5}$,
\newauthor 
Jochen Eisl\"offel\orcidlink{0000-0001-6496-0252}$^{6}$, 
Thomas Urtly$^{7}$$^{\thanks{HOYS Observer}}$, 
Ivan L. Walton$^{7}$$^{\dagger}$, 
Klaas Wiersema$^{8,9}$$^{\dagger}$,
Nick J. Quinn$^{7}$$^{\dagger}$, 
\newauthor 
Georg Piehler$^{10}$$^{\dagger}$, 
Mario Morales Aimar$^{11,12}$$^{\dagger}$, 
Rafael Castillo García$^{11,13,14}$$^{\dagger}$,  
\newauthor 
Tonny Vanmunster$^{15}$$^{\dagger}$,
Francisco C. Sold\'{a}n Alfaro$^{11,13,16}$$^{\dagger}$,
Faustino Garc\'{i}a de la Cuesta$^{17}$$^{\dagger}$,
\newauthor 
Domenico Licchelli$^{18,19}$$^{\dagger}$, 
Alex Escartin Perez$^{11,20,21}$$^{\dagger}$, 
Esteban Fern\'{a}ndez Ma\~{n}anes$^{22}$$^{\dagger}$, 
\newauthor
Noelia Graci\'{a} Ribes$^{22}$$^{\dagger}$, 
Jos\'{e} Luis Salto Gonz\'{a}lez$^{11,23,24}$$^{\dagger}$,
Stephen R.L. Futcher$^{7,13,25,26}$$^{\dagger}$,
\newauthor 
Tim Nelson$^{25,27}$$^{\dagger}$,
Shawn Dvorak$^{13,28}$$^{\dagger}$,
Dawid Mo\'{z}dzierski$^{29}$$^{\dagger}$,
Krzysztof Kotysz$^{29}$$^{\dagger}$,
\newauthor 
Przemys{\l}aw Miko{\l}ajczyk$^{29,30}$$^{\dagger}$, 
George Fleming$^{7}$$^{\dagger}$, 
Mark Phillips$^{7,31}$$^{\dagger}$, 
Tony Vale$^{7,13,32,33,34}$$^{\dagger}$,
\newauthor 
Franky Dubois$^{2,3}$$^{\dagger}$, 
Heinz-Bernd Eggenstein$^{35}$$^{\dagger}$,
Michael A. Heald$^{13}$$^{\dagger}$,  
Pablo Lewin\orcidlink{0000-0003-0828-6368}$^{36}$$^{\dagger}$,
\newauthor 
Derek OKeeffe$^{37}$$^{\dagger}$,
Adam Popowicz$^{38}$$^{\dagger}$, 
Krzysztof Bernacki$^{38}$$^{\dagger}$,
Andrzej Malcher$^{38}$$^{\dagger}$, 
\newauthor 
Slawomir Lasota$^{38}$$^{\dagger}$,  
Jerzy Fiolka$^{38}$$^{\dagger}$, 
Adam Dustor$^{39}$$^{\dagger}$,
Stephen C. Percy$^{40}$$^{\dagger}$,
Pat Devine$^{31}$$^{\dagger}$, 
\newauthor 
Aashini L. Patel\orcidlink{0000-0001-5800-6710}$^{1}$$^{\thanks{Beacon Observer}}$, 
Matthew D. Dickers\orcidlink{0000-0001-9615-9101}$^{1}$$^{\ddagger}$,
Lord Dover$^{1}$$^{\ddagger}$,
Ivana I. Grozdanova\orcidlink{0009-0006-6327-0595}$^{1}$$^{\ddagger}$,
\newauthor 
James S. Urquhart\orcidlink{0000-0002-1605-8050}$^{1}$,
Chris J.R. Lynch$^{1}$,
\\
$^{1}$Centre for Astrophysics and Planetary Science, School of Physics and Astronomy, University of Kent, Canterbury CT2 7NH, UK\\
$^{2}$Public observatory ASTROLAB IRIS, Provinciaal Domein “De Palingbeek”, Verbrandemolenstraat 5, 8902 Zillebeke, Ieper, Belgium\\
$^{3}$Vereniging voor Sterrenkunde, werkgroep veranderlijke sterren, Oostmeers 122 C, 8000 Brugge, Belgium\\
$^{4}$Centre for Mathematical Plasma-Astrophysics, Department of Mathematics, KU Leuven, Celestijnenlaan 200B, 3001 Heverlee, Belgium\\
$^{5}$SUPA, School of Physics \& Astronomy, University of St Andrews, North Haugh, St Andrews KY16 9SS, UK\\
$^{6}$Th\"uringer Landessternwarte, Sternwarte 5, D-07778 Tautenburg, Germany\\
$^{7}$British Astronomical Association, Variable Star Section, PO Box 702, Tonbridge, TN9 9TX, UK\\ 
$^{8}$Centre for Astrophysics Research, University of Hertfordshire, Hatfield, AL10 9AB, UK\\
$^{9}$School of Physics and Astronomy, University of Leicester, University Road, Leicester LE1 7RH, UK\\
$^{10}$Selztal Observatory, Bechtolsheimer Weg 26, D-55278 Friesenheim, Germany\\
$^{11}$Observadores de Supernovas (ObSN)$^{\thanks{\tt \href{https://www.obsn.es/}{Observadores de Supernovas, https://www.obsn.es/}}}$, Spain\\
$^{12}$Observatorio de Sencelles, Sonfred Road 1, 07140 Sencelles, Mallorca, Spain\\
$^{13}$American Association of Variable Star Observers (AAVSO), 185 Alewife Brook Parkway, Suite 410, Cambridge, MA 02138, USA\\ 
$^{14}$Asociacion Astronomica Cruz del Norte, Calle Caceres 18, 28100 Alcobendas, Madrid, Spain\\
$^{15}$CBA Belgium Observatory \& CBA Extremadura Observatory, Walhostraat 1a, B-3401 Landen, Belgium\\
$^{16}$Science Department, Seville University, Av. de la Ciudad Jard\'{i}n, 20-22, 41005 Sevilla, Spain\\
$^{17}$La Vara, Valdes Observatory( MPC~J38), Barrio La Bara, sin n\'{u}mero Mu\~{n}as de Arriba c\'{o}digo postal 33784, Asturias, Spain\\
$^{18}$R.P. Feynman Observatory, Piazzetta del Ges\'{u} 3, 73034, Gagliano del Capo, Italy\\
$^{19}$Center for Backyard Astrophysics (CBA), Piazzetta del Ges\'{u} 3, 73034, Gagliano del Capo,  Italy\\
$^{20}$Belako, Aritz Bidea No 8, 4B Mungia Bizkaia, Basque Country, Spain\\
$^{21}$Hosting Trevinca, Agrupaci\'{o}n Astron\'{o}mica Vizcaina, Bizkaiko Astronomia Elkartea AAVBAE.net, c/ Iparragirre 46, 4º - dpto 2, Bilbo, Basque Country, Spain\\
$^{22}$Estelia Observatory (MCP Y90), Lugar la Riba 91-A, Ladines, Sobrescobio, Asturias, 33993, Spain\\
$^{23}$Sociedad Malague\~{n}a de Astronom\'{i}a (SMA), Centro Cultural Jos\'{e} Mar\'{i}a Guti\'{e}rrez Romero, Cl Rep\'{u}blica Argentina, no 9, Urb. El Limonar, 29016 M\'{a}laga, Spain\\
$^{24}$Cal Maciarol Observatory, Cam\'{i} de l'Observatori S/N, 25691 \`{A}ger, Spain\\
$^{25}$Hampshire Astronomical Group, Hinton Manor Lane, Clanfield, PO8 0QR, UK\\
$^{26}$Royal Astronomical Society, Burlington House, Piccadilly, London W1J 0BQ, UK\\
$^{27}$Horndean Observatory, 6 Falcon Road, Horndean, Waterlooville, Hampshire, PO89BY, UK\\
$^{28}$Rolling Hills Observatory, Clermont, FL 34711, USA\\
$^{29}$Astronomical Institute, University of Wroc{\l}aw, ul. M. Kopernika 11, 51-622 Wroc{\l}aw, Poland\\
$^{30}$Astronomical Observatory, University of Warsaw, Al. Ujazdowskie 4, 00-478 Warsaw, Poland\\
$^{31}$Astronomical Society of Edinburgh, Edinburgh, UK\\
$^{32}$Wiltshire Astronomical Society, The Knoll, Lowden Hill, Chippenham SN15 2BT, UK\\
$^{33}$Bath Astronomers, 19 New King Street, Bath BA1 2BL, UK\\
$^{34}$The Herschel Society, The Herschel Museum of Astronomy, 19 New King Street, Bath BA1 2BL, UK\\
$^{35}$Volkssternwarte Paderborn e.V., Im Schlosspark 13, 33104 Paderborn, Germany \\
$^{36}$The Maury Lewin Astronomical Observatory, 420 N. Grand Ave., Glendora CA 91741, USA\\
$^{37}$Cork Astronomy Club, 15 Ashdale Park, S Douglas rd. Cork, Ireland\\
$^{38}$Department of Electronics, Electrical Engineering and Microelectronics, Silesian University of Technology, Akademicka 16, 44-100 Gliwice, Poland\\
$^{39}$Department of Telecommunications and Teleinformatics, Silesian University of Technology, Akademicka 16, 44-100 Gliwice, Poland\\
$^{40}$The Studios Observatory, Grantham NG31 8SE, UK\\
}

\date{Accepted XXX. Received YYY; in original form ZZZ}

\pubyear{2023}

\begin{document}
\label{firstpage}
\pagerange{\pageref{firstpage}--\pageref{lastpage}}
\maketitle

\begin{abstract} 
We present spot properties on 32 periodic young stellar objects in IC~5070. Long term, $\sim$5~yr, light curves in the $V$, $R$, and $I$-bands are obtained through the HOYS (Hunting Outbursting Young Stars) citizen science project. These are dissected into six months long slices, with 3 months oversampling, to measure 234 sets of amplitudes in all filters. We fit 180 of these with reliable spot solutions. Two thirds of spot solutions are cold spots, the lowest is 2150\,K below the stellar temperature. One third are warm spots that are above the stellar temperature by less than $\sim$2000\,K. Cold and warm spots have maximum surface coverage values of 40~percent, although only 16~percent of warm spots are above 20~percent surface coverage as opposed to 60~percent of the cold spots. Warm spots are most likely caused by a combination of plages and low density accretion columns, most common on objects without inner disc excess emission in $K-W2$. Five small hot spot solutions have $<3$~percent coverage and are 3000~--~5000\,K above the stellar temperature. These are attributed to accretion, and four of them occur on the same object. The majority of our objects are likely to be accreting. However, we observe very few accretion hot spots as either the accretion is not stable on our timescale or the photometry is dominated by other features. We do not identify cyclical spot behaviour on the targets. We additionally identify and discuss a number of objects that have interesting amplitudes, phase changes, or spot properties.
\end{abstract}




\section{Introduction}

Young Stellar Objects (YSOs) display variability on a wide range of timescales. Rotational modulation caused by surface spots occurs when spots rotate in and out of the observer's line of sight, on the timescale of the rotational period. The fast rotation of young stars \citep[see reviews by][]{2007prpl.conf..297H,2014prpl.conf..433B}, hence results in photometric variability on the order of days. The amplitude of variation produced by spots depends on the nature and the characteristics of the spot. 

Cold spots are regions on the photosphere, analogous to sun spots, that are hundreds of degrees cooler than the stellar temperature. These cause flux variations of generally less than ten percent \citep[e.g.][]{2009A&ARv..17..251S}. Cold spots are expected to be present during all stages of the pre-main sequence evolution of stars, whereas hot spots are thought to be exclusive to accreting classical T~Tauri (stage~2) stars with full or transition discs \citep{1994AJ....108.1906H}.

Hot spots are considered to be footprints of accretion with temperatures that are several thousand degrees above the stellar temperature and coverage less than a few percent of the stellar surface \citep{2016ARA&A..54..135H, 1998ApJ...492..743M}. Emission lines as an indicator of accretion require these high temperatures. However, recent surveys present a more diverse picture with low temperature contrast, high coverage regions \citep{2009MNRAS.398..873S, 2012MNRAS.419.1271S, 2016MNRAS.458.3118B}. \citet{2021Natur.597...41E} found low density accretion columns as a possible explanation for these spots, with coverage values up to 20 percent, while \citet{2023MNRAS.526.4885S} found warm regions spatially distinct from the accretion shock on EX~Lup and TW~Hya.

Multi-filter photometry was used to identify surface spots on T~Tauri stars in \citet{1989A&A...211...99B}, and has been used to study spot properties in many surveys since \citep[e.g.][]{1993A&A...272..176B, 1995A&A...299...89B, 2001AJ....121.3160C}.
The strong, active magnetic fields of YSOs cause large, complex structures and rotational variation is produced by asymmetric spot distributions. Photometry is therefore limited when compared to the resolution of methods such as Doppler Imaging \citep{2002AN....323..309S, 2003A&A...408.1103S, 2009ARA&A..47..333D, 2009MNRAS.399.1829S}. However, the comparatively low-cost equipment requirements for optical photometry allow for larger, longer surveys and better number statistics.

In \citet{2023MNRAS.520.5433H} we developed a flux replacement method to identify surface spots using $V$, $R$, and $I$ band data obtained from the Hunting Outbursting Young Stars (HOYS\footnote{\tt \href{https://hoys.space/}{https://hoys.space/}}) citizen science project from an 80~d high cadence section of light curves from objects in IC~5070. We identified 21 objects with cold spots, and six with warm spots with temperatures up to 3000\,K above the stellar temperature. In this work we will apply the same method to a sample of YSOs in the same region to characterise their spots over the entire duration of their light curves since the project started in 2014. This will allow us to track the evolution of the spot properties over time and greatly improve our statistics of the distribution of the typical spot temperatures and coverage values on YSOs. 

In this paper we detail our data in Sect.~\ref{sec: data}. In Sect.\ref{sec: method} we explain our method to divide the light curves, identify periodic objects and determine amplitudes. The results for the period analysis are presented in Sect.~\ref{sec: rotres}. In Sect.~\ref{sec: RESULTS} we discuss our results for the amplitude, phase, and spot property evolution. A number of objects with properties of particular interest are identified and discussed in Sect.~\ref{sec: funky objects}.

\section{Data }
\label{sec: data}
\subsection{HOYS data and photometry \label{subsec: HOYSdata}}

The photometry data for this work were obtained through the HOYS citizen science project \citep{2018MNRAS.478.5091F}. This uses a world wide network of $\approx$~100 amateur, university, and professional telescopes to conduct long term photometric monitoring of 25 nearby ($d < 1$\,kpc) young clusters and star forming regions in optical broad-band filters. The aim of this project is to obtain photometry in all filters for all target regions with a cadence of 12~--~24 hours over many years. 

The data supplied to the project are obtained from a range of instruments. The photometry is calibrated against reference images taken under photometric conditions in $u$, $B$, $V$, $R_C$, and $I_C$ filters. For simplicity we will refer to them as $U$, $B$, $V$, $R$, and $I$, hereafter. The calibration offsets from instrumental to apparent magnitude for these reference images have been obtained from the Cambridge Photometric Calibration Server\footnote{\tt \href{http://gsaweb.ast.cam.ac.uk/followup}{http://gsaweb.ast.cam.ac.uk/followup}}. To account for colour terms in the calibration due to instrumentation (e.g. slightly non-standard filters) or conditions (e.g. cirrus clouds), an internal calibration is applied to all data following \citet{2020MNRAS.493..184E}. We identify non-variable stars in each target region, and use their known magnitudes and colours to determine and correct for the colour terms in each image. This colour calibration also evaluates the photometry uncertainty for each data point and is applied to all HOYS data used in this work.

\subsection{IC~5070 cluster membership}\label{memberselection}

The star forming region IC\,5070 (Pelican Nebula) is part of the W\,80 \hii region alongside NGC\,7000 (North American Nebula). The entire region is frequently referred to as NAP for the combined region names. The NAP region was found to be at $\sim796 \pm 25$\,pc in \citet{2020ApJ...899..128K}, which identified 395 YSOs in six astrometric groups in the region based on Gaia~DR2. Almost all objects were found to be less than $3$\,Myr old, with a majority younger than $1$\,Myr. Using Gaia~DR2 as well, \citet{2020MNRAS.493..184E} determined a distance to the IC~5070 region of 870\,$^{+70}_{-55}$\,pc.

In this current work we have identified potential IC\,5070 cluster members using Gaia DR3 \citep{2016A&A...595A...1G,2023A&A...674A...1G} astrometry. This is part of a larger work characterising the YSO populations in all HOYS fields \citep{10.1093/mnras/stae311}. All Gaia sources within 0.6 degrees of the cluster centre were selected. Sources were removed that had parallax values of less than 0.3\,mas, a signal-to-noise ration (SNR) of the parallax of less than five, and that were fainter than 18 mag in the Gaia $G$-band. We further removed objects with colours indicative of white dwarfs and cataclysmic variables. Thus, sources with $BP-Gmag< -0.2$\,mag, and $Gmag-RP< 0.0$\,mag were excluded. Cluster members were initially manually identified through a histogram of the distances, using a bin width of 20\,pc. Stars in a distance range of 750\,pc to 900\,pc were selected. Within that selection, two groups of stars with a coherent proper motion were identified. For each group the median distance and proper motions and their root mean square (rms) scatter were determined. Candidate cluster members were then selected as all stars that are within three standard deviations around these median values. The potential cluster members were assigned to the group they were closest to in proper motion space. A summary of the astrometric properties of the two groups can be found in Table~\ref{gaiatable}.

\begin{table}
\caption{\label{gaiatable} Properties of the two astrometric groups of young stars in the IC~5070 region, as determined in \citet{10.1093/mnras/stae311}. The columns contain the following information: Group: The group name; $d$, $d^e$, $d^s$: median distance, standard error of the median, rms scatter from the median of the cluster members; $\mu_{\alpha / \delta}$, $\mu_{\alpha / \delta}^e$, $\mu_{\alpha / \delta}^s$: median proper motion in RA/DEC, standard error of the median, rms scatter from the median of the cluster members; N: Number of Gaia~DR3 selected potential members;
}
\centering
\setlength{\tabcolsep}{2.1pt}
\renewcommand{\arraystretch}{1.0}
\begin{tabular} {|c|ccc|ccc|ccc|c|}
\hline
Group & $d$ & $d^e$ & $d^s$ & $\mu_{\alpha}$ & $\mu_{\alpha}^e$ & $\mu_{\alpha}^s$ & $\mu_{\delta}$ & $\mu_{\delta}^e$ & $\mu_{\delta}^s$ & N \\ 
 & \multicolumn{3}{c|}{[pc]} & \multicolumn{3}{c|}{[mas/yr]} & \multicolumn{3}{c|}{[mas/yr]} & \\  \hline
a & 832.4 & 2.7 & 33.6 & -1.328 & 0.028 & 0.347 & -3.076 & 0.026 & 0.320 & 252 \\
b & 824.7 & 4.43 & 36.0 & -0.977 & 0.036 & 0.294 & -4.148 & 0.041 & 0.335 & 114 \\
\hline
\end{tabular}
\end{table}

There were a total of 366 potential cluster members based on the Gaia~DR3 astrometry in the two groups of IC\,5070. Of those 252 are in group a, the other 114 are in group b. Of these 366 cluster members, we have corresponding HOYS light curves for 131 (84, and 47 in groups a and b, respectively) with at least 100 photometry data points in each of the $V$, $R$, and $I$ filters. The majority of the YSOs with HOYS light curves identified in \citet{2021MNRAS.506.5989F} using Gaia~EDR3, were again identified as cluster members. Seven of the objects previously analysed are not included in our list. This is due to a parallax or proper motion value in Gaia~DR3 that is beyond the selection range applied. Two objects that were not previously investigated are now included in our sample.

65 of the 366 potential cluster members have recently been identified as H$_\alpha$ emission line stars in \citet{2023JApA...44...42P}. The overlap between the samples will be examined in more detail in Sect. \ref{sec: spotandstellar}. 

\section{Methodology}
\label{sec: method}

In this section we detail the data analysis used in the paper. To a large extend this work follows our previous work on period identification in HOYS data in \citet{2021MNRAS.506.5989F} and spot property determination from peak to peak amplitude measurements in \citet{2023MNRAS.520.5433H}. We detail the basic principles again below and highlight changes made to our established procedures for the purpose of this work.

\subsection{Sliced Light Curves }
\label{sec: slicey slice}

Our aim is to analyse the properties of surface spots over time. HOYS observations in this field begin in 2014, and the light curves were extracted from our database on October 21st, 2022. We have dissected ({\it sliced}) the $V, R$, and $I$ light curves into blocks of six months in length, every three months. Thus, consecutive slices of light curve data overlap by three months. The starting date for the first slice was chosen as Feb 14, 2014, 12 noon UT. The following slices are numbered consecutively by integers. The start date for the first slice (indexed as zero) is in the first year of HOYS data available for this field, and this day of the year corresponds to the time of worst observability for the IC\,5070 field, given its Right Ascension (i.e. it is at its lowest altitude at midnight). Thus, slices with numbers 3, 7, 11, etc. are centered on February, and hence usually contain the least amount of data. On the other hand, slices numbered 1, 5, 9, etc. are centred on mid August each year and thus generally have the most photometry data points.

For our analyses we require a minimum of photometry data points in each slice. We need at least 50 data points in at least two filters to determine the possible period (see Sect.~\ref{subsec: perdet}), and at least 50 data points in each filter to measure the peak-to-peak amplitudes (see Sect.~\ref{sec: peaktopeak}). Thus, the first slice with sufficient data for the majority of objects is centred on mid May 2018 (MJD\,=\,58162.5), i.e. slice number 16. The final slice with sufficient data in all filters is slice 33, therefore at most there are 17 slices with sufficient data for spot fitting. Hence, our light curves for the region are typically $\sim$5~yr in length. The observational cadence of an object, and thus the number of photometry data points available, is affected by its brightness and distance from the centre of the HOYS field. The majority of images in the HOYS database come from amateur telescopes with varying fields of view and limiting magnitude. Thus, generally fainter objects, further from the central part of the field have less data points available for analysis. Note that in all figures where properties are shown as function of time, the data points are plotted at the start date of their respective slice. 

\subsection{Period Identification}
\label{subsec: perdet}

Identifying periodic objects in IC\,5070 from HOYS data was the focus of study in \citet{2021MNRAS.506.5989F}. A double-blind study of period-finding algorithms was conducted on data from an 80 day period in the summer of 2018 (MJD ranging from 58329.5 to 58409.5). During this time, the IC\,5070 field was the focus of a HOYS observing campaign, leading to a high cadence data set. Photometry data from this period appears in our slices 17 and 18. The campaign was repeated every August in subsequent years, but for the majority of stars this time remains the part of the light curve data with the highest cadence. 

The \citet{2021MNRAS.506.5989F} study concluded that a combination of four period finding algorithms is optimal for completeness in identifying periodic objects in HOYS data. The four methods that were identified are: L1Boot, L1Beta, L2Beta, and the generalised Lomb-Scargle (GLS) periodogram. See the detailed description in \citet{2021MNRAS.506.5989F} and references therein for an in depth discussion of the specifics of each of the methods. The L1Boot, L1Beta, and L2Beta methods are run using the R module {\tt RobPer} \citep{thieler2016robper}. The GLS periodogram is based on \citet{1982ApJ...263..835S,2009A&A...496..577Z} and we are using its implementation in {\tt astropy}.

For the purpose of this work, we are making some small changes to the methodology applied in \citet{2021MNRAS.506.5989F}. First of all, in that work periodic light curves were identified for all stars in the IC\,5070 field. Here, we limit our search to the light curves of stars identified as potential cluster members based on Gaia~DR3 astrometry (see Sect.\,\ref{memberselection}). Furthermore, we search for significant periods in each of the light curve slices independently. This ensures that for objects which show periodic light curve modulation only at certain times, e.g. due to varying surface spot properties, a period is identified. Thus, a more complete sample of rotational variables of the region is obtained. We made some further minor adjustments to the selection of the best period, compared to \citet{2021MNRAS.506.5989F}, which are detailed below.

The four period finding algorithms (L1Boot, L1Beta, L2Beta, and GLS) were applied to the photometry in each slice and filter ($V$, $R$, and $I$), separately. Only slices which had at least 50 data points in two of these filters were considered. For all four period finding methods periodograms with 1000 test periods were obtained. These test periods were sampled homogeneously in frequency space between 0.5\,d and 20\,d. This range was chosen as most appropriate based on the typical cadence of our data and the periods identified in \citet{2021MNRAS.506.5989F}.

In each periodogram the highest peak was chosen as the candidate period. To ensure periods were real, only candidate periods with a peak periodogram power above 0.15 were considered (for L1Boot, L1Beta and L2Beta). This differs from \citet{2021MNRAS.506.5989F}, where a slightly higher threshold of 0.2 was used. The reduction in the detection threshold is justified by the larger amount of data available in the current analysis. Thus, there is more opportunity to identify and remove questionable periods (see below). The GLS equivalent to this peak periodogram power is the False Alarm Probability (FAP). We accepted all periods with a FAP below 0.1 as candidate periods. Despite the geographically distributed nature of the HOYS observations, the typical cadence in the data is approximately one day. Hence, candidate periods identified within one percent of one day were removed. We also removed any candidate periods that were within one percent of 0.5\,d and 20\,d as these are near the edges of the period parameter space we investigated. 

If in a given slice a period was detected matching the above criteria and in at least two filters with a separation of less than 0.05~d, then it was retained as a candidate period for this slice. This generated a list of such candidate periods for each object. In some objects these lists contained periods that were not all identical within the uncertainties. Thus we employed the following methodology to select the most likely correct period in all cases. A histogram of all candidate periods for each object was created with a bin-width of 0.1~d for the entire range of investigated periods (0.5 to 20~d). This resulted in groups of candidate periods in individual or immediately adjacent bins, and distinct groups being separated by one or more bins with no candidate periods within them. Typically these groups only span two adjacent bins (i.e. 0.2~d) and the maximum range was 0.4~d. In all cases, the scatter of the periods within each group was below ten percent of the median. 

For most objects one of these groups contained the vast majority of candidate periods. We visually inspected phase folded light curves determined with the median period for each group. We found that for phase plots from groups with a small number of periods no clear systematic variability could be visually identified. Thus, only groups with more than five candidate periods were kept. For most objects this left only one single group of candidate periods. In all other cases we selected a dominant group if it had more than 2/3rd of the candidate periods for this object. There were only two objects (18, 26) where this was not possible and two almost equal sized groups with different periods remained. In those cases we visually inspected the phase folded light curves and manually selected the most likely group with the most likely correct median period. In both cases the group with the shorter candidate periods was chosen.

Finally, again following the procedure in \citet{2021MNRAS.506.5989F}, we determined the final period for each object in the following way. For each filter, a simple Lomb-Scargle periodogram with 1000 homogeneously distributed test periods was determined for the entire light curve within ten percent of the median of the candidate periods in the dominant group. This resulted in a much higher period resolution. The peak in this periodogram was used as the most likely period of this object in each filter. The median of these periods from the $V$, $R$, and $I$ data has been adopted as the final period of the object. Similarly the standard deviation of these three values is used as the final period uncertainty for the object throughout our analysis. These values are listed in Table~\ref{tbl_pers} for all objects.

\subsection{Peak to peak amplitude identification }    
\label{sec: peaktopeak}

In order to analyse the spot property evolution we require peak to peak amplitudes in $V$, $R$, and $I$ over time. This basically follows the procedure established in \citet{2023MNRAS.520.5433H}. We also use the same notation as in that work. For each periodic object the photometry data slices that were created in Sec.~\ref{sec: slicey slice} were phase folded with the period determined in Sec.~\ref{subsec: perdet}. A minimum of 50 data points per slice and filter were required for this. The running median in the phase plot was calculated, using a smoothing over 0.1 in phase. The threshold of 50 data points is equivalent to a data point every 3.6 days in the slice to ensure a high signal to noise in the measured peak to peak amplitudes.

Following \citet{2023MNRAS.520.5433H} we denote the peak to peak amplitude as $\hat{A}_\lambda^o$, where `$o$' indicates the amplitudes as observed, and $\lambda$ refers to the filter used. It is determined as the difference between the maximum and minimum of the smoothed running median brightness in the phase plot. The associated uncertainties $\sigma \left( \hat{A}_\lambda^o \right)$ are determined as the standard error of the mean of the data used to calculate the maximum and minimum median. We also determine the phase position of the maximum of the phase plot in each filter. 

In \citet{2023MNRAS.520.5433H} we found that a signal to noise ratio (SNR) above three for the peak to peak amplitudes resulted in determined spot properties (spot temperature and coverage) with smaller uncertainties. It further also resulted in more consistent results when adding data from shorter wavelengths into the analysis. Therefore, we discarded all peak to peak amplitudes in slices in which one or more of them had a SNR below three. 

Furthermore, we introduce an additional selection criterion considering the phase of the folded light curve. All peak to peak amplitudes in a particular slice were discarded when the phase position of the maximum or minimum in the folded light curve scattered by more than 45 degrees between the three filters. This removed any slices where there was no consistent periodicity between filters. This mostly occurred where the SNR of the peak to peak amplitudes were just above the threshold of three. The phase of each slice shown later on, is calculated relative to a phase zero point. This has been calculated as the median phase value of the maxima in the phase folded light curves across all slices and filters.

\subsection{Spot property fitting}\label{spot_fitting}

In \citet{2023MNRAS.520.5433H} we established and tested our method for determining spot properties from $\hat{A}_\lambda^o$ values. We apply this method to every slice that contains over 50 data points in $V$, $R$, and $I$. We assume that the $\hat{A}_\lambda^o$ in each six-month slice is generated by a single, uniform spot with temperature $T_S$ on an otherwise un-spotted surface with temperature $T_\star$. We assume that the spot rotates completely in and out of the line of sight and therefore consider the spot to be `in-front' or `behind' of the star. Using a flux replacement model with PHOENIX synthetic spectra, we model for a given stellar temperature $T_\star$, $10^6$ spots homogeneously sampled in the spot temperature $T_S$ range of $2000\,\rm{K} \leq T_S \leq 10000\,\rm{K}$ and spot coverage $f$ range of the visible surface, $0 \leq f \leq 0.5$. The spectra are convolved with the $V$, $R$, and $I$ filter transmission curves, using the {\tt speclite.filters}\footnote{\tt \url{https://github.com/desihub/speclite/blob/master/speclite/filters.py}} package. The magnitude difference in each filter between the spotted and un-spotted surface is taken to be the modelled peak to peak amplitude, $\hat{A}_\lambda^m$. We follow the same notation laid out in \citet{2023MNRAS.520.5433H}. The amplitudes are used together as a set $ \{ \hat{A}_V, \hat{A}_{R}, \hat{A}_{I} \}$, and we define that subscript $\{ V \}$ denotes the shortest wavelength filter and subsequent filters. In this way, $ \hat{A}_{\{V \}} = \{\hat{A}_V, \hat{A}_{R}, \hat{A}_{I}\}$, which is used throughout our analysis.

The best fitting model is established by finding the minimum $RMS$ between the modelled amplitude sets $\hat{A}^m_{\{V \}}$ and the observed amplitudes $\hat{A}^o_{\{V \}}$. This single result will potentially be sensitive to small perturbations in the input peak to peak amplitudes. Thus, the $\hat{A}^o_{\{V \}}$ values are varied within their associated uncertainties for 10000 iterations, and the best fitting models are found for each of them. In the majority of cases there are cold spot and hot spot solutions for the $\hat{A}_{\{V \}}^o$ variations. To remove cases where there is ambiguity between the solutions, we require at least 60$\%$ of the 10000 variations to be either hot or cold spot solutions. If such a 60$\%$ majority is reached, the median and median absolute deviation (MAD) of the majority are taken as the spot properties and associated uncertainties. The ratio of hot spot to cold spot solutions HS:CS$_{\{V\}}$, is used as an indicator of reliability for the solution. 

The spot fitting requires a known stellar temperature for each star. These were taken, if available, from \citet{2020ApJ...904..146F}. The stellar temperatures are rounded to the nearest 50K for the spot fitting. The effect on the spot properties by altering the stellar temperature like this is always less than the statistical uncertainties, for details see \citet{2023MNRAS.520.5433H}. Where there is no stellar temperature reference available the median temperature of the astrometric group (a or b) the star belongs to (as discussed in Sec.~\ref{memberselection}) was used. This is functionally 4100\,K for group a, and 3950\,K for group b. In Table~\ref{tbl_pers} we identify the astrometric group each star belongs to. The exception to this is object 2, which is significantly brighter at 11\,mag in the RP Gaia band. This was identified previously in \citet{2023MNRAS.520.5433H} where an effective temperature of 5000\,K was used to account for its brightness.  In \citet{2023MNRAS.520.5433H} we test the effect of changing the stellar temperature with simulated spots. We found that when the stellar temperature was altered by $\pm200$\,K, the resultant spot properties varied within the statistical uncertainties. 

\begin{table*}

\caption{\label{tbl_pers} Target list of all YSOs investigated in this work. For each object we list our ID number, the J2000 coordinates, proper motions, and parallax from Gaia~DR3, the astrometric group it belong to, the effective temperature used for spot fitting, the determined period and its uncertainty, its $K-W2$ colour (discussed in Sect.~\ref{sec: spotandstellar}),  the Gaia~DR3 ID number, and the ID number of the object in \citet{2021MNRAS.506.5989F}. Objects marked with (*) have no effective temperature reference in \citet{2020ApJ...904..146F} and the median of all astrometric group members were used or 5000\,K for bright objects (see details in text).}
\centering
\setlength{\tabcolsep}{4pt}
\begin{tabular}{|c|cc|cc|c|c|c|cc|c|c|c|}
\hline

Object & RA & Dec & $\mu _\alpha$ & $\mu _\delta$  &Parallax& Group &T$_{\rm{eff}}$ & Period & Period error  & $K-W2$& Gaia DR3 ID & F21 ID \\
ID & \multicolumn{2}{c|}{[deg]} & \multicolumn{2}{c|}{[mas/yr]} & [mas]& & [K] & \multicolumn{2}{c|}{[d]} & [mag]& & \\
\hline
 1 & 312.87736 & 44.06249  & -2.125  & -3.563  &1.270 & a & 4250 & 3.165756 & 0.000050  & 0.817 & 2066866222797779072 & 9321  \\
 2 & 312.87063 & 44.07308  & -1.824  & -2.883  &1.212 & a & 5000* & 4.82799 & 0.00026  & 0.200 & 2066866287224242432 & 9267  \\
 3 & 312.78140 & 44.17626  & -1.870  & -3.665  &1.162 & a & 4100* & 3.53001 & 0.00013  & 0.441 & 2066870689567848192 & 8038  \\
 4 & 312.74312 & 44.24230  & -1.198  & -4.790  &1.308 & b & 4350 & 4.87960 & 0.00028  & 0.710 & 2067058740416252416 & 7422  \\
 5 & 312.74341 & 44.24562  & -1.643  & -3.794  &1.090 & a & 3700 & 3.42336 & 0.00028  & 0.789 & 2067058740416252544 & -  \\
 6 & 312.45490 & 44.17951  & -1.991  & -3.440  &1.209 & a & 4100* & 3.30975 & 0.00019  & 0.306 & 2067061042518587264 & 8025  \\
 7 & 313.09560 & 44.01580  & -1.460  & -3.895  &1.181 & b & 3950* & 3.63383 & 0.00055  & 0.374 & 2162929419847388544 & 9961  \\
 8 & 313.35736 & 44.17924  & -0.815  & -2.950  &1.217 & a & 4000 & 1.4533437 & 0.0000056  & 0.230 & 2162934986124720896 & 7954  \\
 9 & 313.35268 & 44.34278  & -1.226  & -3.110  &1.291 & a & 4000 & 2.779205 & 0.000058  & 0.390 & 2162941273956895872 & 6393  \\
10 & 313.09386 & 44.23338  & -1.234  & -3.571  &1.219 & a & 4300 & 3.00608 & 0.00023  & 0.910 & 2162944916089431168 & 7472  \\
11 & 312.82599 & 44.21893  & -1.046  & -3.998  &1.211 & b & 3950 & 7.87849 & 0.00090  & 1.292 & 2162947596149035648 & 7632  \\
12 & 312.92257 & 44.25196  & -1.278  & -3.644  &1.339 & b & 3800 & 6.7210 & 0.0018  & 1.527 & 2162949382855434496 & -  \\
13 & 312.94394 & 44.37256  & -1.743  & -3.242  &1.164 & a & 3950 & 2.175035 & 0.000017  & 0.261 & 2162950413647417728 & 6149  \\
14 & 313.14528 & 44.23346  & -1.826  & -3.406  &1.209 & a & 4200 & 10.63240 & 0.00042  & 0.245 & 2162950546789495552 & 7465  \\
15 & 313.14322 & 44.29450  & -1.116  & -3.690  &1.216 & b & 3900 & 7.27588 & 0.00066  & 1.508 & 2162951753677142784 & -  \\
16 & 313.07438 & 44.35441  & -1.463  & -3.256  &1.239 & a & 3950 & 3.211 & 0.069  & 0.071 & 2162955086571795712 & 6315  \\
17 & 313.46770 & 44.28486  & -1.437  & -3.392  &1.234 & a & 4950 & 7.3783 & 0.0076  & 0.274 & 2162960481050791936 & -  \\
18 & 313.42432 & 44.36014  & -0.934  & -3.183  &1.226 & a & 4900 & 1.44775 & 0.00013  & 0.174 & 2162964088823157632 & -  \\
20 & 312.75611 & 44.25956  & -1.208  & -3.744  &1.171 & b & 3900 & 5.523 & 0.051  & 1.093 & 2163135573981345280 & -  \\
21 & 312.75654 & 44.26166  & -1.232  & -3.811  &1.239 & b & 4000 & 7.35691 & 0.00057  & 1.633 & 2163135578281583744 & 7181  \\
22 & 312.81307 & 44.30488  & -1.019  & -4.277  &1.197 & b & 3950 & 4.22837 & 0.00032  & 1.382 & 2163136059317926016 & 6813  \\
23 & 312.71903 & 44.27889  & -1.215  & -4.409  &1.267 & b & 4300 & 8.311 & 0.089  & 1.451 & 2163136368555566848 & -  \\
24 & 312.74460 & 44.29188  & -0.746  & -4.149  &1.270 & b & 3900 & 7.2272 & 0.0029  & 1.124 & 2163136402915307136 & 6929  \\
25 & 312.77764 & 44.36131  & -1.538  & -2.702  &1.285 & a & 4750 & 1.39729 & 0.00013  & 0.198 & 2163137261908777472 & 6259  \\
26 & 312.84446 & 44.35210  & -1.062  & -4.050  &1.198 & b & 3950 & 3.882 & 0.021  & 1.176 & 2163137742945115136 & 6337  \\
27 & 312.81884 & 44.38277  & -1.213  & -2.641  &1.176 & a & 4300 & 2.4243735 & 0.0000087  & 0.485 & 2163138601938577024 & 6060  \\
28 & 312.69199 & 44.31941  & -1.146  & -4.597  &1.379 & b & 3500 & 13.690 & 0.262  & 1.200 & 2163139594070911360 & -  \\
29 & 312.93710 & 44.43861  & -1.235  & -3.185  &1.172 & a & 3950 & 3.77296 & 0.00011  & 0.217 & 2163144271295324544 & 5559  \\
30 & 313.10923 & 44.57394  & -1.501  & -2.871  &1.244 & a & 5500 & 1.43272 & 0.00013  & 0.234 & 2163146779556221952 & 4446  \\
31 & 312.84503 & 44.56183  & -2.157  & -3.099  &1.273 & a & 3950* & 4.96447 & 0.00090  & 1.023 & 2163148772421081728 & -  \\
32 & 312.72580 & 44.63561  & -2.087  & -2.686  &1.241& a & 3950 & 9.5466 & 0.0030  & 1.280& 2163156056685634944 & 3988  \\

\hline

19 & 313.36180 & 44.42429  & -1.326  & -2.740  &1.248
& a & 4100* & 0.6261 & 0.0034  & 0.259
& 2162965252757133056 & -  \\
\hline
\end{tabular}
\end{table*}

\section{Rotation Periods}
\label{sec: rotres}

\subsection{Rotation Period Distribution}

Our data set contains light curves in $V$, $R$, and $I$ for 131 
objects. When divided into the six-month slices, 126 
objects had sufficient data in one or more slices to search for a period. For 68 
objects we detected a period in at least one slice. Of those, 31 
objects meet the criteria laid out in Sect.~\ref{subsec: perdet} to determine a final period. We also include an additional object in our sample (19, separate in Table \ref{tbl_pers}) that was removed due to the vicinity of the object to a naked eye star (about three arcminutes away) and the unreliability of the photometry for such objects (see \citet{10.1093/mnras/stae311}). In the case of object 19, the photometry in the $V$ band is affected, but a good quality period was determined prior to removal from $R$ and $I$ band data, and so is included as part of our period discussion but is not followed up on in spot properties. 

From the 32 objects, 22 overlap with the sample published in \citet{2021MNRAS.506.5989F}. The objects where we could not determine a final period were also not included in \citet{2021MNRAS.506.5989F}, with the exception of V1701~Cyg. This source will be discussed in more detail in Sect.~\ref{subsec: object 4766}. The \citet{2021MNRAS.506.5989F} sample contained 40 YSOs. Objects in that work may not be included in our sample because of the change to Gaia~DR3 astrometry, or because we did not have sufficient available $V$ band data for the object. Our requirement for period searching was the availability of over 50 data points in at least two filters. Thus, we did not investigate objects that had little to no $V$-band coverage as they would not be suitable for our spot fitting analysis.

The final periods and their uncertainties are listed in Table~\ref{tbl_pers} together with other source properties. In this table we list our adopted source ID number (based on the sorted order of their Gaia~DR3 ID), the J2000 positions, proper motions, and parallax, the astrometric group they are a member of, the effective temperature used for the spot fitting (from \citet{2020ApJ...904..146F} and discussed in Sect.~\ref{spot_fitting}), the final periods and their uncertainties, their $K-W2$ colour (discussed in Sect. \ref{sec: spotandstellar}), their Gaia~DR3 ID, and the ID numbers for overlapping sources from \citet{2021MNRAS.506.5989F}. 

\begin{figure}
\centering
\includegraphics[width=1.01\columnwidth]{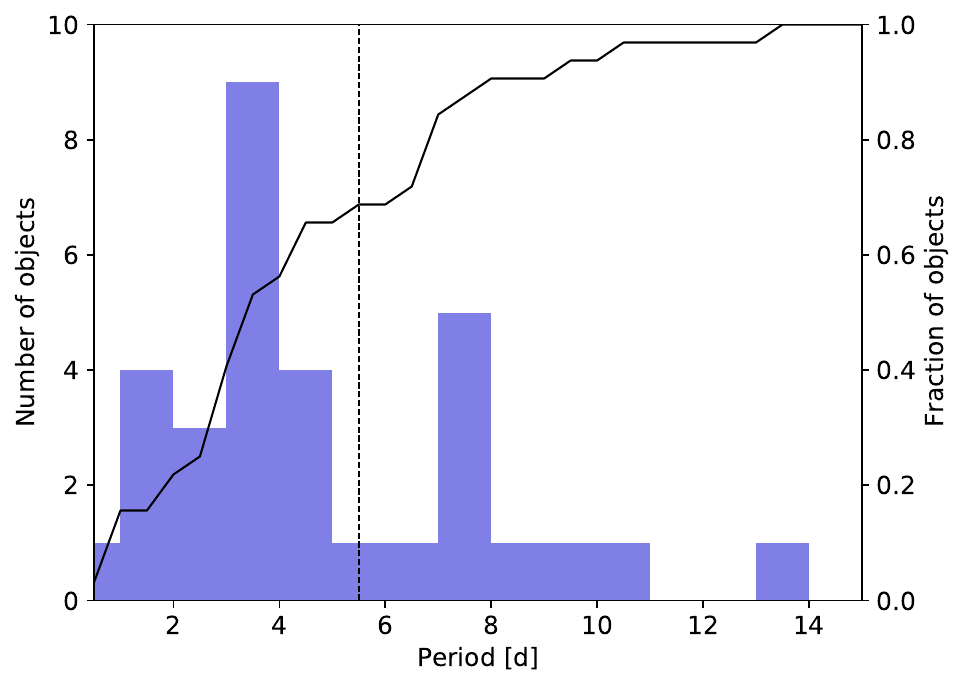} 
\caption{Distribution of measured periods of YSOs in our IC~5070 sample. The solid black line represents the cumulative distribution function. The dashed vertical line placed at a period of 5.5~d separates slow from fast rotators. \label{fig: pers}}
\end{figure}

We show the period distribution of our sample in Fig. \ref{fig: pers}. In blue a histogram of the periods, with 1\,d bins, is displayed. A cumulative distribution function (CDF) is over plotted as a solid black line. The vertical dashed line at $P = 5.5$\,d separates fast and slow rotators. We adopted the same period value for this as in \citet{2021MNRAS.506.5989F} for ease of comparison. The period distribution of our sample is bimodal. This bimodality has been identified in many other samples of YSOs, such as e.g. in ONC, NGC\,2264, and the Perseus molecular cloud \citep{2010A&A...515A..13R, 2007prpl.conf..297H, 2023RAA....23g5015W}. The cause of this bimodality is attributed to star-disc interactions via magnetic fields. Long rotation periods are hence indicating a disc slowing the rotation rate. Conversely, once the disc is dissipated or looses the magnetic connection to the star, the star begins to `spin up', leading to fast rotators. Of our objects, 23 are found to have $P < 5.5$\,d and 9 have $P > 5.5$\,d. The distribution roughly peaks at $\sim$3\,d and $\sim$7\,d, with a gap in-between. This is, within the statistical noise, comparable to \citet{2021MNRAS.506.5989F}, where we found 25 fast and 15 slow rotators. Thus, roughly two thirds of the YSOs in IC~5070 are fast rotators. In \citet{2021MNRAS.506.5989F} a gap in the period distribution in IC~5070 between five and six days was identified. Here we find that Object~20 is the first source that populates this gap with a period of $P = 5.52$\,d. This object is hence in the transition between the slow and fast rotators in its evolution. 

\subsection{Comparison with literature periods}

In Table~\ref{tbl_pers} we show the ID numbers from the \citet{2021MNRAS.506.5989F} study for the objects that were previously identified as periodic. The ten objects that do not have such an ID are hence newly identified rotational variables. They were investigated in \citet{2021MNRAS.506.5989F} but no period was found. Note that this earlier work focused only on an 80~d duration, high cadence part of the HOYS IC~5070 light curves in 2018. These data are roughly in the middle of our slice~17, but that slice contains all the data for the 6-month period from MJD~=~58162.5~$-$~58345.5. For five of the new periodic sources (object~5, 15, 17, 23, 28) we also do not find any period in slice~17. But they have been identified as periodic in other slices. Hence, most likely the amplitudes of their variability were too low during that time for the periodic signal to be detectable. For the remaining five sources (object~12, 18, 19, 20, 31) we do identify a period in the slice~17 data. This can have two reasons: i) The object has shown slightly stronger (higher amplitude) periodic variability outside the 80-day window investigated in \citet{2021MNRAS.506.5989F}, which is included in our slice~17 data. ii) Our slight lowering of the detection threshold in the periodograms (see Sect.~\ref{subsec: perdet}) and the ability to confirm these periods in many other slices.

The periods identified here do not all lie within the uncertainties given in \citet{2021MNRAS.506.5989F}. However, the median separation between our periods and the ones in \citet{2021MNRAS.506.5989F} is 0.27 percent, with a maximum of 1.45 percent. Thus, there are no significant changes in the determined periods between the two works. In our work we have used the entire light curve, and are hence able to verify the period with multiple slices. Therefore, the periods we identify here are more accurate/reliable.

\citet{2019A&A...627A.135B} also investigated variability of YSOs in IC\,5070. Six objects from their work overlap with objects previously identified in \citet{2021MNRAS.506.5989F}, where all but one agree on the period. Our work introduces three additional overlapping objects, namely objects~20, 23, and 28. Object~20 was identified in \citet{2019A&A...627A.135B} as variable but not periodic. Object~23 was identified in \citet{2019A&A...627A.135B} with a period of 8.446\,d, as opposed to our 8.3115~$\pm$~0.089\,d. The earlier value is slightly outside our uncertainty range. Given our longer data set and the identification of this period in seven slices, our period is more reliable. Most noteworthy is object~28 which was identified in \citet{2019A&A...627A.135B} with a period of 0.518\,d. This object, as shown in Table~\ref{tbl_pers}, has the longest period in our sample at $P = 13.690 \pm 0.262$\,d. It also has the highest absolute period uncertainty in the sample (object 16 has the highest relative period uncertainty). This is due to the best period measurement from the $I$ band periodogram being quite far from the results using the $R$ and $V$-Band periodograms (see Sect.~\ref{subsec: perdet}). The amplitudes in $V$ and $R$ for this source are much higher than for the $I$ band, which can explain this difference. Furthermore, in our period search we exclude periods within $1 \%$ of 0.5\,d, as these are on the edge of the investigated parameter space. In order to confirm we did not exclude a period close to 0.5\,d as in \citet{2019A&A...627A.135B}, we investigated the original identified candidate periods, before we made these exclusions. Object~28 had no other significant candidate periods outside of the primary group of 15 detections. That group had a median period of 13.620\,d, and a standard deviation of 0.098\,d. This implies that our period for this source is reliable and the nominal uncertainty is lower than the quoted $\pm 0.2618$\,d.


\begin{figure}
\centering
\includegraphics[width=0.97\columnwidth]{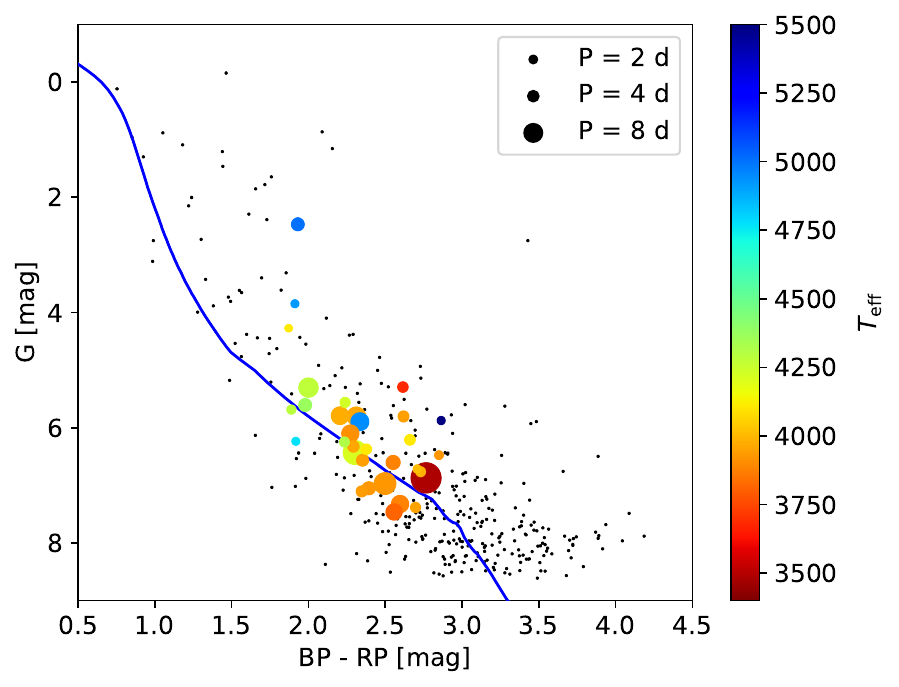} 
\caption{$BP - RP$ vs. $G$ colour magnitude diagram. All selected IC~5070 cluster members as shown as black points and periodic objects are in colour mapped to their effective temperature from \citet{2020ApJ...904..146F}, or the sample median. The marker size is scaled to the period, as shown in the legend. Overlaid in blue is a 1~Myr PARSEC isochrone \citep{2012MNRAS.427..127B}. \label{fig: cmd}}
\end{figure}

\subsection{Gaia Colour Magnitude Diagram}\label{subsec: gaia}

In Fig.~\ref{fig: cmd} we show the Gaia $BP - RP$ versus $G$ colour magnitude diagram (based on Gaia~DR3) for all 336 potential IC~5070 cluster members (groups a and b). The periodic objects in the sample are overlaid as coloured circles. The symbol size is proportional to the period and the colour indicates the $T_\mathrm{eff}$ used in the spot property determination (as in Table~\ref{tbl_pers}). The periodic variables show the same distribution as the other potential cluster members. We find no significant trend between period and colour. They cover a colour range of $1.8$~mag~$< BP - RP < 3.0$~mag, and have a lower limit of $G \sim$~17\,mag. This is due to fainter sources having more noisy HOYS photometry, not allowing the detection of low amplitude periodicity. The brightest source is object~2 at $G \sim$~12\,mag. As discussed in Sec.~\ref{sec: peaktopeak} we use a temperature of 5000\,K for the surface temperature of this source. 

\begin{figure}
\centering
\includegraphics[width=\columnwidth]{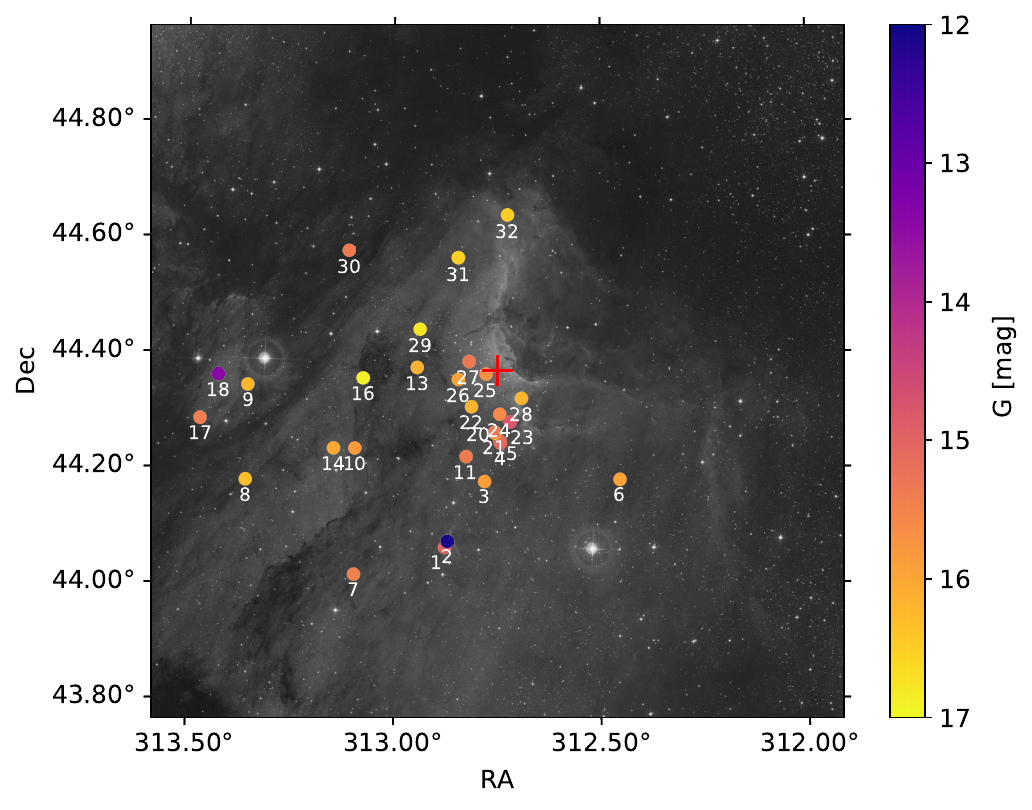}
\caption{DSS Red image, 1.2 degree radius from the centre point. Centre point marked with red cross. Objects in our sample are labelled and color scaled to the Gmag brightness.\label{fig: centrepoint} }
\end{figure}

Object~30 is the hottest object with a temperature of 5500~K. However, it is also the reddest with $BP - RP \sim 3.0$~mag. \citet{2020ApJ...904..146F} provide an extinction of $A_V = 5.0$mag for this source. In \citet{2023MNRAS.520.5433H} our sample contained three objects similar to this. However, in the updated Gaia~DR3 cluster members list, object~30 is the only object that remains. We consider the potential that the object is a contaminant, but it has $d = 803.5$\,pc, $\mu_\alpha = -1.501$, $\mu_\delta = -2.871 $ which places it within group a of the cluster. It has a short period with $P = 1.4327 \pm 0.0001$\,d, which indicates it has lost its connection with its disc. The $K-W2$ colour of 0.23~mag indicates a lack of inner disc material. No $W3 - W4$ colour is available for the object due to a poor signal to noise. 

A digitized sky survey (DSS) image of the region is shown in Fig.\ref{fig: centrepoint}. The positions of all identified periodic objects are over plotted. Object 30 is the only object positioned among one of the dark clouds in the North of the region. It is likely then that this object is strongly reddened by non-local extinction.  

\section{Results and Discussion of Spot Properties}
\label{sec: RESULTS}

This section will discuss the results from the peak to peak amplitude measurements and the spot property determination for our targets. Three of the 32 objects with periods determined from $R$ and $I$-band data have no fitted \vampset. Object 19 is removed to due to unreliable $V$ band photometry, as mentioned in Sect. \ref{sec: rotres}. The other two (12 and 15) are due to an insufficient number (less than 50) of data points in $V$.

\subsection{Object Result Summary Plots}
\label{sec: ampev}

\begin{figure*}
\centering
\includegraphics[width=\columnwidth]{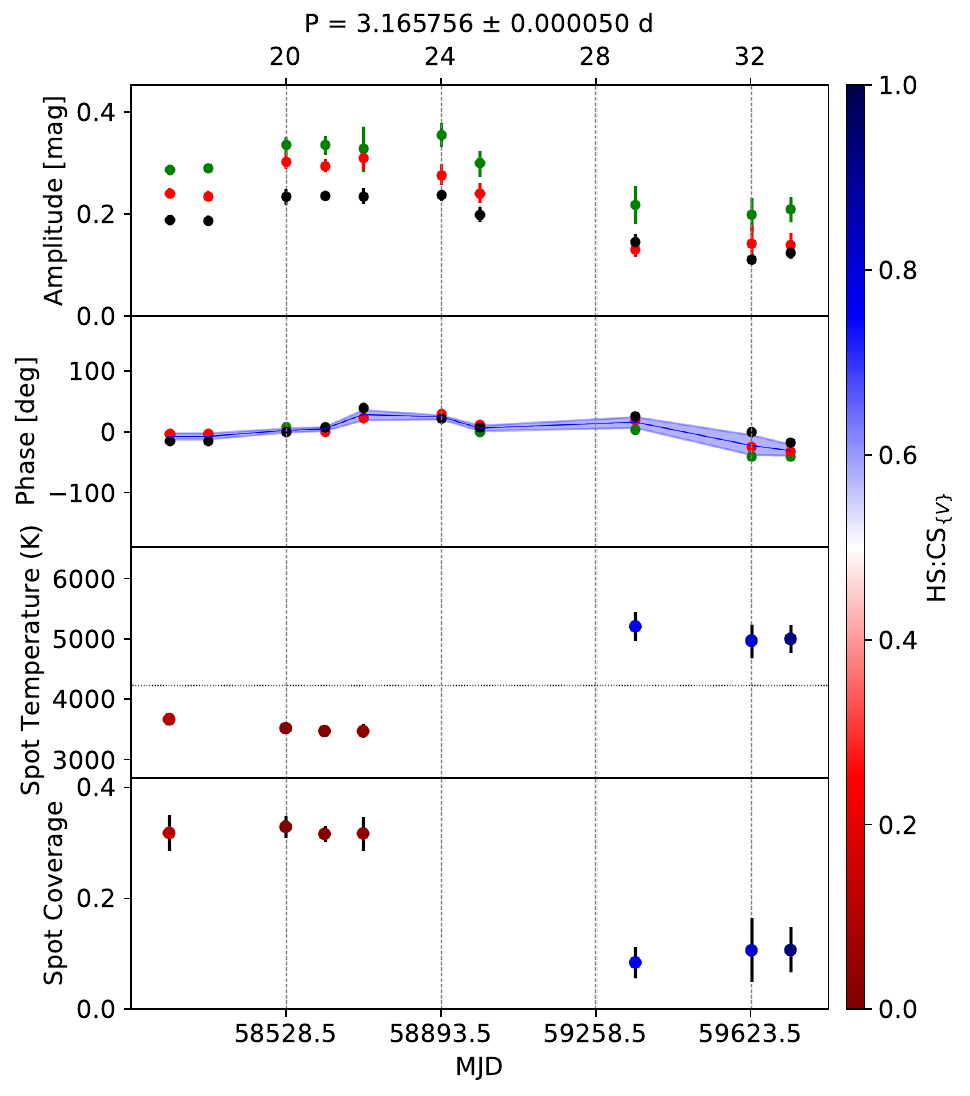} \hfill
\includegraphics[width=\columnwidth]{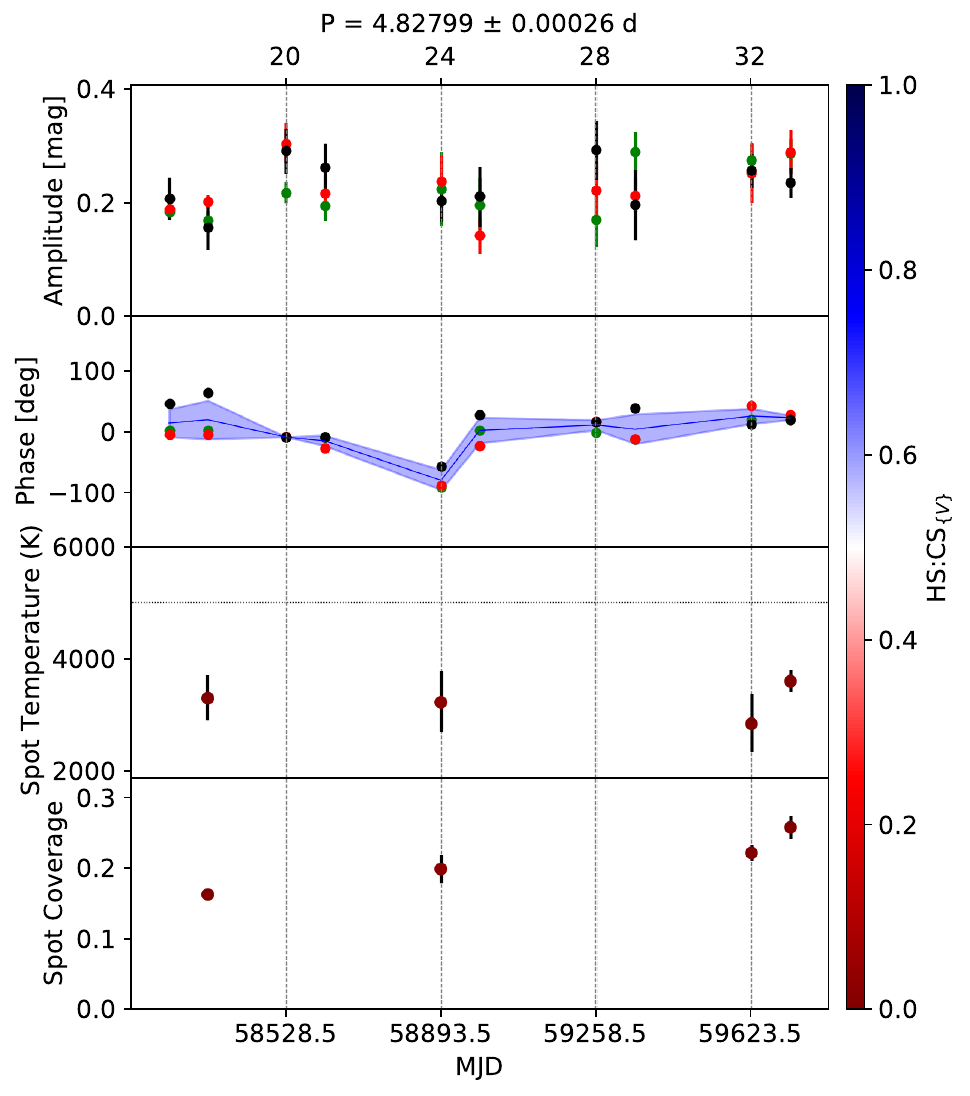} \\
\includegraphics[width=\columnwidth]{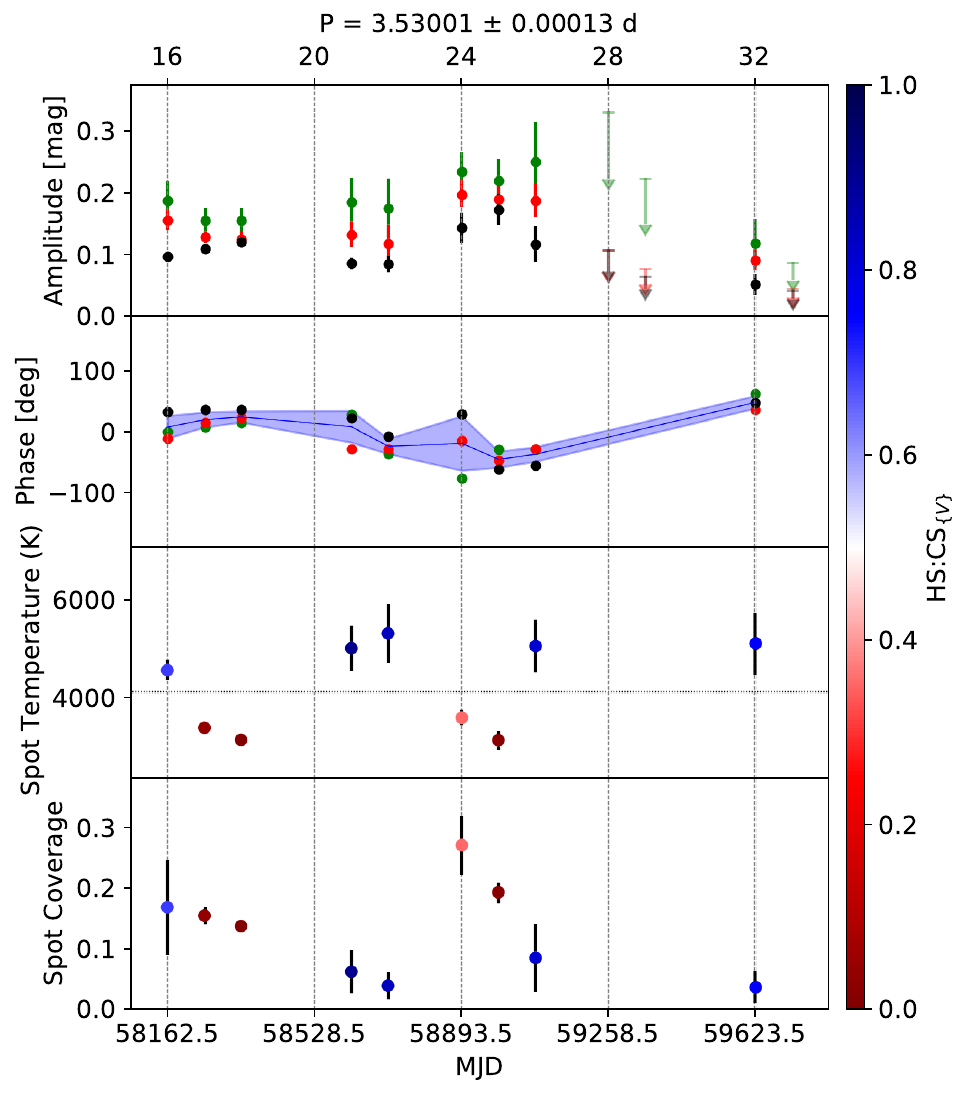} \hfill
\includegraphics[width=\columnwidth]{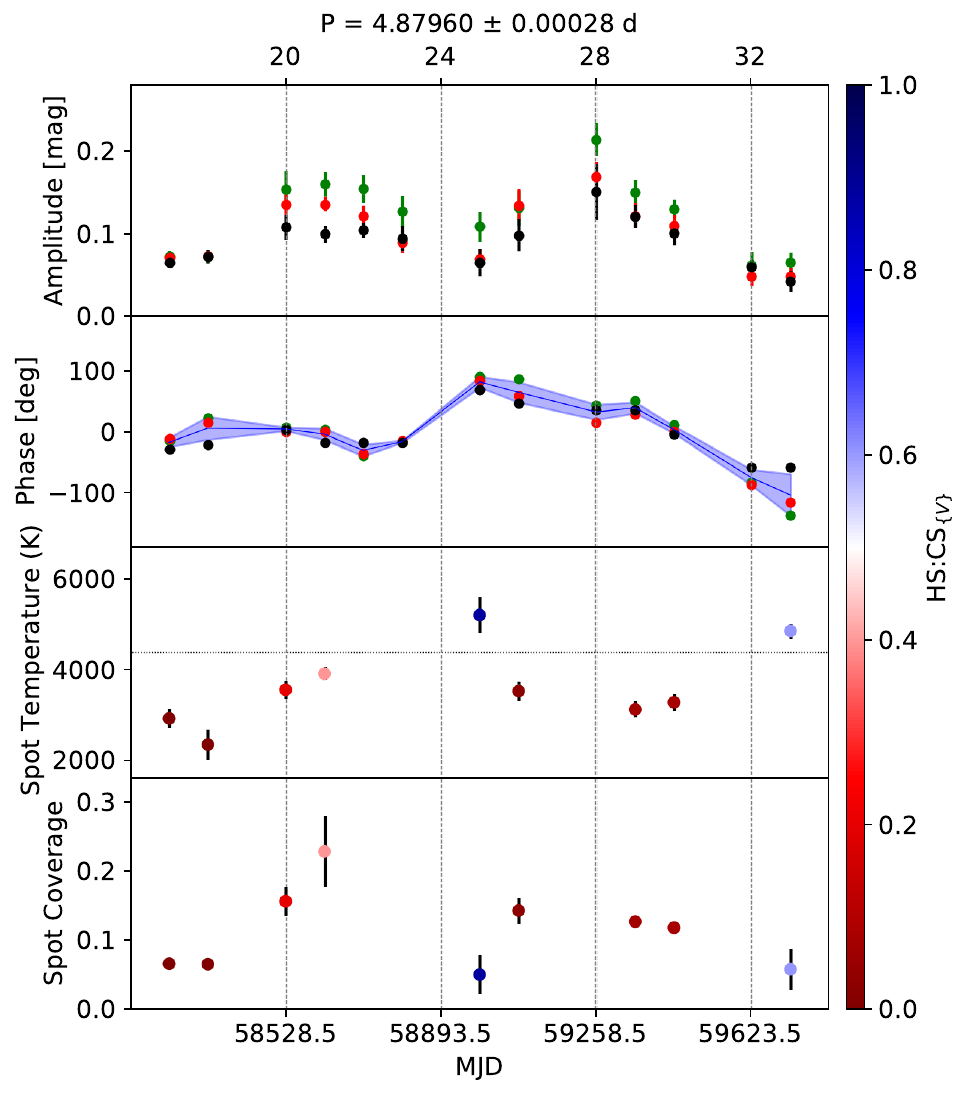} \\
\caption{Result summary plots for the first four objects in our sample. The data for object~1 are shown in the top left panel, object~2 in the top right panel, object~3 in the bottom left panel, and object~4 in the bottom right panel. In each of the summary plots we show from top to bottom the peak to peak amplitudes, the maximum brightness phase positions, the determined spot temperatures, and the spot coverage against time. For more details see the description in Sect.\,\ref{sec: ampev}. \label{fig: ev1}}
\end{figure*}

All measured and determined properties for our objects are summarised in result summary plots, the first of which, for object~1, is shown in the top left panel in Fig.~\ref{fig: ev1}. The plots for all other objects can either be found in the main body of the paper (1,2,3, and 4) if they are discussed in detail, or in the Appendix~\ref{evolfig}. In this section we describe the details of these plots.

The data points in each panel of these figures are plotted at the Modified Julian Date (MJD) of the start of each slice (bottom x-axis), and the slice number is shown at the top x-axis. Above that axis we also note the period for the object. The vertical gray dashed lines indicate one year intervals every start of February. Thus, the slices just left of the gray lines are centred on the times of worst observability of the field and hence usually contain the least amount of data (as discussed in Sect.~\ref{sec: slicey slice}).  

The first (top) panel in these result summary plots (as e.g. Fig.~\ref{fig: ev1}) shows the peak to peak amplitude $\hat{A}^o _V$ (green), $\hat{A}^o _R$ (red), and $\hat{A}^o _I$ (black) values as coloured dots and their uncertainties $\sigma \left( \hat{A}_\lambda^o \right)$. The second panel shows the phase position (in degrees) of the maximum brightness in the phase folded light curve in each filter with the same colour coding as in the top panel. The zero point for these phase values has been set as the median of all determined values for each object. The blue line indicates the running median of the phase positions over all filters in the slice and the light blue shaded area indicates the RMS scatter of the points from that running median. 

As laid out in Sect.~\ref{sec: peaktopeak}, we require a SNR above three for all amplitude measurements and an RMS scatter below 45 degrees for the minimum/maximum brightness phase positions. Thus, only data in slices where these criteria are satisfied are shown as coloured dots in the top two panels. In slices where one or more of the filters do not have a SNR above three for the peak to peak amplitudes, the 3$\sigma$ uncertainties are shown as light-coloured arrows instead and no phase positions are shown. The first example of this is object~3 (see bottom left panel in Fig.~\ref{fig: ev1}). In cases where the amplitudes have the required SNR but the phase position varies too much, the three sigma noise is displayed as light coloured crosses and again no phase positions are shown. This happens e.g. for object~7 (see Fig.~\ref{fig: ev7} in the Appendix). In all slices without any data shown, there are less than 50 data points in at least one of the filters.

The third and fourth (bottom) panels in the result summary plots show the determined spot properties and uncertainties for each slice, if the $\hat{A}_{\{V \}}$ amplitudes and phase positions have the required accuracy. The spot temperatures (in degrees Kelvin) are shown in the third panel. The dashed horizontal line indicates the surface temperature of the star (as listed in Table~\ref{tbl_pers}). The bottom panel shows the spot coverage. The displayed error bars represent the MAD uncertainties, as explained in Sect.~\ref{spot_fitting}. The colour coding of the symbols in the bottom two panels follows the scale shown at the right hand side of the figure. It represents the ratio of the hot spot to cold spot solutions (HS:CS$_{\{V\}}$, see Sect.~\ref{spot_fitting}) during the spot property fitting. Thus, spots that are mostly fitted as cold spots are shown in red, and spots that are mostly classified as hot spots are displayed in blue.

\subsection{Amplitude and Phase evolution}
\label{subsec: consofvar}

A quarter of our objects have \vampset\ values near  continuously from the start to the end of their light curve, only missing individual or pairs of \vampset\ due to insufficient data or low signal to noise. Two objects only contain a single slice with \vampset measurements. The median number of \vampset\ values is nine, and the maximum is 15. The primary reason for slices missing \vampset\ values is insufficient data. As mentioned in Sect.~\ref{sec: slicey slice}, slices centred on February contain the fewest data points, and as such these slices are often missing. A number of objects show longer breaks in \vampset\ values due to low signal to noise values and have hence the 3~$\sigma$ noise shown in their result summary plots. These sources are thus showing longer periods of no or undetectable spots on their surface.  We are unable to observe any activity cycles in our sample as this requires longer light curves.

Note that we are detecting the periodic variability due to rotational modulation over the $\sim180$\,d length of a light curve slice. Not showing \vampset values with our required signal to noise during a slice does not mean that the object has become quiescent. Rotational modulation is caused by asymmetry in the spot distribution. Thus, two similar surface features separated by 180$^\circ$ in longitude, i.e. on opposite `faces' of the star, will not produce a detectable rotational brightness modulation. Additionally, a significant change of the spot properties (temperature, coverage) or position (phase) during the 180\,d in a slice, will not allow us to detect the periodic brightness changes. Such slices will hence usually only have the noise shown in the result summary plots. 

Examining the phase position over the length of the light curves provides a further check of the quality of our period determination. In case there is a systematic, linear trend in the phase position over time, then this can be considered as an indicator of a small systematic error in the period of the object. None of our objects shows such a behaviour over the typically 4~--~5~yr length of the light curves. Furthermore, the phase position tracks the longitudinal position of the surface feature on the star. Thus, changes in phase can track the relative longitudinal position of the spot on the surface relative to the start of the light curve. These changes can either be caused by actual movement of the spots on the surface, or the appearance/disappearance of spots at different longitudes. The majority of objects show no significant phase shifts over the length of their light curve. There are some exceptions to this, and two examples are discussed in Sects.~\ref{disc-obj3} and \ref{disc-obj4}. An in depth study of the spot lifetimes and their appearance at different longitudes requires much longer datasets and hence is beyond the scope of this work.

\begin{figure*}
\centering
\includegraphics[width=0.97\columnwidth]{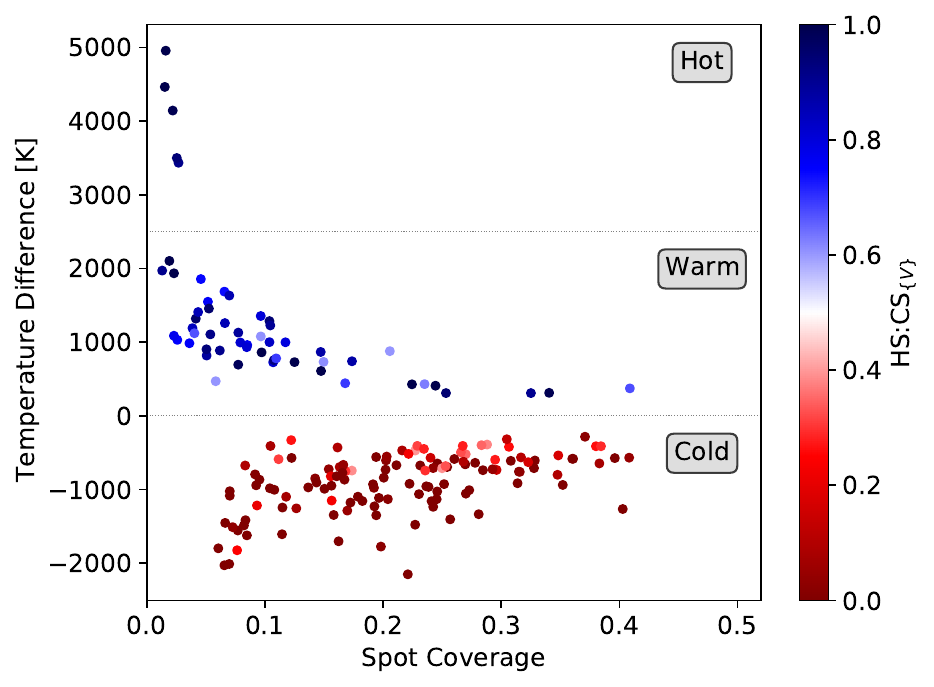} \hfill
\includegraphics[width=1.02\columnwidth]{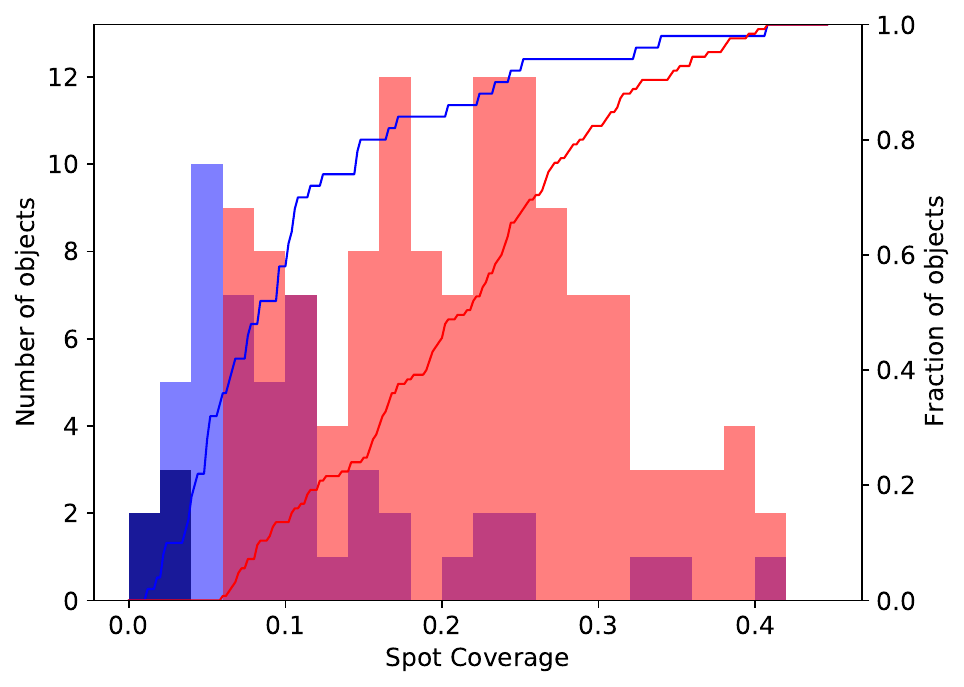}
\caption{  {\bf Left:}  Spot temperature difference $T_S - T_\star$ versus spot coverage. Horizontal dotted lines mark $T_S - T_\star = 0$~K and $T_S - T_\star = 2500$\,K, which separate the cold, warm and hot spot solutions. The markers are colour coded according to the HS:CS$_{\{V\}}$ ratio. {\bf Right:} Distribution of spot coverage of hot (dark blue), warm (blue), and cold spots (red). The solid lines are the CDFs of the warm (blue) and the cold spots (red). \label{fig: spot_prop}}
\end{figure*}

\subsection{Spot Property Distribution} \label{sec: spotev}

Our previous work \citep{2023MNRAS.520.5433H} investigated spots on YSOs in IC~5070 during an 80~d period based on periodic YSOs identified in \citet{2021MNRAS.506.5989F}. We identified 21 stars that had cold spots and six that had warm/hot spots.  Thus, the cold spot objects outnumbered the hot spot objects by a factor of 3.5. For about 13 percent (4 stars) of the periodic objects we could either not find a valuable spot solution (2 stars) or they are potential AA\,Tau-like contaminants (2 stars). 

We now consider the determined spot properties from the observed \vampset\ values for all slices in our periodic objects. Each of the 234 slices with high signal to noise peak to peak amplitudes is considered in the statistical analysis. We note, however, that neighbouring slices do overlap by 3 months. In those 234 amplitude sets we can reliably fit 180 as spots. Of these we identify 125 as cold spot solutions and 55 solutions are warm or hot spots. Thus, cold spot solutions outnumber the warm/hot spot solutions by a factor of about 2.3, slightly smaller than in \citet{2023MNRAS.520.5433H}. In the remaining 54 slices no reliable spot solution can be found (21) or the best solution is at the edge of the investigated parameter space (33). This fraction of about 23 percent is about twice as high as in \citet{2023MNRAS.520.5433H}. 

The difference between the spot temperature and stellar surface temperature $T_S - T_\star$ for all our solutions is plotted against the spot coverage in the left hand panel of Fig.~\ref{fig: spot_prop}. The colour coding of the symbols refers to the HS:CS$_{\{V\}}$ ratio. We see that there is a `gap' in the temperature difference distribution between about two and three thousand Kelvin. We hence refer to spots with a temperature difference above 0~K and below 2500~K as warm spots, and to all spots that have more than 2500~K temperature difference as hot spots. The differences between these groups will be discussed in Sect.~\ref{subsec: warmhot}. In the right hand panel of Fig.~\ref{fig: spot_prop} we show the distribution of the spot coverage for hot, warm, and cold spots as histograms, as well as cumulative distribution functions (CDFs) for the latter two. 

\subsubsection{Cold spots}

In \citet{2023MNRAS.520.5433H} we simulated periodic variability due to extinction from interstellar dust, and found that if we try to fit this as spots, these spots clustered at the lower temperature boundary of the PHOENIX models (2000~K), or at the $f = 0.5$ boundary for the coverage. We therefore remove spots that are fitted as implausibly large or cold, i.e. within 10 percent of $f = 0.5$ and $T_S = 2000$\,K. 27 spots were removed because of their temperature, all but two were within 2.5 percent of the lower temperature limit. Six further spots were removed for being too close to $f=0.5$. Although we adopted a ten percent margin near this value, all spots that were removed were within three percent of it. The largest spot remaining in our sample has a coverage of $f = 0.408$. The clustering around the parameter space boundaries indicates that the peak to peak amplitudes for these objects cannot be fit with our simple spot model. The remaining objects which are considered valid solutions for the spot properties are significantly away from the parameter space borders.

Among the cold spots 90 percent are between 0\,K and 1500\,K below the stellar surface temperature. The coldest spots with temperatures more than 1500\,K below the surface temperature, have a spot coverage of less than $f = 0.25$. We do not observe cold spots with a coverage above $f=  0.25$ and more than 1500\,K, below the surface temperature, which has several contributing factors. These objects likely have a complex network of surface features, and the model simplifies it to one uniform spot temperature and a fixed surface coverage. The coverage we calculate is the asymmetrical part of the spot distribution, and so it represents a lower limit. The distribution of the cold spot coverage values appears homogeneous from the lower limit of $f = 0.06$ to $f \sim 0.3$, where it tails off. A two sided KS-test indicates a probability of 34.4 percent that the observed coverage distribution in this range is homogeneous. This is well above the 5 percent threshold for rejection of the null hypothesis although not conclusively positive with limited statistics. 

Spots are modelled up to 0\,K temperature difference compared to the star and zero spot coverage. However, there is a clear observational bias just above and below $T_S - T_\star$ = 0\,K and for very small coverage values. There the temperature difference and/or spot coverage are not sufficient to produce a SNR above three for the peak to peak amplitudes. This is demonstrated in the left panel of Fig.~\ref{fig: spot_prop}, where for the cold spots the minimum coverage is $f =  0.061$ and the smallest temperature difference is 283~K below the stellar temperature. Thus, we are unable to detect objects above the upper envelope of points in the $T_S - T_\star$ vs. $f$ plot (Fig. \ref{fig: spot_prop}, left). The lack of objects at larger temperature differences and large coverage values is however real. We note that our simulations in \citet{2023MNRAS.520.5433H} have shown that there are small systematic effects in our methodology. These cause the spot properties near the upper envelope to be determined as slightly smaller and at a slightly larger temperature difference. However, these systematic shifts are always smaller than the MAD uncertainties for the spot temperature and coverage.

\subsubsection{Warm and hot spots}
\label{subsec: warmhot}

For spot solutions with temperatures above the surface temperature, the spot parameters cover a wide range of temperatures and sizes. 90 percent of these spot solutions have temperatures that are less than $\sim$2000\,K above the stellar temperature. We also find a dearth of spots between 2000\,K and 3000\,K above the stellar temperature. Due to this large apparent gap in the distribution, we use a spot temperature of 2500\,K above the stellar temperature to separate warm and hot spots, and discuss these separately. The number of warm/hot spots in the earlier sample from \citet{2023MNRAS.520.5433H} was very small. Only six warm/hot spot solutions were found. Five of these were warm spots and one object had a spot temperature of 2650~$\pm$~970~K, i.e was situated in our apparent gap of temperature differences. {bf In the future we intend to investigate the YSOs in all HOYS fields and will thus have a much better statistical picture if this apparent gap is indeed real.}

We identify five hot spot solutions that have a temperature difference of more than 3000\,K above the stellar temperature. The upper limit for the temperature in the spot models is 10000\,K. There is no single case of any spot solutions approaching this limit. This is most likely due to the simplicity of our model (it averages the spot temperature over the coverage) and the unavailability of any shorter wavelength (B and U-band) peak to peak amplitudes. All of the hot spots are very small, with a maximum coverage of $f = 0.027$. This is expected for accretion column footprints \citep[e.g.][]{2016ARA&A..54..135H, 1998ApJ...492..743M}. Four of the five hot spot solutions occur on a single object (30, see Fig.~\ref{fig: ev30}). We will discuss this star in more detail in Sect.~\ref{disc_obj30}. We currently have very limited statistics about the hot spot solutions. 

Similar to the cold spots, there is a bias against the detection of small warm spots with low temperature differences compared to the stellar temperature. The smallest temperature difference measured in the warm spot regime is 310\,K above the stellar temperature. However, due to the temperatures and filters involved, the regions of undetectable warm spots (defined by the lower envelope in the $T_S - T_\star$ vs. $f$ plot in the left panel of Fig.~\ref{fig: spot_prop}) are much smaller. Similarly we find again that higher temperature differences between spot and stellar surface usually correspond to lower a coverage. However, small a coverage does not necessarily correspond to a high temperature difference, as a number of warm spots also occupy the low temperature difference and coverage range. Furthermore, there is a much better defined upper envelope in the $T_S - T_\star$ vs. $f$ plot, above which objects do not seem to exist. 

Compared to the hot spot solutions, the warm spot solutions cover a wide range of sizes. The largest warm spot is as large as the largest cold spot. However, unlike the cold spots the distribution in coverage is not homogeneous. The histogram in the right hand panel in Fig.~\ref{fig: spot_prop} shows a clear peak at lower coverage values. We have performed a two-sided KS-test of the coverage values for the warm spots with coverage values between zero and 0.2. We find a probability of 43.3 percent that the distribution follows a normal distribution in this range. However, we note that there is a clear tail of much larger coverage values and that the low coverage values are influenced by our detection bias. There are eight warm spots (16 percent of total warm spots) that have a coverage above $f=0.2$, as opposed to a 60 percent of the cold spots. The largest warm spot in our previous sample from \citet{2023MNRAS.520.5433H} was $f=0.125$. 

\subsubsection{Rapidly changing spot properties \label{sec: spotjump}}

If one looks through the result summary plots for all objects (e.g. Fig.~\ref{fig: ev1} or in Appendix~\ref{evolfig}), one finds several examples where a spot solution changes from a warm to a cold spot (or the other way) in neighbouring slices. Usually these cases create single-slice warm spots on otherwise cold spot dominated objects. Note that there is not a single case of three consecutive slices where the middle one has a different classification (warm/cold) from the two others. Here we briefly discuss these changes in the classification from one slice to the next, and how they relate to changes in the HS:CS$_{\{V\}}$ ratio.

In our spot fitting methodology \citep[see Sect.~\ref{spot_fitting} and details in][]{2023MNRAS.520.5433H} we vary the measured \vampset\ values within their uncertainties 10000 times, to estimate the MAD errors for the spot properties, and to determine the HS:CS$_{\{V\}}$ ratio. This ratio indicates the fraction of all solutions that are hot/warm spots compared cold spots. We have introduced a somewhat arbitrary cut-off for these HS:CS$_{\{V\}}$ values between 0.4 and 0.6 for which we consider the spot fitting solutions ambiguous. Examining the relationship of the spot parameters (in particular the temperature) and HS:CS$_{\{V\}}$ over time, we find that in many occasions towards a switch from a cold to a warm spot or visa versa, the HS:CS$_{\{V\}}$ value does approach the boundary for ambiguous solutions.

\begin{figure*}
\centering
\includegraphics[width=\columnwidth]{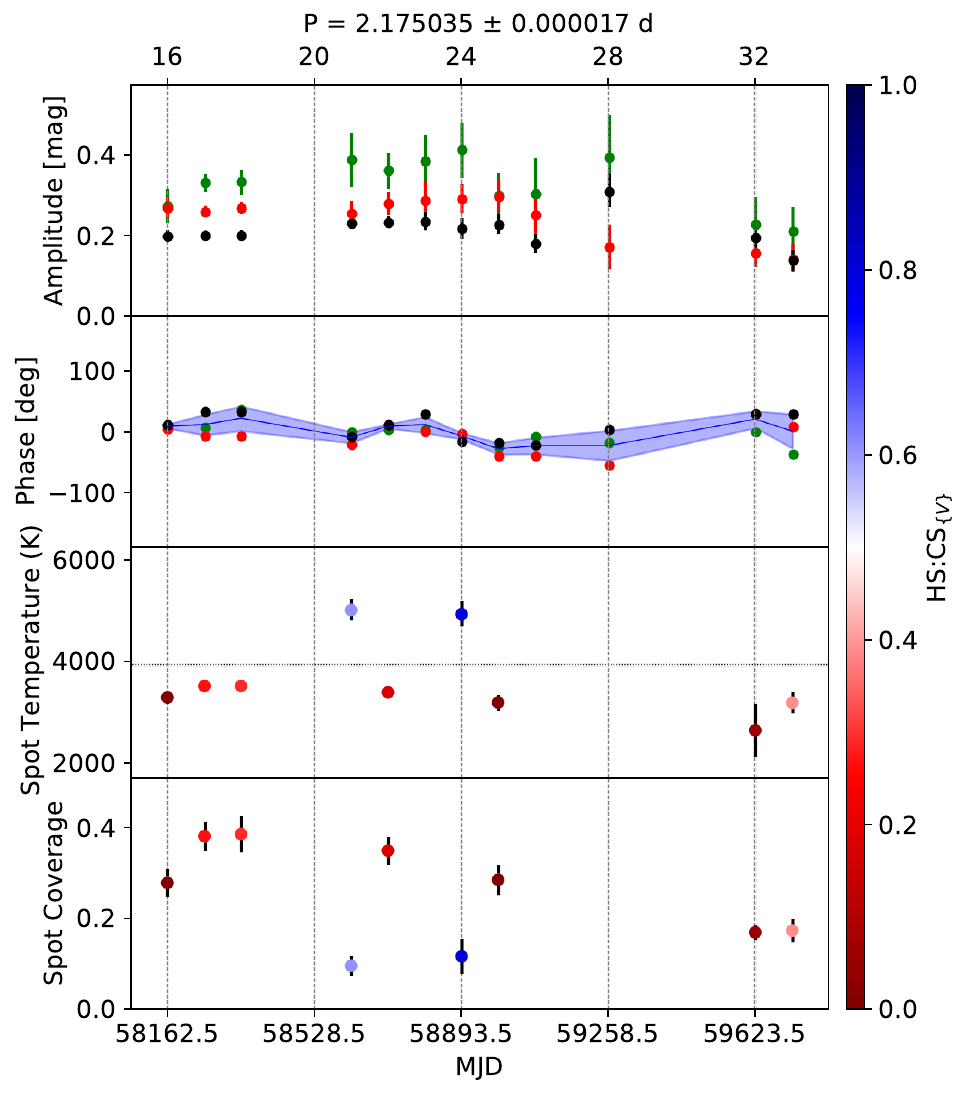} \hfill
\includegraphics[width=\columnwidth]{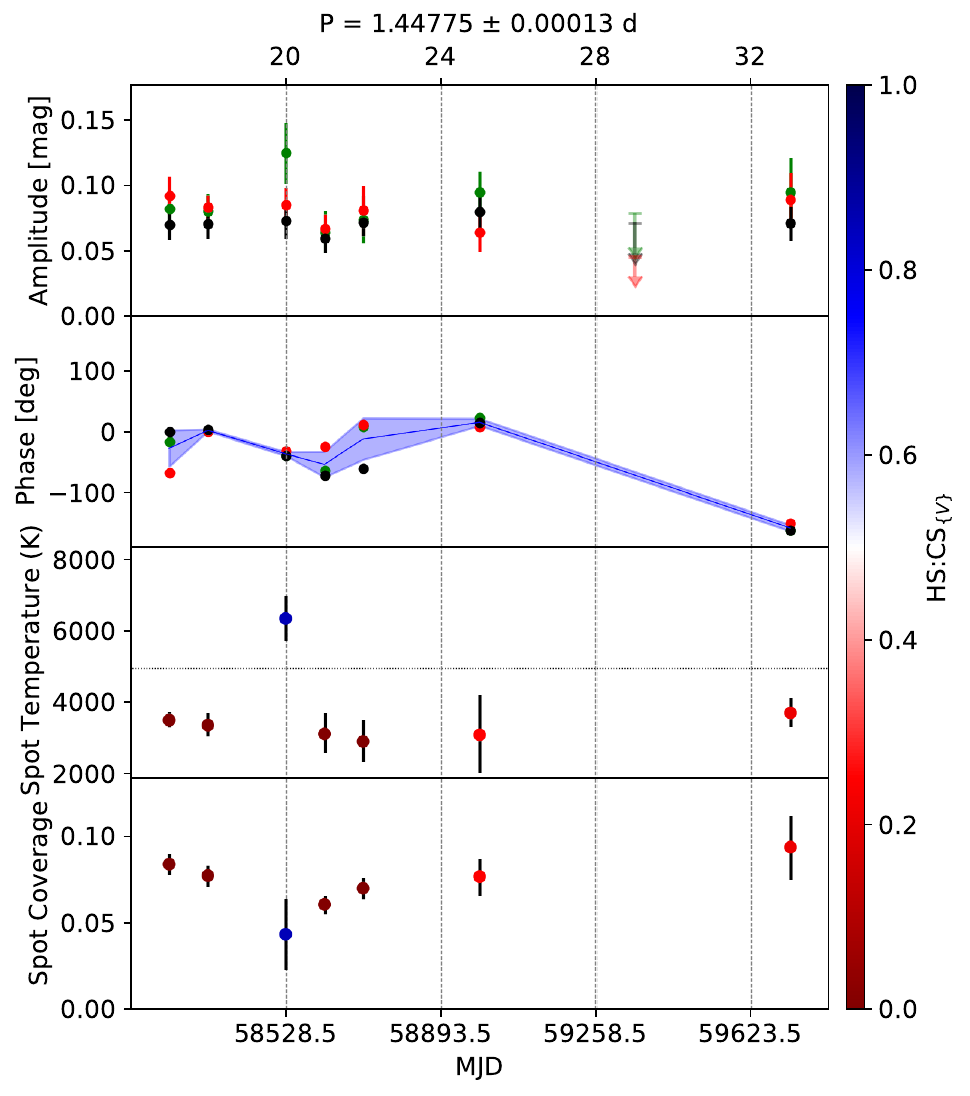} \\
\caption{As Fig.~\ref{fig: ev1}, but for the objects 13 (left) and 18 (right). \label{fig: ev2}}
\end{figure*}

In our rigorous testing of the spot fitting methodology in \citet{2023MNRAS.520.5433H} we have shown that a small hot spot can make a significant contribution to the peak to peak amplitudes. In the 180\,d long slices used here, the asymmetric part of the spot distribution on the stars can evolve. In particular contributions from accretion hot spots, caused by material flowing along magnetospheric structures which might be short lived compared to the duration of a slice \citep[e.g.][]{2012MNRAS.425.2948D}, need to be stable throughout the slice to create regular periodic light curves \citep{2014EPJWC..6404004K}, and hence allowing us to reliably determine their resulting spot properties. Thus, evolving spots will most likely lead to HS:CS$_{\{V\}}$ values changing or moving into the range where no clear solution can be found. Given the very small fraction (5/234, i.e. 2.1 percent) of hot spot solutions, we can conclude that it is highly unlikely that the footprints of the accretion columns are stable over timescales of half a year. We discuss object~30, which seems to be the exception to this in Sect.~\ref{disc-obj3}. On the other hand, the much larger fractions of reliable warm and cold spot solutions found indicate that the physical causes (see Sect.~\ref{sec: spotandstellar}) for these are typically stable for longer than half a year.
 
In Fig.~\ref{fig: ev2} we show two examples (object~13 left panel, and object~18 right panel) of these changes for the classification from a warm to a cold spot solution within the duration of one slice. Object 13 shows predominately cold spots, with exceptions in slices 21 and 24. The slices next to them, which contain overlapping data show cold spot solutions. The peak to peak amplitudes surrounding this pattern do not show any corresponding significant change. For slice 21, the HS:CS$_{\{V\}}$ ratio is 0.608, very close to our cut-off. The spot properties for slice 23 are ambiguous since the HS:CS$_{\{V\}}$ ratio is 0.423, and in slice 24 the ratio is 0.795. Thus, during this entire period the surface features on this objects probably evolve on timescales shorter than half a year and thus no clear, indisputable spot solution with our simple model can be found. Similarly, slice 20 in object~18 shows a brief change in classification (warm spot compared to cold spots throughout), accompanied by a HS:CS$_{\{V\}}$ ratio of 0.848. Note that in this slice the peak to peak amplitudes are clearly different, especially in the $V$-band, and that there is no data in the preceding slice 19. Thus, the statistical interpretation of the distribution of the warm and cold spot solutions, needs to take the respective HS:CS$_{\{V\}}$ ratio into consideration. In other words, the more lightly coloured spot solutions in the left panel of Fig.~\ref{fig: spot_prop} should be considered slightly less reliable than the darker coloured ones.

\subsection{Relation of Spot and Stellar Properties}\label{sec: spotandstellar}

In this section we briefly examine how the determined spot properties relate to the stellar properties. We reiterate that these objects are likely to exhibit multiple, potentially evolving, surface features at once. As discussed in Sect.~\ref{subsec: consofvar},  rotational modulation is caused by the asymmetric part of the spot distribution. In order to represent the spot properties of individual objects, we will refer to {\it dominant spot properties} throughout this section. These are defined as the proportion of reliable spot property solutions for an object, that are above (warm/hot) or below (cold) the stellar temperature. 

\subsubsection{Stellar Rotation Periods and Disc Excess Emission}
 
\begin{figure}
\centering
\includegraphics[width=\columnwidth] {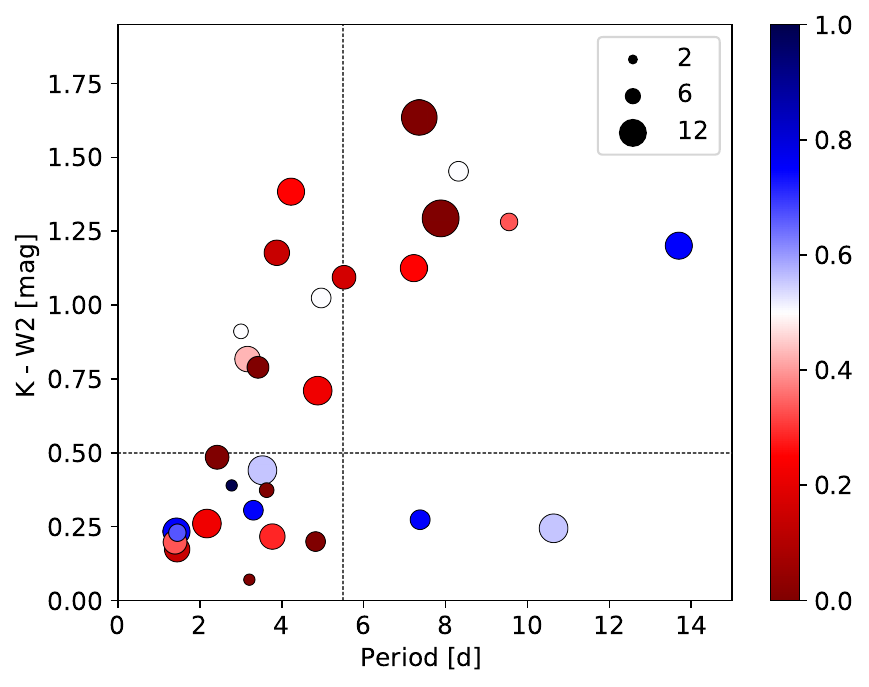}
\caption{Period against $K-W2$ for our objects. The horizontal dashed line separates objects with a detectable inner disc (above) from sources without (below). The vertical dashed line separates fast and slow rotators. Markers are coloured based on the proportion of warm/hot spot measurements on them, according to the colour scale on the right. Values of zero (dark red) indicate that all spot solutions are cold, values of one (dark blue) mean all spot solutions are warm/hot. The marker size is scaled to number of spot property measurements, as shown in the legend. \label{fig: domprop_kw2per}}
\end{figure}

In Fig.~\ref{fig: domprop_kw2per} we show how these dominant spot properties relate to the period and $K - W2$ colours for the objects. The marker colour in the figure indicates the proportion of the spot properties that are warm/hot spots. One (dark blue symbols) indicates that all reliable spot solutions are warm/hot and zero (dark red) means they are all cold. This is not to be confused with the ratio HS:CS$_{\{V\}}$, which is part of the result of the spot fitting process for an individual slice. There are three objects that show equal numbers of warm/hot and cold spots. The marker size in the plot is scaled to the number of spot measurements, i.e. larger circles represent sources with more reliable spot solutions. Note that this is not equal to the number of \vampset\ values, as some of them do result in ambiguous spot solutions for which no spot temperature and coverage can be reliably determined. 

In Fig.~\ref{fig: domprop_kw2per} we show a horizontal dashed line at $K - W2 = 0.5$~mag \citep{2012A&A...540A..83T} which separates sources with a detectable warm inner disc (above) and without (below). The $W2$ data is obtained in all cases from the WISE All-sky catalogue \citep{2013yCat.2328....0C}. For the K-band magnitude we generally use 2MASS \citep{2006AJ....131.1163S}, with the exception of object~17 for which this is not available. We utilise the UKIDSS GPS data \citep{2008MNRAS.391..136L} for this source. The absence of excess $K-W2$ emission typically indicates a lack of warm dust in the inner ($\sim$1~AU) disc. The vertical dashed line at 5.5~d in Fig.~\ref{fig: domprop_kw2per} separates fast (left) from slow (right) rotators. The exact value for this has been adopted for consistency with \citet{2021MNRAS.506.5989F}, who found a clear gap in the period distribution in IC~5070 at this value.

We find that 15 objects of our sample do not show a disc excess emission and 14 objects do. This disc fraction of about 50 percent is consistent with the value for this cluster determined in \citet{10.1093/mnras/stae311}, and also with the typical age of about 1~Myr of the objects in the region \citep{2020ApJ...899..128K}. Based on this age it is unlikely that the non-disc objects have evolved debris discs. It is more likely that they have cleared their inner discs due to the start of planet formation, and in later stages, photo-evaporation. The objects that have cleared their inner disc are known as transition disc sources. An alternative for the non-disc objects is that they retain the inner disc edge, but the disc contains a cavity in the inner part. These are referred to as pre-transition discs.

Observations at longer wavelengths can be used to probe the presence of disc material further from the star. Unfortunately  almost half of our objects do not have a detection with a sufficient SNR in the W3 and W4 bands to characterise the cooler material of the outer disc. There are 17 objects, however, that have a better than 5$\sigma$ detection in all four WISE filters. We determine the slope of the spectral energy distribution ($\alpha_{SED}$) following  \citep{2013Ap&SS.344..175M} for all of these. We find that amongst the 17 objects, one is classified as a Class~1 protostar (28), 15 are Class~2 objects, and one (7) is a Class~3 source. Thus, if these objects are representative of the entire sample, we find that 88 percent of our objects are Class~2 sources, in good agreement with the typical age of the stars in the region. 

The bimodal period distribution in our sample has been discussed in Sect.~\ref{sec: rotres}. A longer period may indicate that the central star still has magnetic ties to the inner disc, slowing rotation. Faster rotators have been spinning up due to a lack of this disc-breaking mechanism \citep{2007prpl.conf..297H}. The majority (20/29, i.e. 70 percent) of our objects are fast rotators. Of those, a majority of 13/20 (65 percent) show no detectable inner disc. There is a loose positive correlation of the rotation period and disc excess emission, with the exception of a few select objects (14 and 17), which have a longer period without inner disc emission. The very fast rotators ($P < 3$\,d) do not show evidence of an inner disc. 

\subsubsection{Causes of the Surface Spots}

In Fig.~\ref{fig: domprop_kw2per} we can see that the majority of our objects are dominated by cold spots. This is expected as 70 percent of our reliable spot solutions are cold. The presence of these cold spots on young stars is well understood as being caused by magnetic activity on the stellar surfaces, driven by the rotation of the stars (see reviews by \citet{2007prpl.conf..297H} and \citet{2014prpl.conf..433B}). These spots are known to have a range of temperatures and can reach up to half the visible surface \citep{1992ASPC...34...39S, 1995A&A...299...89B}. 

Our sample also includes eight warm/hot spot dominated objects. One of these consists of only a single warm spot measurement. We also find that objects with a higher number of reliable spot solutions (larger circles) are more likely to be dominated by cold spots. Investigating the result summary plots for all sources in our sample shows that cold spot dominated objects are more `consistent' in their behaviour. In other words they are more likely to show cold spot solutions throughout, or for longer continuous periods. On the other hand, warm/hot spot solutions on objects tend to be shorter lived. Only four individual objects in the sample contain four or more warm/hot spot solutions. Thus, a more reliable statistical analysis of these warm/hot spot dominated sources will require a larger sample.

All but one of the warm/hot spot dominated objects have $K-W2$ colour below 0.5~mag, indicating a lack of an inner disc. Can these mostly warm/hot spot sources be caused by accretion, despite the lack of a detectable inner disc? Accretion on such transitional disc objects has been the source of multiple studies. In similarly aged regions the proportion of accreting transition discs is 70~--~80 percent \citep{2009ApJ...700.1017K, 2010ApJ...712..925C, 2010ApJ...718.1200M}. \citet{2010ApJ...710..597S} found that when transition disc objects were accreting the median accretion rate was comparable to full discs. Similarly, \citet{2014A&A...568A..18M} studied transition disc objects in a similar age range to IC~5070 and found that among the transition disc objects that are still accreting, accretion occurs at a similar rates to full discs. The discs tended to be gas-rich close to the inner edge while being dust depleted. Although, \citet{2015MNRAS.450.3559N} found that when compared with stellar and disc mass, transition disc objects have lower accretion rates. Furthermore, eleven of our objects are identified in \citet{2023JApA...44...42P} as H$\alpha$ emission sources. Not all our sources are in the surveyed region. The slit-less spectroscopy data for that study was obtained in December 2012 and January 2013, and so unfortunately are not concurrent and the accretion rate may have varied since 2012/2013.

The above reinforces that many of our hot/warm spot dominated objects are still accreting despite their lack of detectable inner discs. Can the determined hot/warm spot properties be explained by accretion? Hot spots as the footprints of accretion are predicted to have temperatures of 8000-10000\,K and a small surface coverage of less than a few percent \citep{2016ARA&A..54..135H, 1998ApJ...492..743M}. Recent works have identified a wider range of lower temperatures and higher coverage values \citep{2009MNRAS.398..873S, 2012MNRAS.419.1271S, 2016MNRAS.458.3118B}. These lower limits of temperature do, however, not explain the lines identified with spectroscopy which require higher energies \citep{2003ApJ...592..266M, 2005ApJ...625..906M}. In \citet{2021Natur.597...41E} the dense line producing region has a coverage of 0.1 percent, but the cooler, extended region has a coverage in the range of 10-20 percent of the stellar surface. When the high density region disappears, the UV emission dramatically decreases but the optical peaks remain. Thus, a large fraction of our warm spot solutions could be explained by low density accretion columns.

Furthermore, \citet{2023MNRAS.526.4885S} identified stable accretion hot spots on EX~Lup and TW~Hya with emission line spectroscopy. The photometry of the objects was dominated by spots that are 500~--~2000\,K above the stellar surface temperatures. These spots were found to vary in longitudinal position distinctly from the line emission region. In the case of EX~Lup, which has a cleaner sinusoidal rotational modulation than TW~Hya, the warm spot was sometimes ahead and sometimes behind of the line emission hot spot. Over the course of 55 days the spot reduced in size from 88 percent to 10 percent coverage while simultaneously increasing in temperature from 4700\,K to 5200\,K. This behaviour is consistent with the rapidly changing spot properties discussed in Sect.~\ref{sec: spotjump}.

Accretion is not the only possibility for warm spots on young stars. \citet{2020MNRAS.497.4602F} identified warm spots on the active young star V5198~Cyg, an object which shows no evidence of accretion or mass transfer. Chromospheric active regions known as plages or faculae have been identified on stars of varying ages and types, from active main-sequence stars \citep{1993A&A...268..671S}, to late type stars \citep{2014NewA...32....1Z} and commonly on close binaries \citep{2008A&A...479..557F}. However, there is limited data or studies of these on young stars. The young solar type star HD~171488, aged 30~--~50~Myr, has been observed to have chromospheric plages and spots with a 20$^\circ$ difference in phase \citep{2010A&A...518A..48F}.  

Several studies have also identified photospheric warm spots on WTTSs \citep{2023MNRAS.520.3964X, 2017ApJ...836..200G, 2014MNRAS.444.3220D}. These are also referred to as plages but are distinct from chromospheric plages. 
\citet{2019A&A...625A..27S} observed warm spots alongside cool spots with opposing polarities, each with a coverage of a few percent, spatially distinct with longitudinal separations of 10$^\circ$ to 50$^\circ$. These are presumed to be caused by mass flow between the regions of opposing polarities. The warm regions in this case would always be accompanied by cool regions. In our methodology, a smaller warm region will dominate the photometric variability in the light curves. This scenario hence may provide an explanation of the behaviour seen in many of our objects (see Sect.~\ref{sec: spotjump}), where spot temperatures change (from warm to cold and vice versa) without detectable phase changes. 

Photospheric plages have been observed to have coverage values on the smaller side of our warm spot distribution. \citet{2017MNRAS.465.3343D} observes a coverage of six percent for warm spots but also clarifies that Zeeman Doppler Imaging constrains the lower boundary for coverage. A number of our low coverage warm spots may hence also be explained by photospheric plages. Low density accretion columns as in \citet{2021Natur.597...41E} have an upper limit of 20 percent coverage. However, the very large warm features observed on EX~Lup in \citet{2023MNRAS.526.4885S} reveal a possible explanation for the small number of larger ($f > 0.2$) warm spots in our distribution. Object~28, which accounts for five of these eight large warm spots, is discussed in Sect.~\ref{disc_obj28}.

\section{Discussion of Individual Objects}
\label{sec: funky objects}

\subsection{The variable star V1701~Cyg \label{subsec: object 4766}}

The variable star V1701~Cyg was identified in \citet{2021MNRAS.506.5989F} (object ID: 4766) with a period of 11.7162\,d. This period was not found again in our work. This source is the only object in the entire sample for which this occurred. Slice~17, which contains the majority of the 2018 high cadence data used in \citet{2021MNRAS.506.5989F}, resulted in a candidate period of 6.55\,d. We found five distinct candidate periods for this object in different slices. The most common being 6.55\,d, which was identified in two different slices. The additional four candidate periods were 7.67\,d, 9.69\,d, 12.61\,d, and 14.58\,d. 

The spot fitting in \citet{2023MNRAS.520.5433H} identified V1701~Cyg as a likely AA-Tau type contaminant. A $K-W2$ colour of 1.6~mag showed the presence of an inner disc, and the slope of the spectral energy distribution ($\alpha_{sed} = -0.62$) from WISE classified it as a Class~II object. It is therefore possible that the lack of consistent periods indicates dust occultations at varying distances from the central star over time. Hence, while not suited to our analysis here, this object is interesting and should be followed up in detail. 

\subsection{Very Cold Spots or Dust Extinction on Object~2}\label{disc_obj2}

Object~2 (see result summary in top right panel of Fig.~\ref{fig: ev1}) contains ten \vampset\ measurements. Yet only four reliable spot solutions are found. The remaining six slices were removed as solutions were within 10~K of 2000\,K. Four of these slices are `pairs' with overlapping data. The causes of these very cold spots are hence either dust extinction from the disc, or cold spots below the 2000\,K lower limit.  

This object was discussed in Sect.~\ref{subsec: gaia}, as the brightest object in the sample at G~$\approx$~12\,mag. The stellar temperature of 5000\,K was chosen to account for this, and yet the majority of the spot solutions are at a temperature difference of 3000\,K below stellar surface. We have considered whether an underestimation of the stellar temperature would lead to the spot solutions changing. In \citet{2023MNRAS.520.5433H} we tested the effect of altering stellar temperature by $\pm$200\,K, but this altered the spot temperatures by less than the statistical uncertainties. The hottest object in the sample is object~30 (see Sect.~\ref{disc_obj30}) which has a temperature of 5500\,K. It cannot be directly compared with object~2 as it undergoes extreme reddening due to non local extinction. While near a bright rimmed globule (see Fig.~\ref{fig: centrepoint}) it is not located within an obvious dark cloud. Even if the true temperature is 5500\,K, then this would still not effect the spot temperatures by more than the typical statistical uncertainties. Object 2 would therefore require a significantly higher temperature in order to increase the temperature of the resultant spot properties.

The object has $K - W2 = 0.20$\,mag and $\alpha_{SED} = -1.39$. This classifies it as a Class~2 object with a transitional disc. For dust to produce AA-Tau like variability it has to be close to the co-rotation radius, and the $K-W2$ value implies that this is not the case. Furthermore, we have shown in \citet{2023MNRAS.520.5433H} that dust extinction causing spot solutions to approach the lower temperature limit are caused by achromatic extinction. The \vampset\ values in object~2, however, are not achromatic. The highest filter with the highest amplitude varies throughout the light curve. In general, $\hat{A}^o_V$ is expected to be the highest and $\hat{A}^o_I$ the lowest. The light curve modulation is strong, with an amplitude (0.2~mag) that is higher than the median of our sample, but the modulation in colour is erratic.

The alternative solution is that the spots can indeed be colder than 2000\,K at times. In either case, the source warrants more detailed further investigation.

\subsection{Phase shifts on Object~3}\label{disc-obj3}

Object~3, shown in the bottom left panel in Fig.~\ref{fig: ev1} has $\sim$2.5~yr of detectable rotational variability at the start of the light curve. During that time the phase position remains constant within the measurement uncertainties. Note that the dominant spot temperature changes between warm and cold during that period. This is discussed in Sect.~\ref{sec: spotjump}. After slice 26 there is break in detectable \vampset\ values for about 18~months. Two slices in the break have sufficient data but the \vampset\ values have a SNR below three (due to high noise in the $V$ band). When the \vampset\ values resume, the amplitudes are reduced and there is a 90$^\circ$ shift in phase. The higher noise in the gap may be due to more complex spot activity. When the sinusoidal modulation resumes after the gap, a surface spot at a different part of the surface compared to the earlier dominating spot might have formed.  

\subsection{Potential Spot Motion on Object~4}\label{disc-obj4}

The evolution of object~4 is shown in the bottom right panel in Fig.~\ref{fig: ev1}. This source only has four gaps in the \vampset\ values, mostly in the slices centred on the worst observability of the source. Between slices 17 to 23 the amplitudes increase and decrease slowly, while showing a near constant phase position. This changes from slice 25 onward. A warm spot is found with a phase shift of 90$^\circ$ compared to the earlier cold spots. Together with the lack of successful spot property fitting for slices 22 and 23, this indicates a more complex situation of no clear single dominating spot for the duration of a slice (half a year). Between slices 25 and 34 the \vampset\ values increase and then decreases (similar to the first two years). During this time the phase position changes gradually. Again, in some slices no spot properties can be fitted, indicating a constant change in the nature of the dominating spot. However, the longitudinal position of the most asymmetric part of the spot distribution on the surface of the star is systematically moving at a rate of about 90$^\circ$ per year.

\subsection{Very Cold Spots on Object~5}\label{disc_obj5}

Similar to object~2 (Sect.~\ref{disc_obj2}) object~5 has eight spot temperature solutions too close to the lower 2000~K limit. Indeed these two objects account for nearly half of these cases. Object~5 has an effective temperature of 3686~K \citep{2020ApJ...904..146F}, $K - W2 = 0.79$~mag, and $\alpha_{sed} = -1.02$. Thus, it is a Class~2 object with an inner disc. The spot temperatures (albeit low) and coverage values for the reliable solutions do not stand out compared to the values found in the rest of the sample. Thus, similar to object~2, this source might have some of the lowest temperature spots and should be investigated in more detail.

\subsection{Large Warm Spots on Object~28}\label{disc_obj28}

Object~28 displays five of the eight warm spot measurements with coverage values over $f = 0.2$. It has the longest period of the sample at $P = 13.69$\,d and $K - W2 = 1.20$\,mag. It hence has an inner disc, and it is therefore likely that this object is accreting. However, it is not identified as an emission line star in \citet{2023JApA...44...42P}, but this does not exclude accretion as the data is not contemporary to our observations. It also has $\alpha_{sed} = 0.32$\,mag, making it the one Class~1 protostar in our sample. 

In \citet{2023MNRAS.520.5433H} we tested the outcome of simulated variability induced by dust with $R_V = 5$ and $R_V = 3.1$. This could in principle result in large warm spots. However, this can be excluded for this source as the colour modulation is not in agreement with such extinction. Object~28 has above average peak to peak amplitudes in our sample, with the largest amplitudes in the $R$ band. Coverage values above $f = 0.2$ are not expected from low density accretion columns \citep{2021Natur.597...41E}. Large warm regions on EX~Lup \citep{2023MNRAS.526.4885S} show that on a stably accreting object it is possible for the photometry to be dominated by large features with low temperature contrast. 

Object~28 is one of the two objects in our sample with three consecutive warm spot solutions in slices 21~--~23. Slice 24 has insufficient data, and then slices 25 and 26 have two further large warm spot solutions. This period of extreme stability implies large warm features as observed on EX~Lup. 

The stability seen on object 28 could partly be due to its early evolutionary stage. As pre-main sequence stars develop a radiative core their magnetic fields become increasingly complex, and their dipole field becomes weaker \citep{2012ApJ...755...97G, 2011MNRAS.417..472D}. This object is unlikely to have had sufficient time to develop a radiative core, therefore it is likely to host a strong, simple magnetic field. This supports the stability of the spot observed here. 

\subsection{Stable hot spots on Object~30}\label{disc_obj30}

Object~30 is discussed in Sect.~\ref{sec: rotres}, as the hottest star in our sample and with high reddening due to non-local extinction. Both, the period and $K - W2$ indicate that this object has no inner disc. Object~30 accounts for four of the five hot spot solutions in our sample. These spot solutions have a temperature of more than 3000\,K above the stellar temperature. The object has a short period of 1.43272\,d and  $K - W2 = 0.234$\,mag. It is not included in the emission line star catalogue of \citet{2023JApA...44...42P} as it is outside of their search area.

The spot solutions for the object start with two cool spots in slices 17 and 18. This correspond to the high cadence period of the light curve in 2018. The HS:CS$_{\{V\}}$ are 0.27 and 0.34 indicating these solutions are credible but not definitive on the side of cold spots. From slice 21 to the end of the light curve, every \vampset\ identified leads to a strong warm/hot spot classification, with the lowest HS:CS$_{\{V\}}$ being 0.75. Four of these spot solutions have temperatures above 8900~K and very small coverage values. These align with the expectations for accretion based hot spots.

There are two `pairs' of hot spots in slices 21/22 and 32/33, with overlapping data. There are two spots on this object which are not 2500\,K hotter than the stellar surface and are hence classified as warm spots. These have temperatures of 6700\,K and 7300\,K, which is not out of the range of accretion based activity. Hence, we can reasonably consider this object to be accreting, and all of the warm/hot spot solutions can be considered accretion based hot spots. 

\section{Conclusions}

\citet{10.1093/mnras/stae311} identified 366 YSOs in IC~5070 from Gaia~DR3 data. Of these, 131 have light curves with at least 100 data points in $V$, $R$, and $I$ from the HOYS citizen science project. We have sliced these light curves into six month long sections, every three months. In this way, half of each slice overlaps with the previous. We searched for a period within each individual slice that contained at least 50 data points in two of the three filters.

Following the method established in \citet{2023MNRAS.520.5433H} we have used the peak to peak amplitudes in $V$, $R$, and $I$ to measure spot properties. Out of the 234 amplitude sets, we find 180 reliable spot solutions. Amplitude sets that are not fit as spots are either near the edge of our parameter space, or create ambiguous solutions between cold and warm spots. The latter is most likely caused by complex spot structures, such as multiple spots with different temperature contrasts, which cannot be fit by our simple model. We are not able to observe any cyclical behaviour in our objects, as this requires longer light curves than currently available.

We measured a wide range of spot properties. Roughly two thirds of all reliable spot solutions are cold. We find the coverage values of cold spots are homogeneously distributed between our lower detection limit of 6 percent and 30 percent, with maximum coverage values of around 40 percent. 90 percent of the cold spot solutions are between 0\,K and 1500\,K below the surface temperature. The minimum temperature difference is $\sim$~280\,K and the maximum temperature difference is $\sim$~2150\,K. We do not consider spot solutions that are within ten percent of our lower temperature limit (2000\,K), as these measurements are potentially non-spot contamination. Alternatively, the spot temperatures are indeed below the lower temperature limit of our models. We discuss two objects (2, 5) that display the majority of these spot solutions and conclude that in these cases it is likely that the spot temperatures are indeed lower than we are able to fit. 

We find that a significant proportion (almost one third) of our spot solutions have temperatures above the stellar temperature, but by less than $\sim$2000\,K. We find no spot temperatures between $\sim$2000\,K and 3000\,K above the stellar temperature. Thus, we suggest a limit at 2500\,K above the stellar temperature, separating warm and hot spots. Warm spots have a temperature contrast between 0\,K and 2500\,K and hot spots are more than 2500\,K above the stellar temperature. 

Warm spots have a wide range of surface coverage values. The maximum coverage is around 40 percent, although 84 percent of warm spots have a coverage of less than 20 percent. The coolest warm spots have a temperature contrast of $\sim$~310\,K above the stellar temperature. Low temperature contrast, large coverage spots have been observed in other objects \citep{2009MNRAS.398..873S, 2012MNRAS.419.1271S, 2016MNRAS.458.3118B}. Accretion hot spots are expected to have temperatures of 8000~--~10000\,K and coverage values less than a few percent \citep{2016ARA&A..54..135H}. The line emission associated with accretion requires these high temperatures. However, recent works by \citet{2021Natur.597...41E} and \citet{2023MNRAS.526.4885S} have shown that the photometry of stably accreting objects is not necessarily dominated by the small, hot accretion shock region due to extended, low density accretion columns or spatially distinct warm regions. We show that these warm surface regions are common in our sample. $V$, $R$, and $I$ band data does not cover the peak of the spectral energy distribution for high temperature accretion shocks, and therefore our methodology is more sensitive to low temperature contrast warm spots.  

We find five hot spot solutions that are $\sim$3000\,K above the stellar temperature, with coverage values of less than three percent. These align with expectations for accretion column footprints. Four of these five hot spot solutions are on the same object (30), which is a YSO without a detectable inner disc (in $K-W2$). We consider this object to be actively accreting over the gap in the inner disc. The rarity (5/180, $\sim$3 percent) of such hot spot solutions shows that such accretion column footprints are either i) usually not stable on the timescale (6~months) at which we are able to investigate the spot properties, or ii) the photometry is not dominated by the accretion shock region as is discussed with the warm spot regime.  

Our entire YSO sample has an inner disc fraction of about 50 percent, based on the $K - W2$ infrared excess. The objects in our sample are dominated by cold spot variability, which is expected considering that approximately two thirds of our spot solutions are cold. We find that objects with more reliable spot solutions overall tend to have a higher proportion of cold spot solutions. Warm/hot spot dominated objects tend to have fewer reliable spot solutions. This implies that cold spots are longer lived, or more likely to produce rotational variability over longer timescales. Increasing our sample size will allow us to investigate this relationship. 

We find eight warm/hot spot dominated objects. Only one of these has a detectable inner disc (28). Accretion is common over an inner disc gap, although whether the rate is equivalent with full disc object is subject to debate, and so these warm/hot spots may be accretion related features. Alternatively, warm photospheric plages have been observed on WTTSs with coverage values of around six percent \citep{2017MNRAS.465.3343D}. The low coverage warm spots may be accretion related warm regions or plages, whereas the larger regions are more likely accretion related. The single warm/hot spot dominated object with a detectable inner disc is the only Class~1 protostar in our sample. It displays large warm spots in consecutive slices indicating extreme stability in the spot behaviour and therefore accretion.

In the future, we intend to apply our methodology to YSOs in the remaining HOYS fields, significantly increasing the number of investigated YSOs. This will allow for a much better statistical investigation of in particular the warm/hot spot solutions, the relationship between disc presence and spot properties, and the investigation of spot lifetimes.

\section*{Acknowledgements}

We would like to thank all contributors of observational data for their efforts towards the success of the Hunting Outbursting Young Stars project. 
CH is supported by the Science and Technology Facilities Council under grant number STFC Kent 2020 DTP ST/V50676X/1.
This research has been partially supported by the supercomputing infrastructure of the NLHPC (ECM-02) in Chile.  S. Vanaverbeke acknowledges the help of the IT support team of the NLHPC while working on the calculations described in this paper.


\section*{Data Availability Statement}

The data underlying this article are available in the HOYS database at http://astro.kent.ac.uk/HOYS-CAPS/.


\bibliographystyle{mnras}
\bibliography{bibliography} 

\begin{thebibliography}{}
\makeatletter
\relax
\def\mn@urlcharsother{\let\do\@makeother \do\$\do\&\do\#\do\^\do\_\do\%\do\~}
\def\mn@doi{\begingroup\mn@urlcharsother \@ifnextchar [ {\mn@doi@}
  {\mn@doi@[]}}
\def\mn@doi@[#1]#2{\def\@tempa{#1}\ifx\@tempa\@empty \href
  {http://dx.doi.org/#2} {doi:#2}\else \href {http://dx.doi.org/#2} {#1}\fi
  \endgroup}
\def\mn@eprint#1#2{\mn@eprint@#1:#2::\@nil}
\def\mn@eprint@arXiv#1{\href {http://arxiv.org/abs/#1} {{\tt arXiv:#1}}}
\def\mn@eprint@dblp#1{\href {http://dblp.uni-trier.de/rec/bibtex/#1.xml}
  {dblp:#1}}
\def\mn@eprint@#1:#2:#3:#4\@nil{\def\@tempa {#1}\def\@tempb {#2}\def\@tempc
  {#3}\ifx \@tempc \@empty \let \@tempc \@tempb \let \@tempb \@tempa \fi \ifx
  \@tempb \@empty \def\@tempb {arXiv}\fi \@ifundefined
  {mn@eprint@\@tempb}{\@tempb:\@tempc}{\expandafter \expandafter \csname
  mn@eprint@\@tempb\endcsname \expandafter{\@tempc}}}

\bibitem[\protect\citeauthoryear{{Bhardwaj}, {Panwar}, {Herczeg}, {Chen}  \&
  {Singh}}{{Bhardwaj} et~al.}{2019}]{2019A&A...627A.135B}
{Bhardwaj} A.,  {Panwar} N.,  {Herczeg} G.~J.,  {Chen} W.~P.,   {Singh} H.~P.,
  2019, \mn@doi [\aap] {10.1051/0004-6361/201935418}, \href
  {https://ui.adsabs.harvard.edu/abs/2019A&A...627A.135B} {627, A135}

\bibitem[\protect\citeauthoryear{{Bouvier} \& {Bertout}}{{Bouvier} \&
  {Bertout}}{1989}]{1989A&A...211...99B}
{Bouvier} J.,  {Bertout} C.,  1989, \aap, \href
  {https://ui.adsabs.harvard.edu/abs/1989A&A...211...99B} {211, 99}

\bibitem[\protect\citeauthoryear{{Bouvier}, {Cabrit}, {Fernandez}, {Martin}  \&
  {Matthews}}{{Bouvier} et~al.}{1993}]{1993A&A...272..176B}
{Bouvier} J.,  {Cabrit} S.,  {Fernandez} M.,  {Martin} E.~L.,   {Matthews}
  J.~M.,  1993, \aap, \href
  {https://ui.adsabs.harvard.edu/abs/1993A&A...272..176B} {272, 176}

\bibitem[\protect\citeauthoryear{{Bouvier}, {Covino}, {Kovo}, {Martin},
  {Matthews}, {Terranegra}  \& {Beck}}{{Bouvier}
  et~al.}{1995}]{1995A&A...299...89B}
{Bouvier} J.,  {Covino} E.,  {Kovo} O.,  {Martin} E.~L.,  {Matthews} J.~M.,
  {Terranegra} L.,   {Beck} S.~C.,  1995, \aap, \href
  {https://ui.adsabs.harvard.edu/abs/1995A&A...299...89B} {299, 89}

\bibitem[\protect\citeauthoryear{{Bouvier}, {Matt}, {Mohanty}, {Scholz},
  {Stassun}  \& {Zanni}}{{Bouvier} et~al.}{2014}]{2014prpl.conf..433B}
{Bouvier} J.,  {Matt} S.~P.,  {Mohanty} S.,  {Scholz} A.,  {Stassun} K.~G.,
  {Zanni} C.,  2014, in {Beuther} H.,  {Klessen} R.~S.,  {Dullemond} C.~P.,
  {Henning} T.,  eds, Protostars and Planets VI. p.~433 (\mn@eprint {arXiv}
  {1309.7851}), \mn@doi{10.2458/azu\_uapress\_9780816531240-ch019}

\bibitem[\protect\citeauthoryear{{Bozhinova}, {Scholz}  \&
  {Eisl{\"o}ffel}}{{Bozhinova} et~al.}{2016}]{2016MNRAS.458.3118B}
{Bozhinova} I.,  {Scholz} A.,   {Eisl{\"o}ffel} J.,  2016, \mn@doi [\mnras]
  {10.1093/mnras/stw455}, \href
  {https://ui.adsabs.harvard.edu/abs/2016MNRAS.458.3118B} {458, 3118}

\bibitem[\protect\citeauthoryear{{Bressan}, {Marigo}, {Girardi}, {Salasnich},
  {Dal Cero}, {Rubele}  \& {Nanni}}{{Bressan}
  et~al.}{2012}]{2012MNRAS.427..127B}
{Bressan} A.,  {Marigo} P.,  {Girardi} L.,  {Salasnich} B.,  {Dal Cero} C.,
  {Rubele} S.,   {Nanni} A.,  2012, \mn@doi [\mnras]
  {10.1111/j.1365-2966.2012.21948.x}, \href
  {https://ui.adsabs.harvard.edu/abs/2012MNRAS.427..127B} {427, 127}

\bibitem[\protect\citeauthoryear{{Carpenter}, {Hillenbrand}  \&
  {Skrutskie}}{{Carpenter} et~al.}{2001}]{2001AJ....121.3160C}
{Carpenter} J.~M.,  {Hillenbrand} L.~A.,   {Skrutskie} M.~F.,  2001, \mn@doi
  [\aj] {10.1086/321086}, \href
  {https://ui.adsabs.harvard.edu/abs/2001AJ....121.3160C} {121, 3160}

\bibitem[\protect\citeauthoryear{{Cieza} et~al.,}{{Cieza}
  et~al.}{2010}]{2010ApJ...712..925C}
{Cieza} L.~A.,  et~al., 2010, \mn@doi [\apj] {10.1088/0004-637X/712/2/925},
  \href {https://ui.adsabs.harvard.edu/abs/2010ApJ...712..925C} {712, 925}

\bibitem[\protect\citeauthoryear{{Cutri} \& {et al.}}{{Cutri} \& {et
  al.}}{2013}]{2013yCat.2328....0C}
{Cutri} R.~M.,  {et al.} 2013, VizieR Online Data Catalog, \href
  {http://adsabs.harvard.edu/abs/2013yCat.2328....0C} {2328}

\bibitem[\protect\citeauthoryear{{Donati} \& {Landstreet}}{{Donati} \&
  {Landstreet}}{2009}]{2009ARA&A..47..333D}
{Donati} J.~F.,  {Landstreet} J.~D.,  2009, \mn@doi [\araa]
  {10.1146/annurev-astro-082708-101833}, \href
  {https://ui.adsabs.harvard.edu/abs/2009ARA&A..47..333D} {47, 333}

\bibitem[\protect\citeauthoryear{{Donati} et~al.,}{{Donati}
  et~al.}{2011}]{2011MNRAS.417..472D}
{Donati} J.~F.,  et~al., 2011, \mn@doi [\mnras]
  {10.1111/j.1365-2966.2011.19288.x}, \href
  {https://ui.adsabs.harvard.edu/abs/2011MNRAS.417..472D} {417, 472}

\bibitem[\protect\citeauthoryear{{Donati} et~al.,}{{Donati}
  et~al.}{2012}]{2012MNRAS.425.2948D}
{Donati} J.~F.,  et~al., 2012, \mn@doi [\mnras]
  {10.1111/j.1365-2966.2012.21482.x}, \href
  {https://ui.adsabs.harvard.edu/abs/2012MNRAS.425.2948D} {425, 2948}

\bibitem[\protect\citeauthoryear{{Donati} et~al.,}{{Donati}
  et~al.}{2014}]{2014MNRAS.444.3220D}
{Donati} J.~F.,  et~al., 2014, \mn@doi [\mnras] {10.1093/mnras/stu1679}, \href
  {https://ui.adsabs.harvard.edu/abs/2014MNRAS.444.3220D} {444, 3220}

\bibitem[\protect\citeauthoryear{{Donati} et~al.,}{{Donati}
  et~al.}{2017}]{2017MNRAS.465.3343D}
{Donati} J.~F.,  et~al., 2017, \mn@doi [\mnras] {10.1093/mnras/stw2904}, \href
  {https://ui.adsabs.harvard.edu/abs/2017MNRAS.465.3343D} {465, 3343}

\bibitem[\protect\citeauthoryear{{Espaillat}, {Robinson}, {Romanova},
  {Thanathibodee}, {Wendeborn}, {Calvet}, {Reynolds}  \&
  {Muzerolle}}{{Espaillat} et~al.}{2021}]{2021Natur.597...41E}
{Espaillat} C.~C.,  {Robinson} C.~E.,  {Romanova} M.~M.,  {Thanathibodee} T.,
  {Wendeborn} J.,  {Calvet} N.,  {Reynolds} M.,   {Muzerolle} J.,  2021,
  \mn@doi [\nat] {10.1038/s41586-021-03751-5}, \href
  {https://ui.adsabs.harvard.edu/abs/2021Natur.597...41E} {597, 41}

\bibitem[\protect\citeauthoryear{{Evitts} et~al.,}{{Evitts}
  et~al.}{2020}]{2020MNRAS.493..184E}
{Evitts} J.~J.,  et~al., 2020, \mn@doi [\mnras] {10.1093/mnras/staa158}, \href
  {https://ui.adsabs.harvard.edu/abs/2020MNRAS.493..184E} {493, 184}

\bibitem[\protect\citeauthoryear{{Fang}, {Hillenbrand}, {Kim}, {Findeisen},
  {Herczeg}, {Carpenter}, {Rebull}  \& {Wang}}{{Fang}
  et~al.}{2020}]{2020ApJ...904..146F}
{Fang} M.,  {Hillenbrand} L.~A.,  {Kim} J.~S.,  {Findeisen} K.,  {Herczeg}
  G.~J.,  {Carpenter} J.~M.,  {Rebull} L.~M.,   {Wang} H.,  2020, \mn@doi
  [\apj] {10.3847/1538-4357/abba84}, \href
  {https://ui.adsabs.harvard.edu/abs/2020ApJ...904..146F} {904, 146}

\bibitem[\protect\citeauthoryear{{Frasca}, {Biazzo}, {Ta{\c{s}}}, {Evren}  \&
  {Lanzafame}}{{Frasca} et~al.}{2008}]{2008A&A...479..557F}
{Frasca} A.,  {Biazzo} K.,  {Ta{\c{s}}} G.,  {Evren} S.,   {Lanzafame} A.~C.,
  2008, \mn@doi [\aap] {10.1051/0004-6361:20077915}, \href
  {https://ui.adsabs.harvard.edu/abs/2008A&A...479..557F} {479, 557}

\bibitem[\protect\citeauthoryear{{Frasca}, {Biazzo}, {K{\H{o}}v{\'a}ri},
  {Marilli}  \& {{\c{C}}ak{\i}rl{\i}}}{{Frasca}
  et~al.}{2010}]{2010A&A...518A..48F}
{Frasca} A.,  {Biazzo} K.,  {K{\H{o}}v{\'a}ri} Z.,  {Marilli} E.,
  {{\c{C}}ak{\i}rl{\i}} {\"O}.,  2010, \mn@doi [\aap]
  {10.1051/0004-6361/201014460}, \href
  {https://ui.adsabs.harvard.edu/abs/2010A&A...518A..48F} {518, A48}

\bibitem[\protect\citeauthoryear{{Froebrich} et~al.,}{{Froebrich}
  et~al.}{2018}]{2018MNRAS.478.5091F}
{Froebrich} D.,  et~al., 2018, \mn@doi [\mnras] {10.1093/mnras/sty1350}, \href
  {https://ui.adsabs.harvard.edu/abs/2018MNRAS.478.5091F} {478, 5091}

\bibitem[\protect\citeauthoryear{{Froebrich}, {Scholz}, {Eisl{\"o}ffel}  \&
  {Stecklum}}{{Froebrich} et~al.}{2020}]{2020MNRAS.497.4602F}
{Froebrich} D.,  {Scholz} A.,  {Eisl{\"o}ffel} J.,   {Stecklum} B.,  2020,
  \mn@doi [\mnras] {10.1093/mnras/staa2275}, \href
  {https://ui.adsabs.harvard.edu/abs/2020MNRAS.497.4602F} {497, 4602}

\bibitem[\protect\citeauthoryear{{Froebrich} et~al.,}{{Froebrich}
  et~al.}{2021}]{2021MNRAS.506.5989F}
{Froebrich} D.,  et~al., 2021, \mn@doi [\mnras] {10.1093/mnras/stab2082}, \href
  {https://ui.adsabs.harvard.edu/abs/2021MNRAS.506.5989F} {506, 5989}

\bibitem[\protect\citeauthoryear{Froebrich et~al.,}{Froebrich
  et~al.}{2024}]{10.1093/mnras/stae311}
Froebrich D.,  et~al., 2024, \mn@doi [MNRAS] {10.1093/mnras/stae311}, 529, 1283

\bibitem[\protect\citeauthoryear{{Gaia Collaboration} et~al.,}{{Gaia
  Collaboration} et~al.}{2016}]{2016A&A...595A...1G}
{Gaia Collaboration} et~al., 2016, \mn@doi [\aap]
  {10.1051/0004-6361/201629272}, \href
  {https://ui.adsabs.harvard.edu/abs/2016A&A...595A...1G} {595, A1}

\bibitem[\protect\citeauthoryear{{Gaia Collaboration} et~al.,}{{Gaia
  Collaboration} et~al.}{2023}]{2023A&A...674A...1G}
{Gaia Collaboration} et~al., 2023, \mn@doi [\aap]
  {10.1051/0004-6361/202243940}, \href
  {https://ui.adsabs.harvard.edu/abs/2023A&A...674A...1G} {674, A1}

\bibitem[\protect\citeauthoryear{{Gregory}, {Donati}, {Morin}, {Hussain},
  {Mayne}, {Hillenbrand}  \& {Jardine}}{{Gregory}
  et~al.}{2012}]{2012ApJ...755...97G}
{Gregory} S.~G.,  {Donati} J.~F.,  {Morin} J.,  {Hussain} G.~A.~J.,  {Mayne}
  N.~J.,  {Hillenbrand} L.~A.,   {Jardine} M.,  2012, \mn@doi [\apj]
  {10.1088/0004-637X/755/2/97}, \href
  {https://ui.adsabs.harvard.edu/abs/2012ApJ...755...97G} {755, 97}

\bibitem[\protect\citeauthoryear{{Gully-Santiago} et~al.,}{{Gully-Santiago}
  et~al.}{2017}]{2017ApJ...836..200G}
{Gully-Santiago} M.~A.,  et~al., 2017, \mn@doi [\apj]
  {10.3847/1538-4357/836/2/200}, \href
  {https://ui.adsabs.harvard.edu/abs/2017ApJ...836..200G} {836, 200}

\bibitem[\protect\citeauthoryear{{Hartmann}, {Herczeg}  \& {Calvet}}{{Hartmann}
  et~al.}{2016}]{2016ARA&A..54..135H}
{Hartmann} L.,  {Herczeg} G.,   {Calvet} N.,  2016, \mn@doi [\araa]
  {10.1146/annurev-astro-081915-023347}, \href
  {https://ui.adsabs.harvard.edu/abs/2016ARA&A..54..135H} {54, 135}

\bibitem[\protect\citeauthoryear{{Herbert}, {Froebrich}  \& {Scholz}}{{Herbert}
  et~al.}{2023}]{2023MNRAS.520.5433H}
{Herbert} C.,  {Froebrich} D.,   {Scholz} A.,  2023, \mn@doi [\mnras]
  {10.1093/mnras/stac3051}, \href
  {https://ui.adsabs.harvard.edu/abs/2023MNRAS.520.5433H} {520, 5433}

\bibitem[\protect\citeauthoryear{{Herbst}, {Herbst}, {Grossman}  \&
  {Weinstein}}{{Herbst} et~al.}{1994}]{1994AJ....108.1906H}
{Herbst} W.,  {Herbst} D.~K.,  {Grossman} E.~J.,   {Weinstein} D.,  1994,
  \mn@doi [\aj] {10.1086/117204}, \href
  {https://ui.adsabs.harvard.edu/abs/1994AJ....108.1906H} {108, 1906}

\bibitem[\protect\citeauthoryear{{Herbst}, {Eisl{\"o}ffel}, {Mundt}  \&
  {Scholz}}{{Herbst} et~al.}{2007}]{2007prpl.conf..297H}
{Herbst} W.,  {Eisl{\"o}ffel} J.,  {Mundt} R.,   {Scholz} A.,  2007, in
  {Reipurth} B.,  {Jewitt} D.,   {Keil} K.,  eds, Protostars and Planets V.
  p.~297 (\mn@eprint {arXiv} {astro-ph/0603673})

\bibitem[\protect\citeauthoryear{{Kim} et~al.,}{{Kim}
  et~al.}{2009}]{2009ApJ...700.1017K}
{Kim} K.~H.,  et~al., 2009, \mn@doi [\apj] {10.1088/0004-637X/700/2/1017},
  \href {https://ui.adsabs.harvard.edu/abs/2009ApJ...700.1017K} {700, 1017}

\bibitem[\protect\citeauthoryear{{Kuhn}, {Hillenbrand}, {Carpenter}  \& {Avelar
  Menendez}}{{Kuhn} et~al.}{2020}]{2020ApJ...899..128K}
{Kuhn} M.~A.,  {Hillenbrand} L.~A.,  {Carpenter} J.~M.,   {Avelar Menendez}
  A.~R.,  2020, \mn@doi [\apj] {10.3847/1538-4357/aba19a}, \href
  {https://ui.adsabs.harvard.edu/abs/2020ApJ...899..128K} {899, 128}

\bibitem[\protect\citeauthoryear{{Kurosawa} \& {Romanova}}{{Kurosawa} \&
  {Romanova}}{2014}]{2014EPJWC..6404004K}
{Kurosawa} R.,  {Romanova} M.~M.,  2014, in European Physical Journal Web of
  Conferences. p. 04004, \mn@doi{10.1051/epjconf/20136404004}

\bibitem[\protect\citeauthoryear{{Lucas} et~al.,}{{Lucas}
  et~al.}{2008}]{2008MNRAS.391..136L}
{Lucas} P.~W.,  et~al., 2008, \mn@doi [\mnras]
  {10.1111/j.1365-2966.2008.13924.x}, \href
  {https://ui.adsabs.harvard.edu/abs/2008MNRAS.391..136L} {391, 136}

\bibitem[\protect\citeauthoryear{{Majaess}}{{Majaess}}{2013}]{2013Ap&SS.344..175M}
{Majaess} D.,  2013, \mn@doi [\apss] {10.1007/s10509-012-1308-y}, \href
  {https://ui.adsabs.harvard.edu/abs/2013Ap&SS.344..175M} {344, 175}

\bibitem[\protect\citeauthoryear{{Manara}, {Testi}, {Natta}, {Rosotti},
  {Benisty}, {Ercolano}  \& {Ricci}}{{Manara}
  et~al.}{2014}]{2014A&A...568A..18M}
{Manara} C.~F.,  {Testi} L.,  {Natta} A.,  {Rosotti} G.,  {Benisty} M.,
  {Ercolano} B.,   {Ricci} L.,  2014, \mn@doi [\aap]
  {10.1051/0004-6361/201323318}, \href
  {https://ui.adsabs.harvard.edu/abs/2014A&A...568A..18M} {568, A18}

\bibitem[\protect\citeauthoryear{{Mer{\'\i}n} et~al.,}{{Mer{\'\i}n}
  et~al.}{2010}]{2010ApJ...718.1200M}
{Mer{\'\i}n} B.,  et~al., 2010, \mn@doi [\apj] {10.1088/0004-637X/718/2/1200},
  \href {https://ui.adsabs.harvard.edu/abs/2010ApJ...718.1200M} {718, 1200}

\bibitem[\protect\citeauthoryear{{Muzerolle}, {Calvet}  \&
  {Hartmann}}{{Muzerolle} et~al.}{1998}]{1998ApJ...492..743M}
{Muzerolle} J.,  {Calvet} N.,   {Hartmann} L.,  1998, \mn@doi [\apj]
  {10.1086/305069}, \href
  {https://ui.adsabs.harvard.edu/abs/1998ApJ...492..743M} {492, 743}

\bibitem[\protect\citeauthoryear{{Muzerolle}, {Hillenbrand}, {Calvet},
  {Brice{\~n}o}  \& {Hartmann}}{{Muzerolle} et~al.}{2003}]{2003ApJ...592..266M}
{Muzerolle} J.,  {Hillenbrand} L.,  {Calvet} N.,  {Brice{\~n}o} C.,
  {Hartmann} L.,  2003, \mn@doi [\apj] {10.1086/375704}, \href
  {https://ui.adsabs.harvard.edu/abs/2003ApJ...592..266M} {592, 266}

\bibitem[\protect\citeauthoryear{{Muzerolle}, {Luhman}, {Brice{\~n}o},
  {Hartmann}  \& {Calvet}}{{Muzerolle} et~al.}{2005}]{2005ApJ...625..906M}
{Muzerolle} J.,  {Luhman} K.~L.,  {Brice{\~n}o} C.,  {Hartmann} L.,   {Calvet}
  N.,  2005, \mn@doi [\apj] {10.1086/429483}, \href
  {https://ui.adsabs.harvard.edu/abs/2005ApJ...625..906M} {625, 906}

\bibitem[\protect\citeauthoryear{{Najita}, {Andrews}  \& {Muzerolle}}{{Najita}
  et~al.}{2015}]{2015MNRAS.450.3559N}
{Najita} J.~R.,  {Andrews} S.~M.,   {Muzerolle} J.,  2015, \mn@doi [\mnras]
  {10.1093/mnras/stv839}, \href
  {https://ui.adsabs.harvard.edu/abs/2015MNRAS.450.3559N} {450, 3559}

\bibitem[\protect\citeauthoryear{{Panwar}, {Jose}  \& {Rishi}}{{Panwar}
  et~al.}{2023}]{2023JApA...44...42P}
{Panwar} N.,  {Jose} J.,   {Rishi} C.,  2023, \mn@doi [Journal of Astrophysics
  and Astronomy] {10.1007/s12036-023-09935-x}, \href
  {https://ui.adsabs.harvard.edu/abs/2023JApA...44...42P} {44, 42}

\bibitem[\protect\citeauthoryear{{Rodr{\'\i}guez-Ledesma}, {Mundt}  \&
  {Eisl{\"o}ffel}}{{Rodr{\'\i}guez-Ledesma} et~al.}{2010}]{2010A&A...515A..13R}
{Rodr{\'\i}guez-Ledesma} M.~V.,  {Mundt} R.,   {Eisl{\"o}ffel} J.,  2010,
  \mn@doi [\aap] {10.1051/0004-6361/200913494}, \href
  {https://ui.adsabs.harvard.edu/abs/2010A&A...515A..13R} {515, A13}

\bibitem[\protect\citeauthoryear{{Scargle}}{{Scargle}}{1982}]{1982ApJ...263..835S}
{Scargle} J.~D.,  1982, \mn@doi [\apj] {10.1086/160554}, \href
  {https://ui.adsabs.harvard.edu/abs/1982ApJ...263..835S} {263, 835}

\bibitem[\protect\citeauthoryear{{Scholz}, {Xu}, {Jayawardhana}, {Wood},
  {Eisl{\"o}ffel}  \& {Quinn}}{{Scholz} et~al.}{2009}]{2009MNRAS.398..873S}
{Scholz} A.,  {Xu} X.,  {Jayawardhana} R.,  {Wood} K.,  {Eisl{\"o}ffel} J.,
  {Quinn} C.,  2009, \mn@doi [\mnras] {10.1111/j.1365-2966.2009.15021.x}, \href
  {https://ui.adsabs.harvard.edu/abs/2009MNRAS.398..873S} {398, 873}

\bibitem[\protect\citeauthoryear{{Scholz}, {Stelzer}, {Costigan}, {Barrado},
  {Eisl{\"o}ffel}, {Lillo-Box}, {Riviere-Marichalar}  \& {Stoev}}{{Scholz}
  et~al.}{2012}]{2012MNRAS.419.1271S}
{Scholz} A.,  {Stelzer} B.,  {Costigan} G.,  {Barrado} D.,  {Eisl{\"o}ffel} J.,
   {Lillo-Box} J.,  {Riviere-Marichalar} P.,   {Stoev} H.,  2012, \mn@doi
  [\mnras] {10.1111/j.1365-2966.2011.19781.x}, \href
  {https://ui.adsabs.harvard.edu/abs/2012MNRAS.419.1271S} {419, 1271}

\bibitem[\protect\citeauthoryear{{Sicilia-Aguilar}, {Henning}  \&
  {Hartmann}}{{Sicilia-Aguilar} et~al.}{2010}]{2010ApJ...710..597S}
{Sicilia-Aguilar} A.,  {Henning} T.,   {Hartmann} L.~W.,  2010, \mn@doi [\apj]
  {10.1088/0004-637X/710/1/597}, \href
  {https://ui.adsabs.harvard.edu/abs/2010ApJ...710..597S} {710, 597}

\bibitem[\protect\citeauthoryear{{Sicilia-Aguilar} et~al.,}{{Sicilia-Aguilar}
  et~al.}{2023}]{2023MNRAS.526.4885S}
{Sicilia-Aguilar} A.,  et~al., 2023, \mn@doi [\mnras] {10.1093/mnras/stad3029},
  \href {https://ui.adsabs.harvard.edu/abs/2023MNRAS.526.4885S} {526, 4885}

\bibitem[\protect\citeauthoryear{{Skelly}, {Unruh}, {Barnes}, {Lawson},
  {Donati}  \& {Collier Cameron}}{{Skelly} et~al.}{2009}]{2009MNRAS.399.1829S}
{Skelly} M.~B.,  {Unruh} Y.~C.,  {Barnes} J.~R.,  {Lawson} W.~A.,  {Donati}
  J.~F.,   {Collier Cameron} A.,  2009, \mn@doi [\mnras]
  {10.1111/j.1365-2966.2009.15411.x}, \href
  {https://ui.adsabs.harvard.edu/abs/2009MNRAS.399.1829S} {399, 1829}

\bibitem[\protect\citeauthoryear{{Skrutskie}, {Cutri}, {Stiening}, {Weinberg},
  {Schneider}, {Carpenter}  \& {et al.}}{{Skrutskie}
  et~al.}{2006}]{2006AJ....131.1163S}
{Skrutskie} M.~F.,  {Cutri} R.~M.,  {Stiening} R.,  {Weinberg} M.~D.,
  {Schneider} S.,  {Carpenter} J.~M.,   {et al.} 2006, \mn@doi [\aj]
  {10.1086/498708}, \href {http://adsabs.harvard.edu/abs/2006AJ....131.1163S}
  {131, 1163}

\bibitem[\protect\citeauthoryear{{Strassmeier}}{{Strassmeier}}{1992}]{1992ASPC...34...39S}
{Strassmeier} K.~G.,  1992, in {Filippenko} A.~V.,  ed.,  Astronomical Society
  of the Pacific Conference Series Vol. 103, Robotic Telescopes in the 1990s.
  pp 39--52

\bibitem[\protect\citeauthoryear{{Strassmeier}}{{Strassmeier}}{2002}]{2002AN....323..309S}
{Strassmeier} K.~G.,  2002, \mn@doi [Astronomische Nachrichten]
  {10.1002/1521-3994(200208)323:3/4\textless{}309::AID-ASNA309\textgreater{}3.0.CO;2-U},
  \href {https://ui.adsabs.harvard.edu/abs/2002AN....323..309S} {323, 309}

\bibitem[\protect\citeauthoryear{{Strassmeier}}{{Strassmeier}}{2009}]{2009A&ARv..17..251S}
{Strassmeier} K.~G.,  2009, \mn@doi [\aapr] {10.1007/s00159-009-0020-6}, \href
  {https://ui.adsabs.harvard.edu/abs/2009A&ARv..17..251S} {17, 251}

\bibitem[\protect\citeauthoryear{{Strassmeier}, {Rice}, {Wehlau}, {Hill}  \&
  {Matthews}}{{Strassmeier} et~al.}{1993}]{1993A&A...268..671S}
{Strassmeier} K.~G.,  {Rice} J.~B.,  {Wehlau} W.~H.,  {Hill} G.~M.,
  {Matthews} J.~M.,  1993, \aap, \href
  {https://ui.adsabs.harvard.edu/abs/1993A&A...268..671S} {268, 671}

\bibitem[\protect\citeauthoryear{{Strassmeier}, {Kratzwald}  \&
  {Weber}}{{Strassmeier} et~al.}{2003}]{2003A&A...408.1103S}
{Strassmeier} K.~G.,  {Kratzwald} L.,   {Weber} M.,  2003, \mn@doi [\aap]
  {10.1051/0004-6361:20031029}, \href
  {https://ui.adsabs.harvard.edu/abs/2003A&A...408.1103S} {408, 1103}

\bibitem[\protect\citeauthoryear{{Strassmeier}, {Carroll}  \&
  {Ilyin}}{{Strassmeier} et~al.}{2019}]{2019A&A...625A..27S}
{Strassmeier} K.~G.,  {Carroll} T.~A.,   {Ilyin} I.~V.,  2019, \mn@doi [\aap]
  {10.1051/0004-6361/201834906}, \href
  {https://ui.adsabs.harvard.edu/abs/2019A&A...625A..27S} {625, A27}

\bibitem[\protect\citeauthoryear{{Teixeira}, {Lada}, {Marengo}  \&
  {Lada}}{{Teixeira} et~al.}{2012}]{2012A&A...540A..83T}
{Teixeira} P.~S.,  {Lada} C.~J.,  {Marengo} M.,   {Lada} E.~A.,  2012, \mn@doi
  [\aap] {10.1051/0004-6361/201015326}, \href
  {https://ui.adsabs.harvard.edu/abs/2012A&A...540A..83T} {540, A83}

\bibitem[\protect\citeauthoryear{Thieler, Fried, Rathjens  et~al.}{Thieler
  et~al.}{2016}]{thieler2016robper}
Thieler A.~M.,  Fried R.,  Rathjens J.,   et~al., 2016, Journal of Statistical
  Software, 69, 1

\bibitem[\protect\citeauthoryear{{Wang}, {Fang}, {Herczeg}, {Gao}, {Tian},
  {Zhou}, {Zhang}  \& {Chen}}{{Wang} et~al.}{2023}]{2023RAA....23g5015W}
{Wang} X.-L.,  {Fang} M.,  {Herczeg} G.~J.,  {Gao} Y.,  {Tian} H.-J.,  {Zhou}
  X.-Y.,  {Zhang} H.-X.,   {Chen} X.-P.,  2023, \mn@doi [Research in Astronomy
  and Astrophysics] {10.1088/1674-4527/acd58b}, \href
  {https://ui.adsabs.harvard.edu/abs/2023RAA....23g5015W} {23, 075015}

\bibitem[\protect\citeauthoryear{{Xiang}, {Gu}, {Donati}, {Hussain}, {Collier
  Cameron}  \& {MaTYSSE Collaboration}}{{Xiang}
  et~al.}{2023}]{2023MNRAS.520.3964X}
{Xiang} Y.,  {Gu} S.,  {Donati} J.~F.,  {Hussain} G.~A.~J.,  {Collier Cameron}
  A.,   {MaTYSSE Collaboration} 2023, \mn@doi [\mnras] {10.1093/mnras/stad363},
  \href {https://ui.adsabs.harvard.edu/abs/2023MNRAS.520.3964X} {520, 3964}

\bibitem[\protect\citeauthoryear{{Zechmeister} \& {K{\"u}rster}}{{Zechmeister}
  \& {K{\"u}rster}}{2009}]{2009A&A...496..577Z}
{Zechmeister} M.,  {K{\"u}rster} M.,  2009, \mn@doi [\aap]
  {10.1051/0004-6361:200811296}, \href
  {https://ui.adsabs.harvard.edu/abs/2009A&A...496..577Z} {496, 577}

\bibitem[\protect\citeauthoryear{{Zhang}, {Pi}, {Zhu}, {Zhang}  \&
  {Li}}{{Zhang} et~al.}{2014}]{2014NewA...32....1Z}
{Zhang} L.,  {Pi} Q.,  {Zhu} Z.,  {Zhang} X.,   {Li} Z.,  2014, \mn@doi [\na]
  {10.1016/j.newast.2014.02.010}, \href
  {https://ui.adsabs.harvard.edu/abs/2014NewA...32....1Z} {32, 1}

\makeatother
\end{thebibliography}


\clearpage
\newpage

\appendix

\begin{figure}
\section{Object Property plots} \label{evolfig}
\centering
\includegraphics[width=\columnwidth]{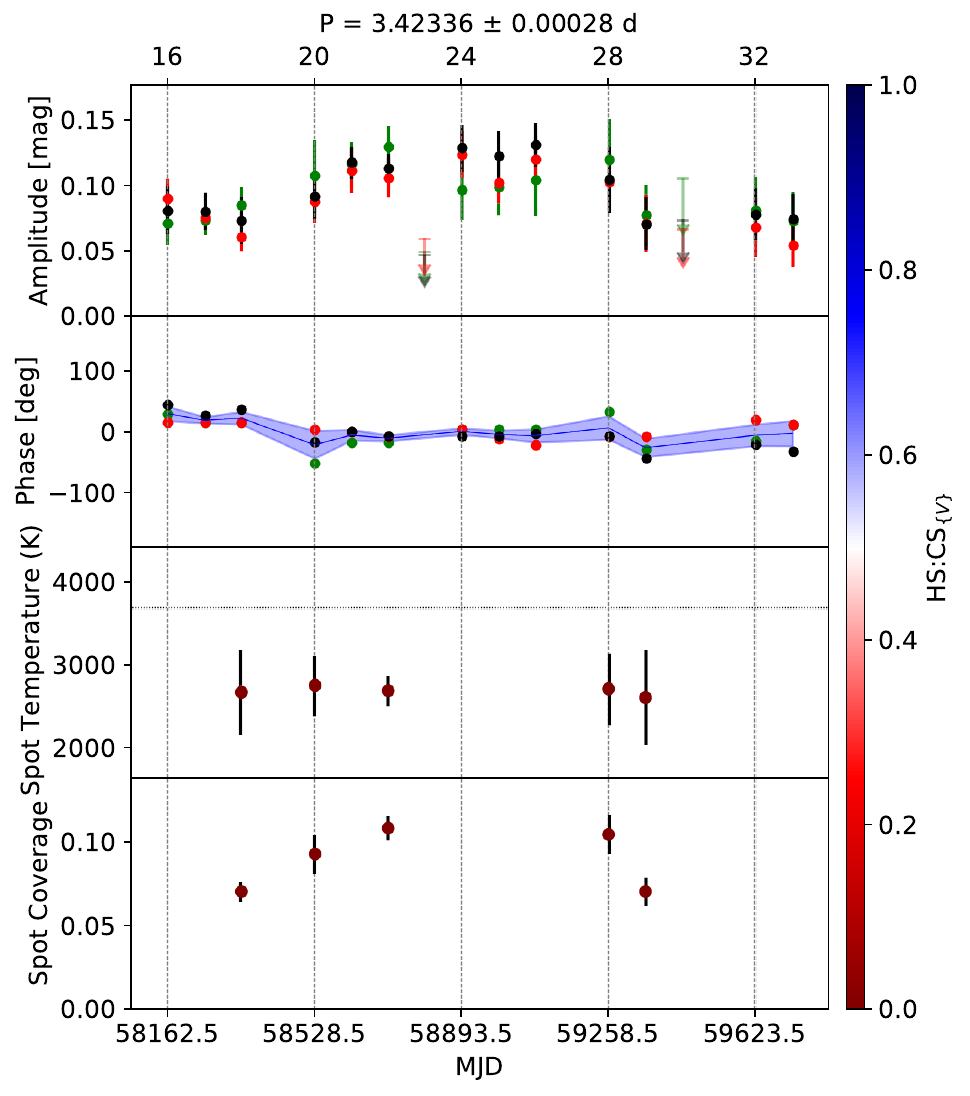}
\caption{As Fig. \ref{fig: ev1} for object 5 \label{fig: ev5}}
\end{figure}

\begin{figure}
\centering
\includegraphics[width=\columnwidth]{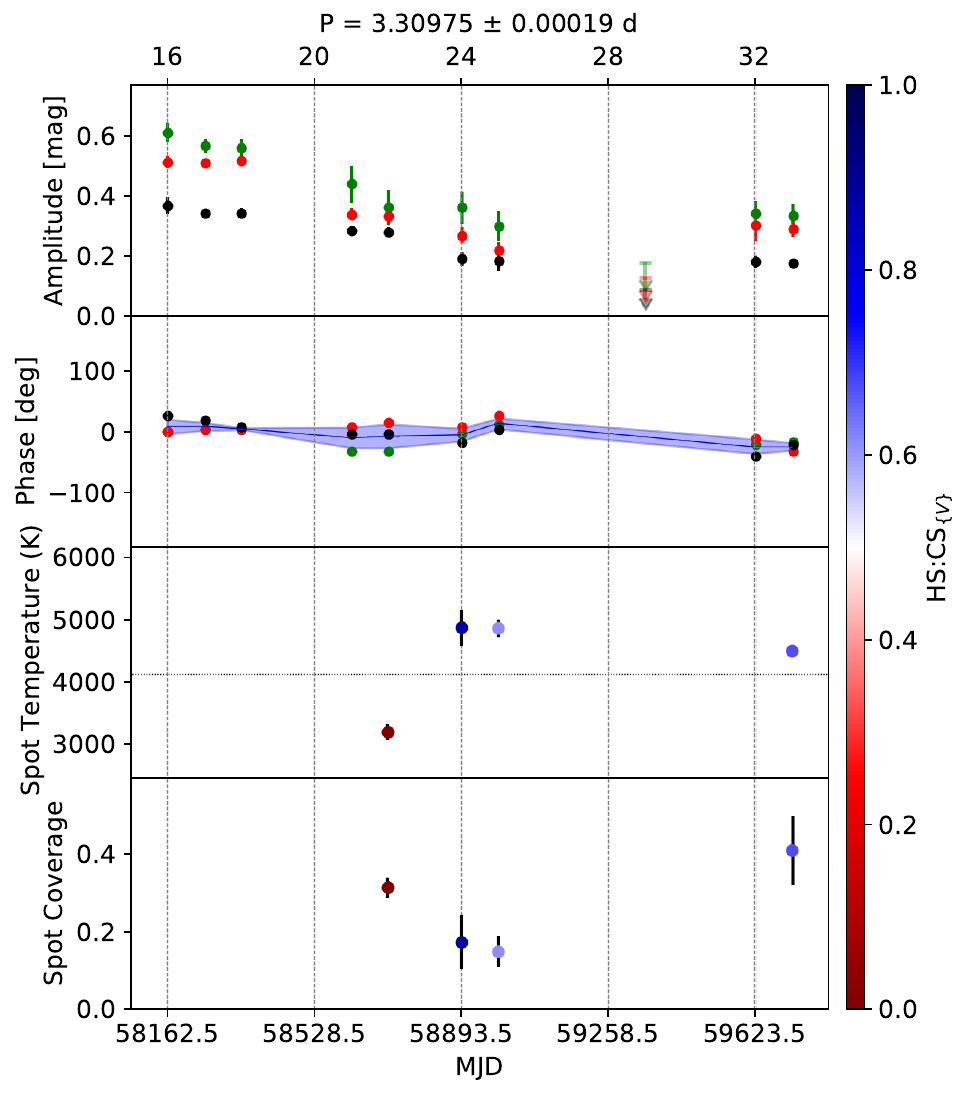}
\caption{As Fig. \ref{fig: ev1} for object 6 \label{fig: ev6}}
\end{figure}

\begin{figure}
\centering
\includegraphics[width=\columnwidth]{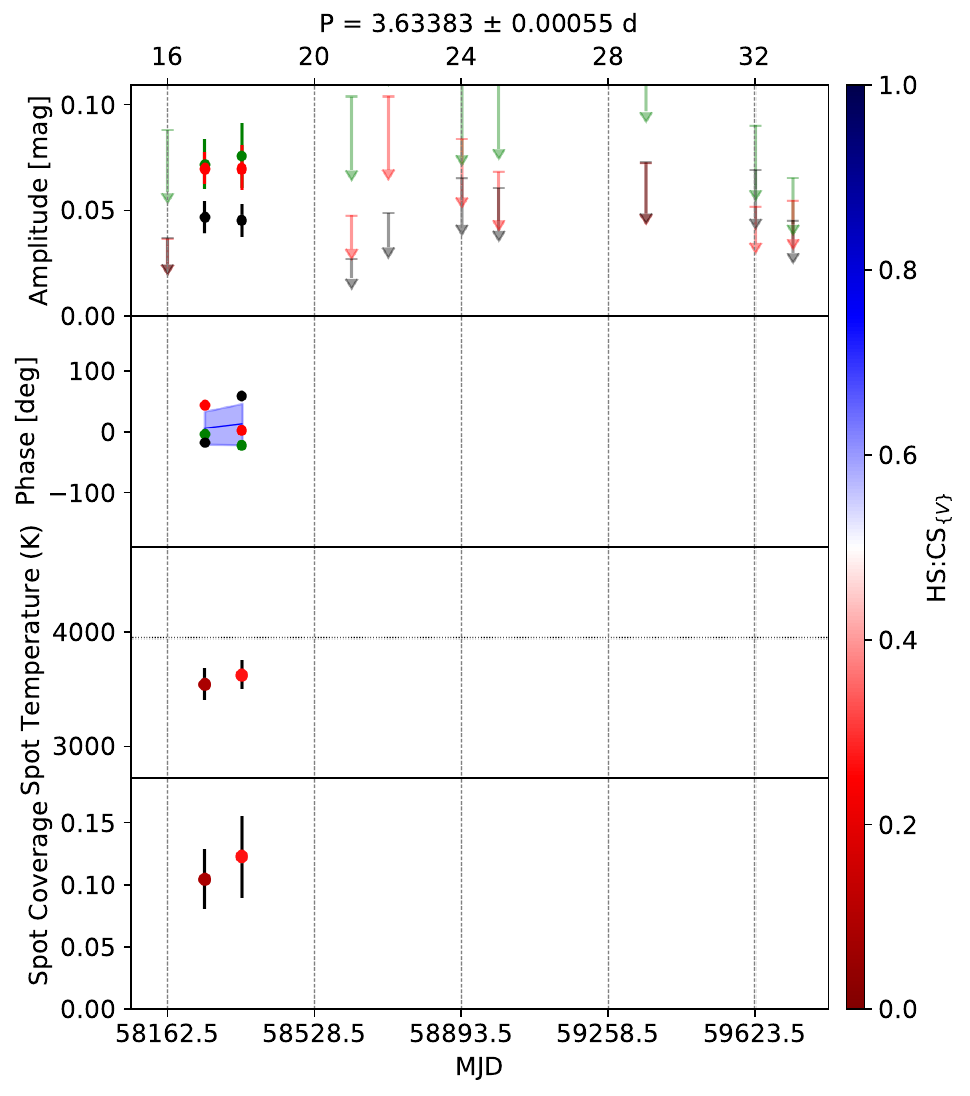}
\caption{As Fig. \ref{fig: ev1} for object 7 \label{fig: ev7}}
\end{figure}

\begin{figure}
\centering
\includegraphics[width=\columnwidth]{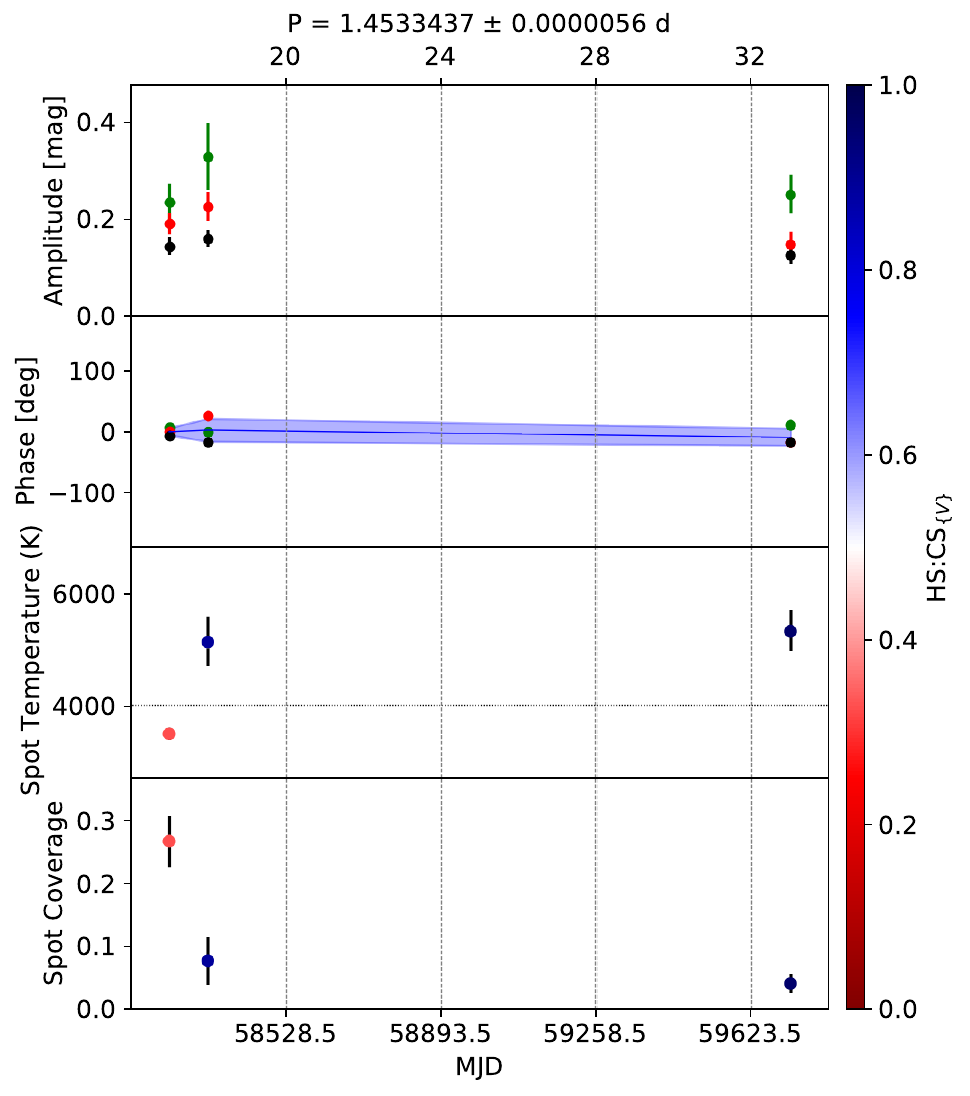}
\caption{As Fig. \ref{fig: ev1} for object 8 \label{fig: ev8}}
\end{figure}

\begin{figure}
\centering
\includegraphics[width=\columnwidth]{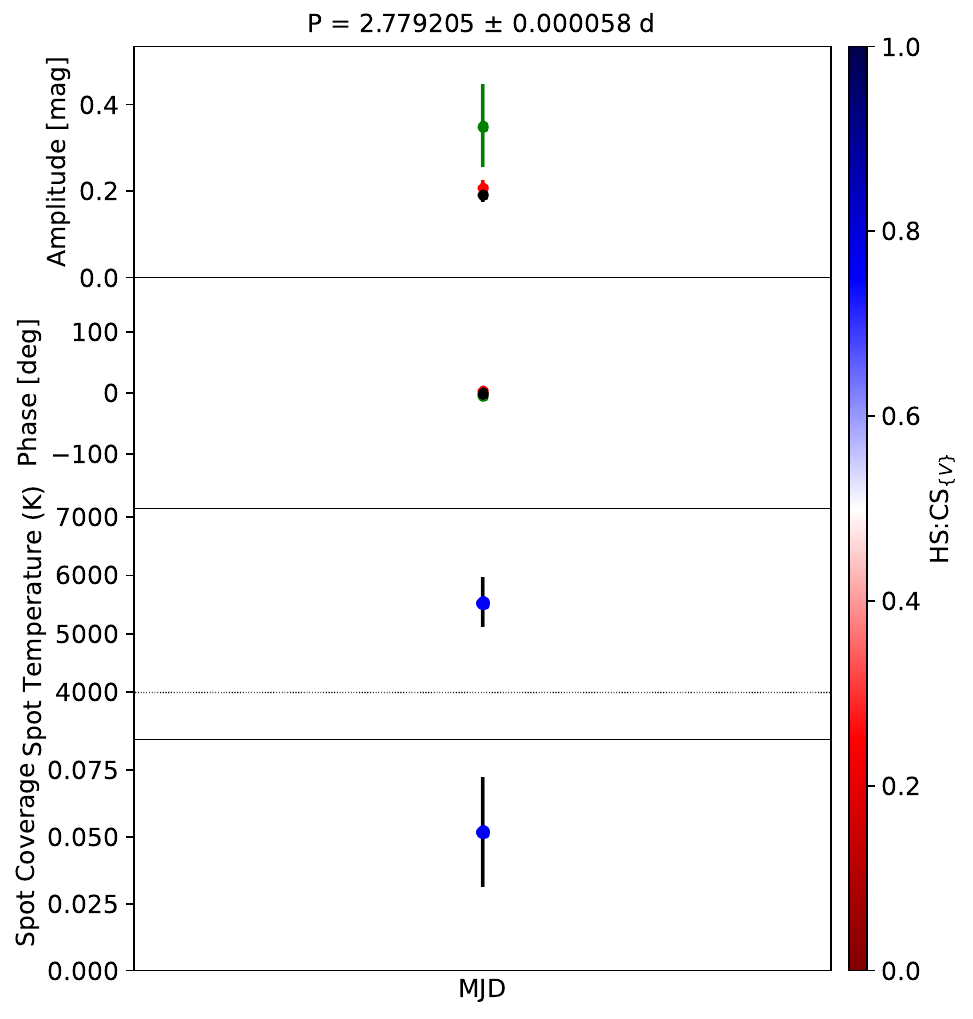}
\caption{As Fig. \ref{fig: ev1} for object 9 \label{fig: ev9}}
\end{figure}

\begin{figure}
\centering
\includegraphics[width=\columnwidth]{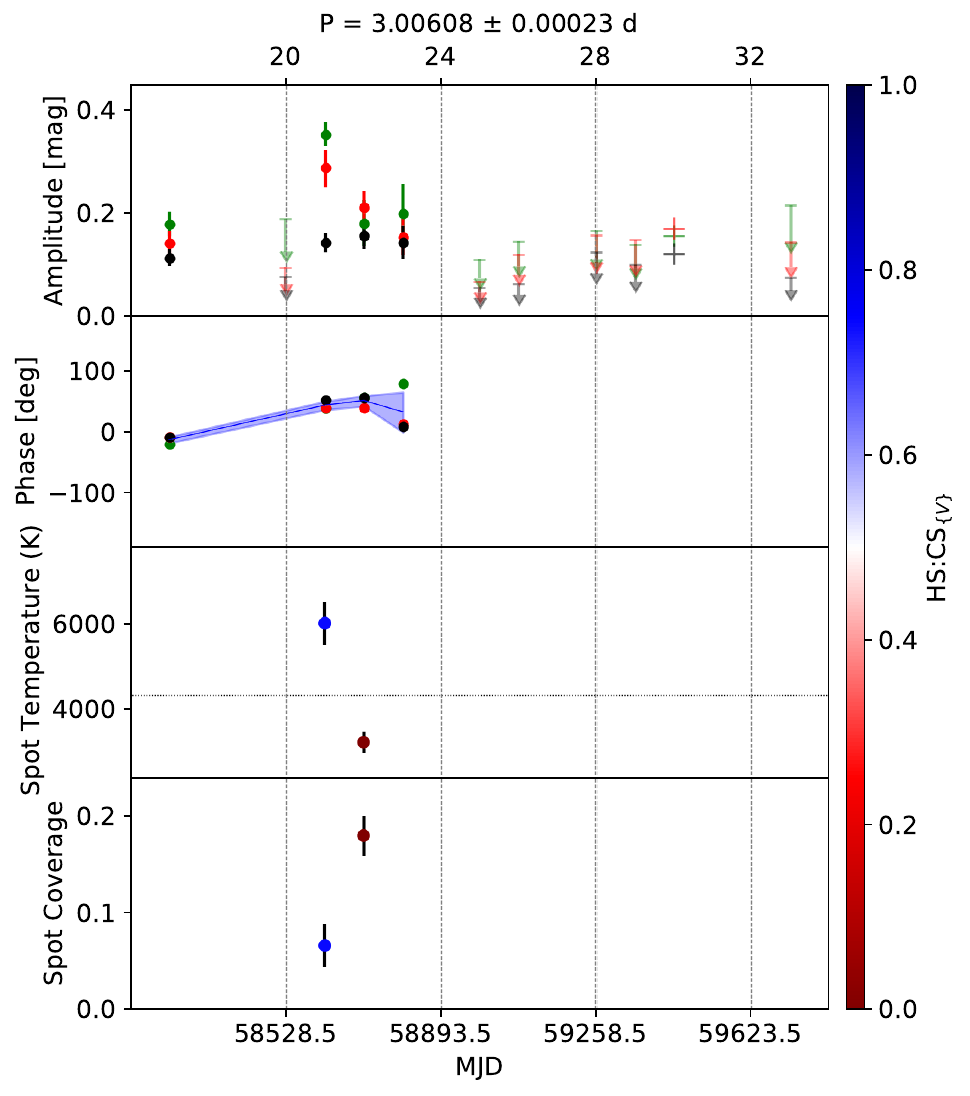}
\caption{As Fig. \ref{fig: ev1} for object 10 \label{fig: ev10}}
\end{figure}

\begin{figure}
\centering
\includegraphics[width=\columnwidth]{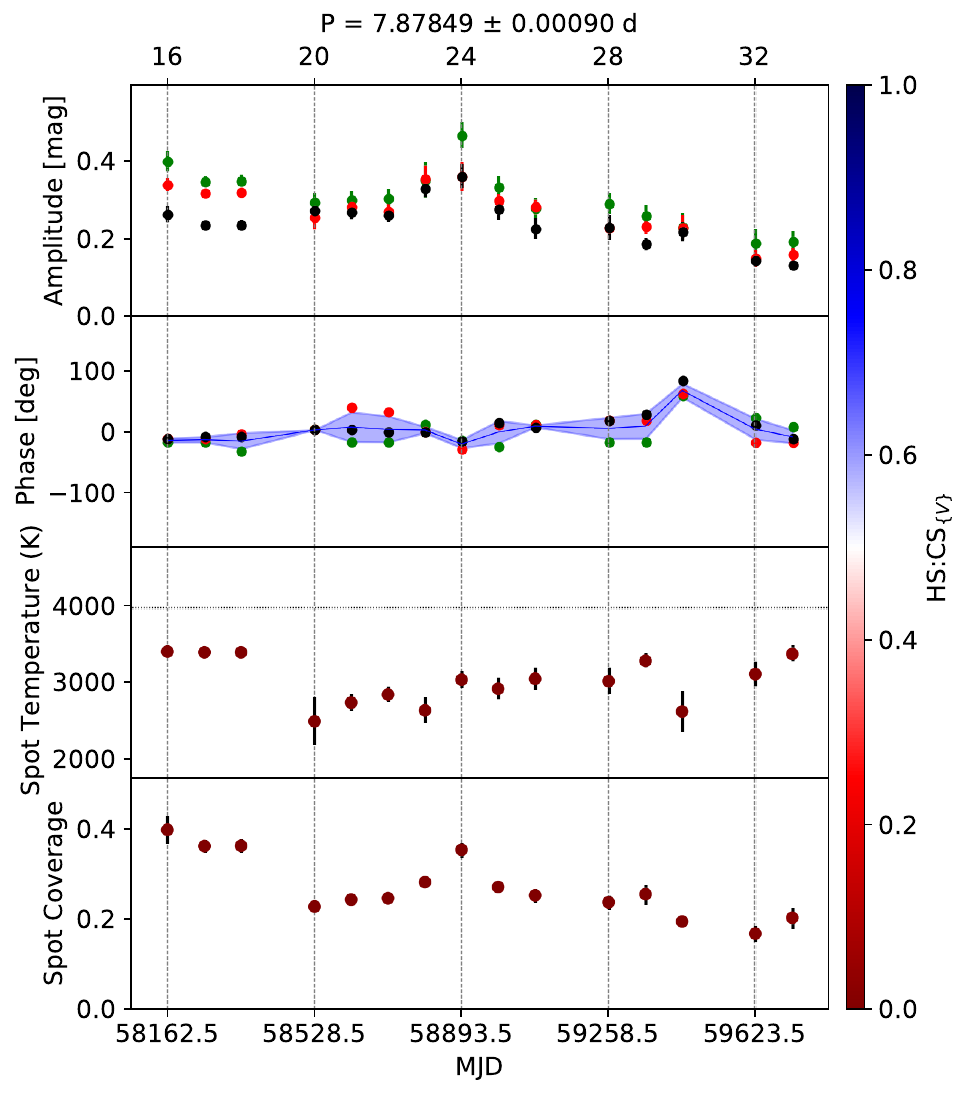}
\caption{As Fig. \ref{fig: ev1} for object 11 \label{fig: ev11}}
\end{figure}

\begin{figure}
\centering
\includegraphics[width=\columnwidth]{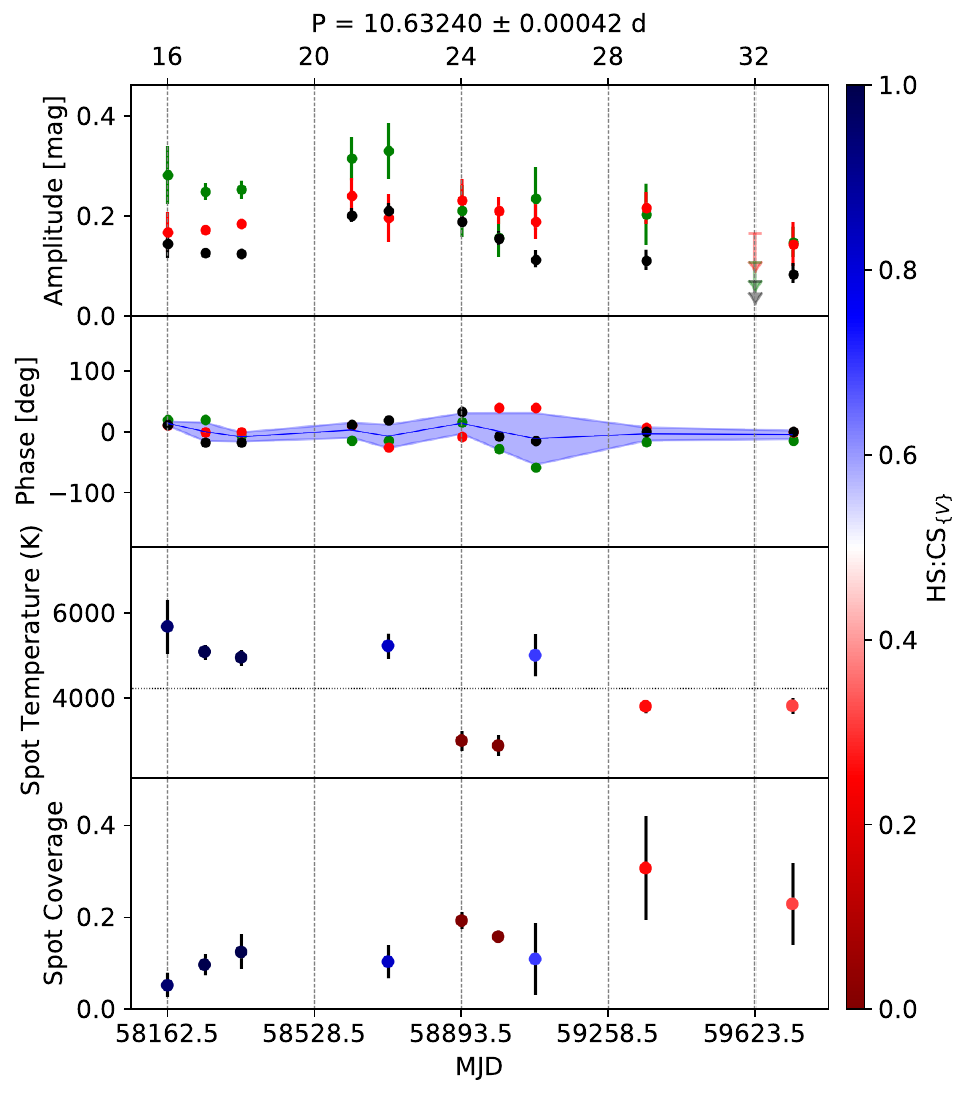}
\caption{As Fig. \ref{fig: ev1} for object 14 \label{fig: ev14}}
\end{figure}

\begin{figure}
\centering
\includegraphics[width=\columnwidth]{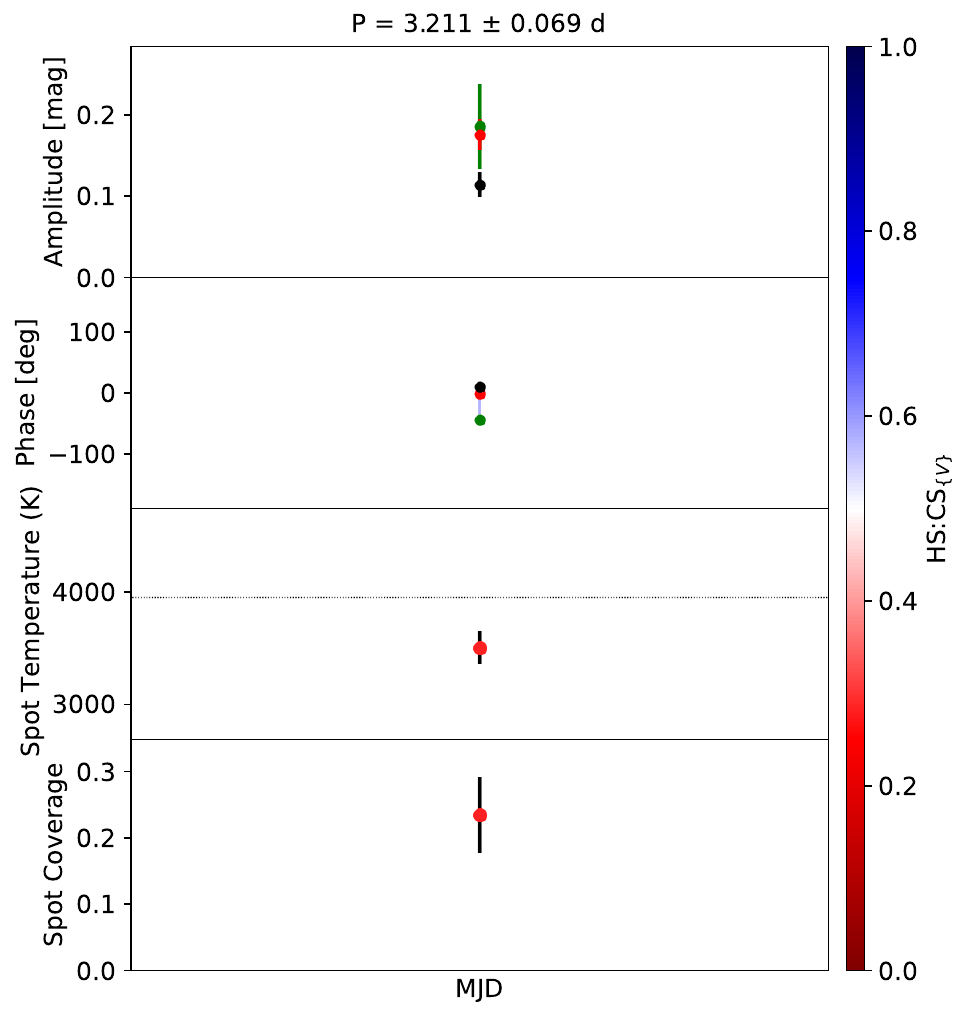}
\caption{As Fig. \ref{fig: ev1} for object 16 \label{fig: ev16}}
\end{figure}

\begin{figure}
\centering
\includegraphics[width=\columnwidth]{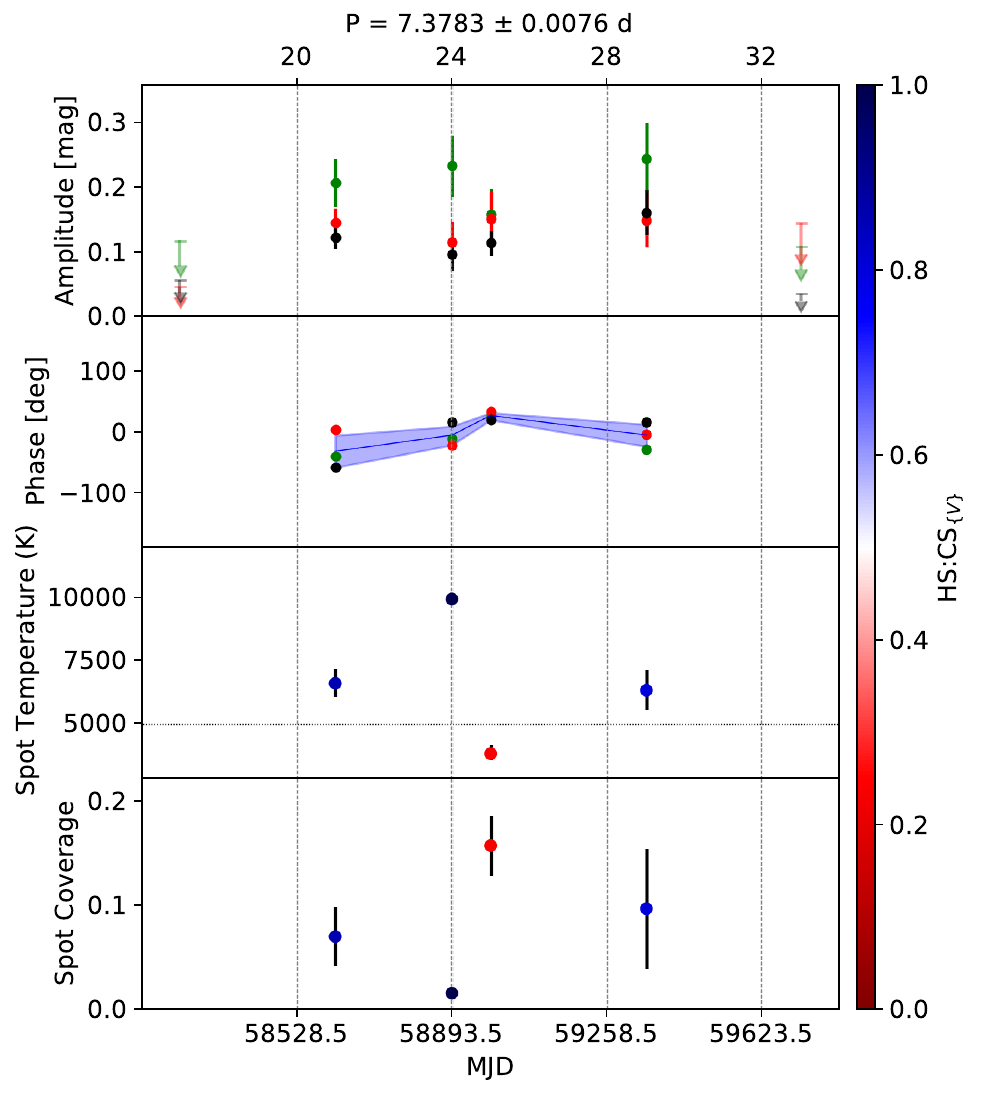}
\caption{As Fig. \ref{fig: ev1} for object 17 \label{fig: ev17}}
\end{figure}

\begin{figure}
\centering
\includegraphics[width=\columnwidth]{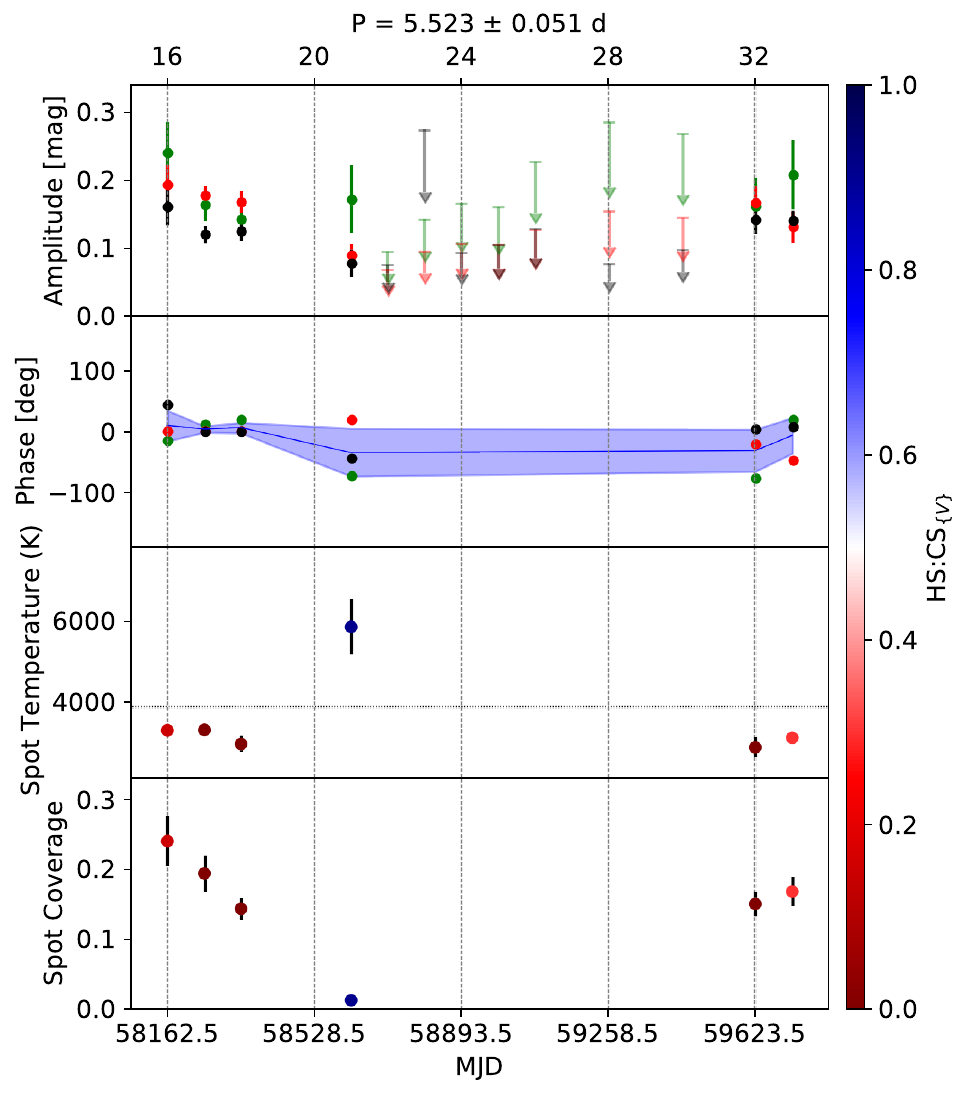}
\caption{As Fig. \ref{fig: ev1} for object 20 \label{fig: ev20}}
\end{figure}

\begin{figure}
\centering
\includegraphics[width=\columnwidth]{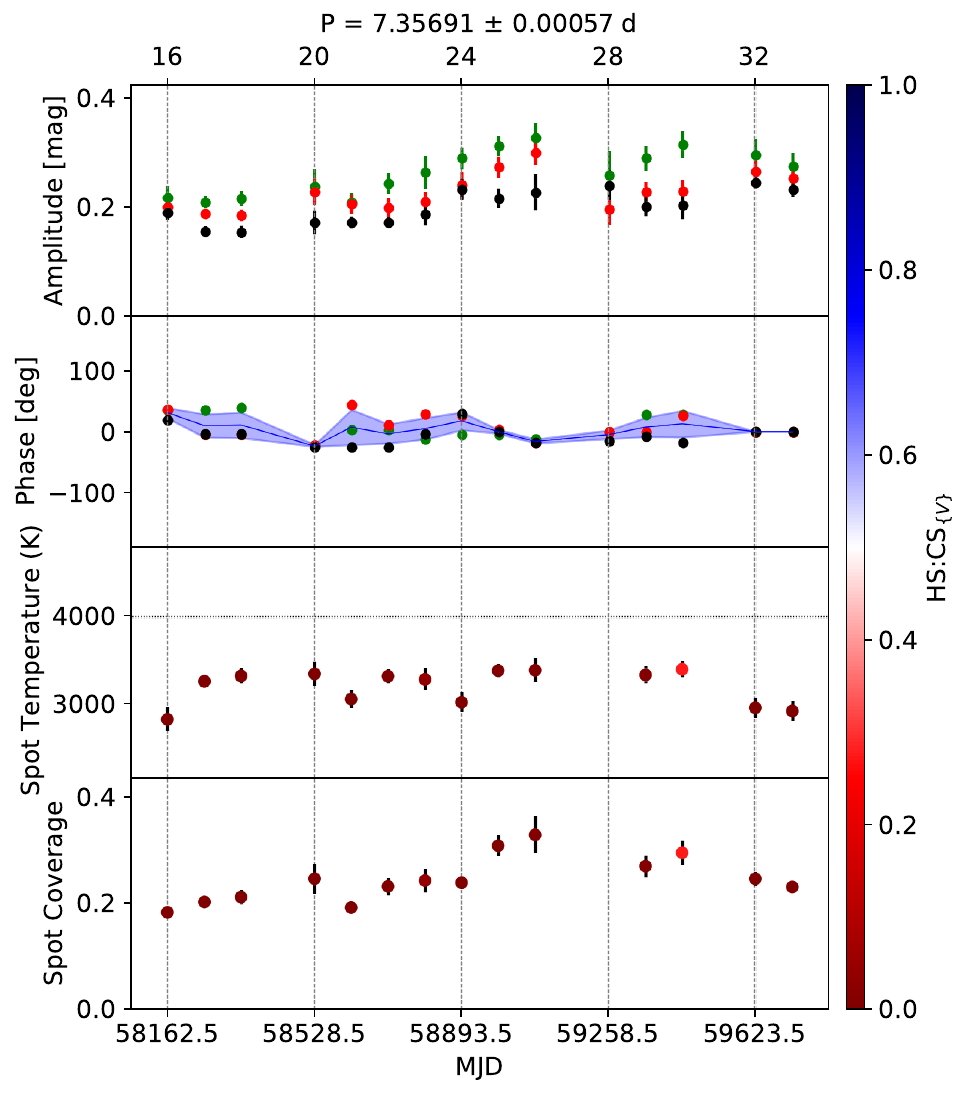}
\caption{As Fig. \ref{fig: ev1} for object 21 \label{fig: ev21}}
\end{figure}

\begin{figure}
\centering
\includegraphics[width=\columnwidth]{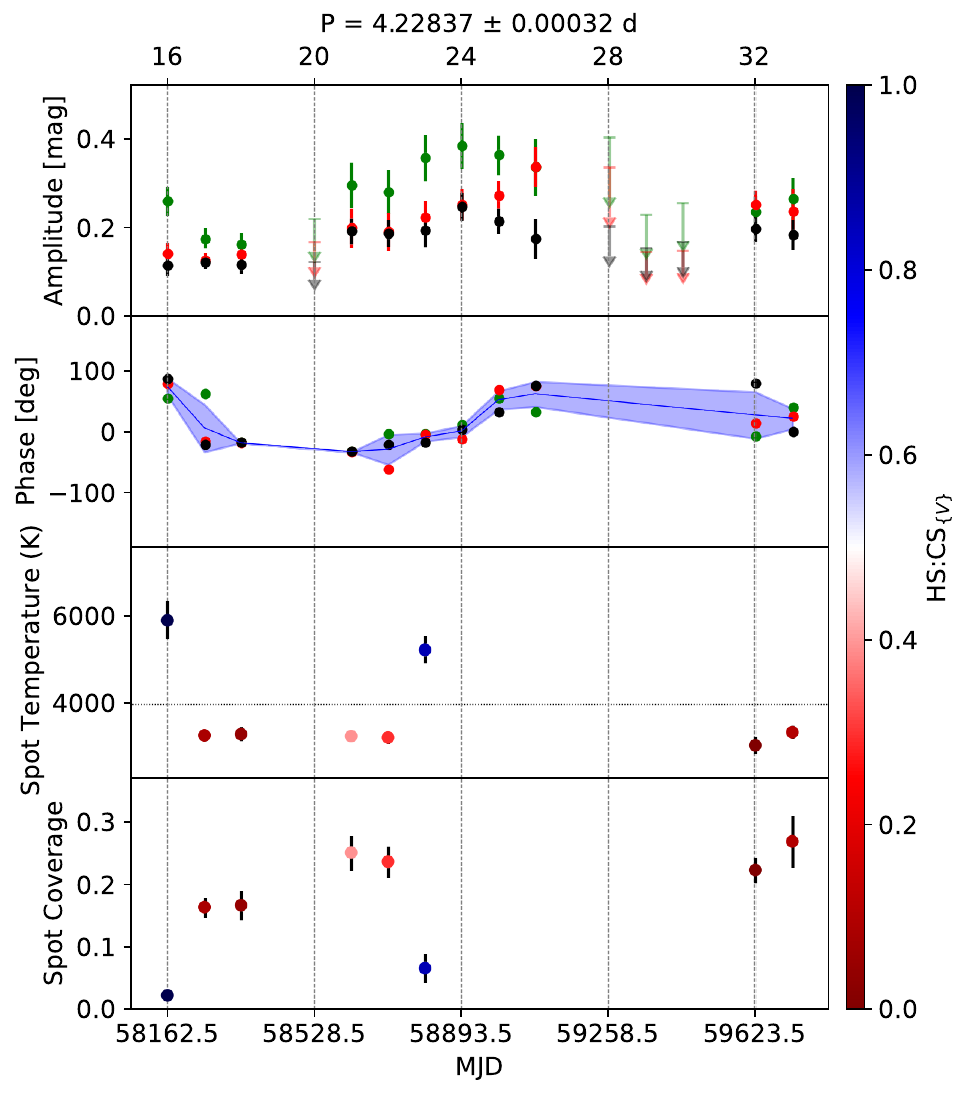}
\caption{As Fig. \ref{fig: ev1} for object 22 \label{fig: ev22}}
\end{figure}

\begin{figure}
\centering
\includegraphics[width=\columnwidth]{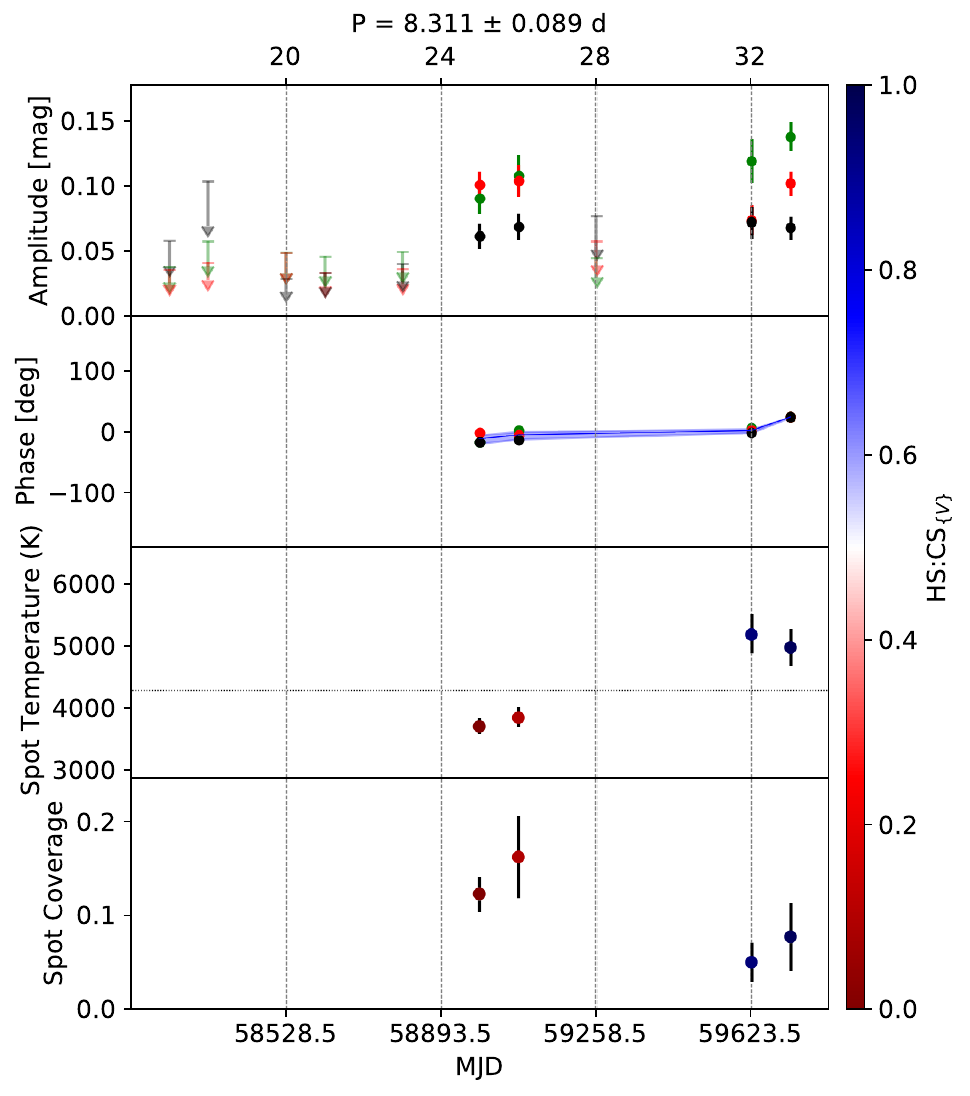}
\caption{As Fig. \ref{fig: ev1} for object 23 \label{fig: ev23}}
\end{figure}

\begin{figure}
\centering
\includegraphics[width=\columnwidth]{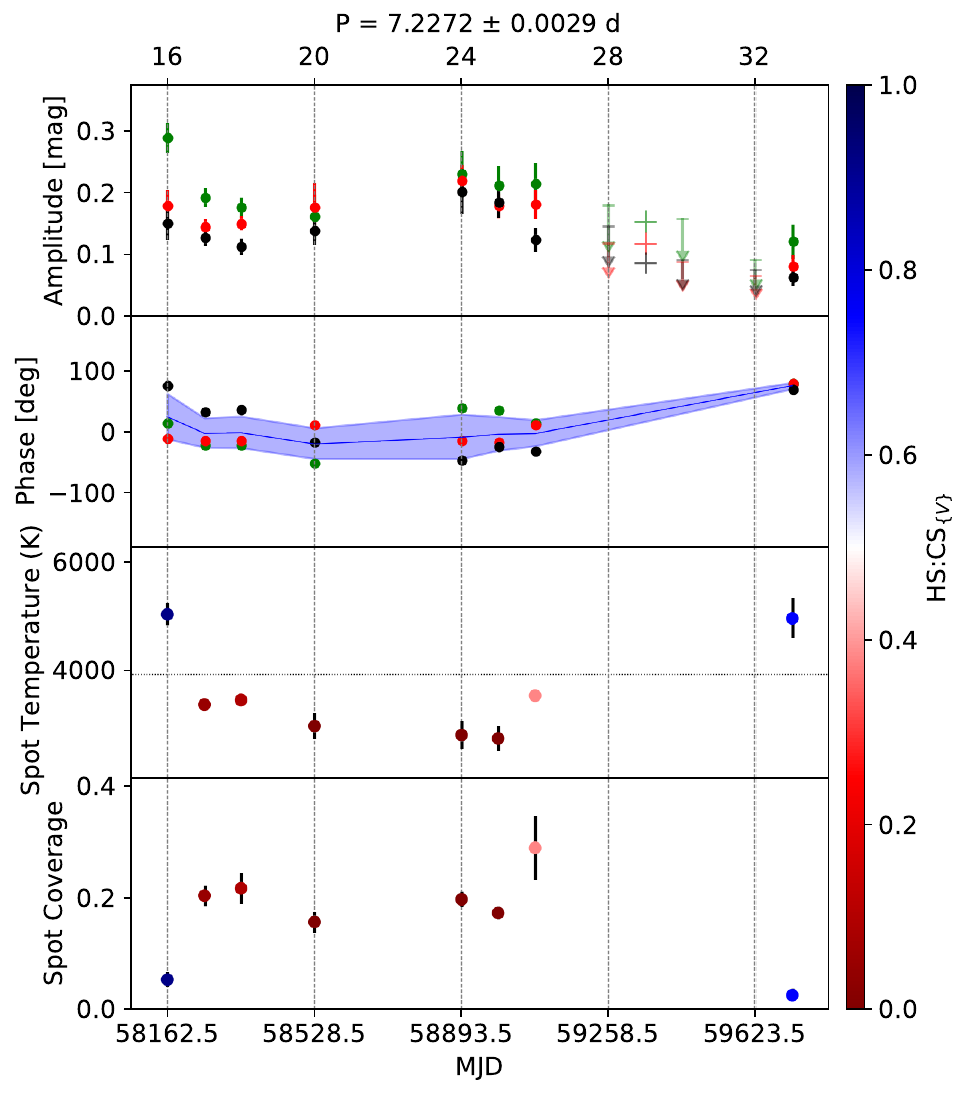}
\caption{As Fig. \ref{fig: ev1} for object 24 \label{fig: ev24}}
\end{figure}

\begin{figure}
\centering
\includegraphics[width=\columnwidth]{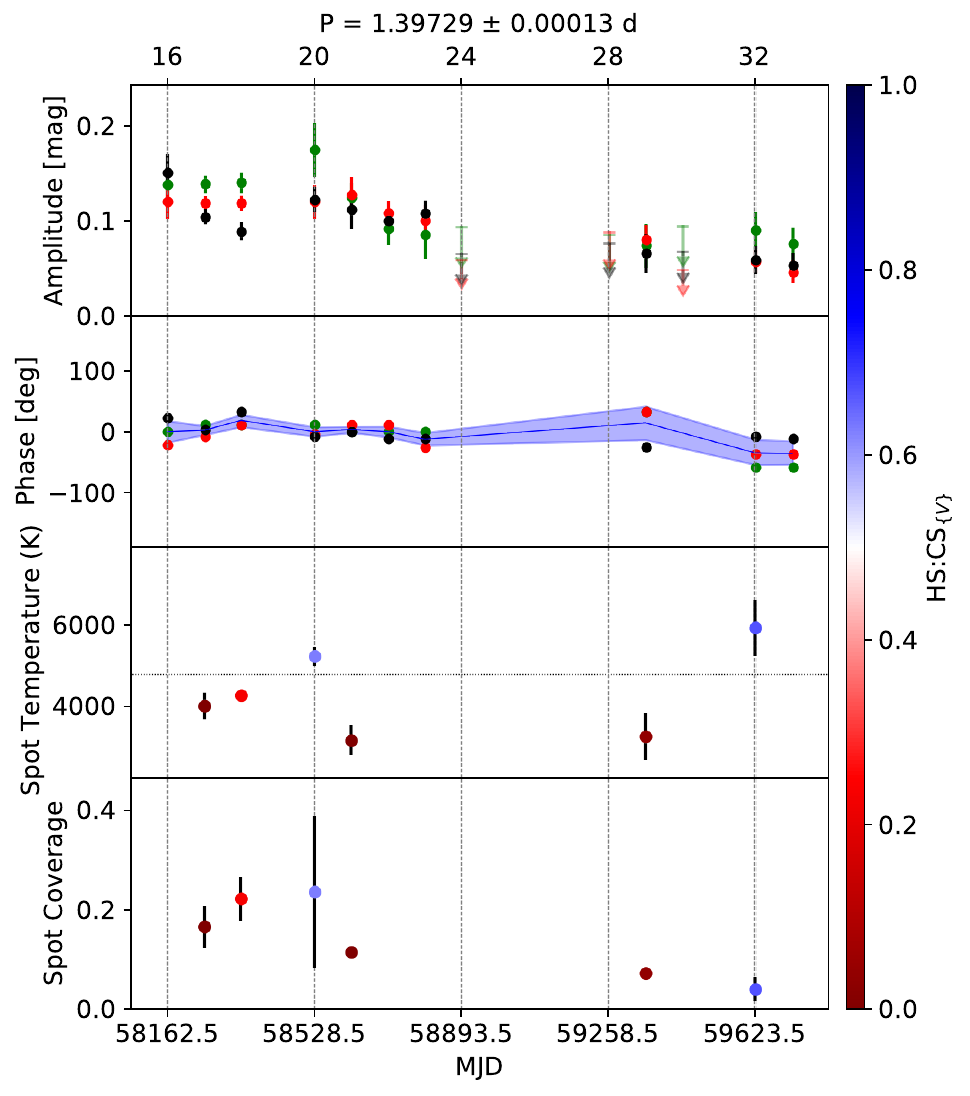}
\caption{As Fig. \ref{fig: ev1} for object 25 \label{fig: ev25}}
\end{figure}

\begin{figure}
\centering
\includegraphics[width=\columnwidth]{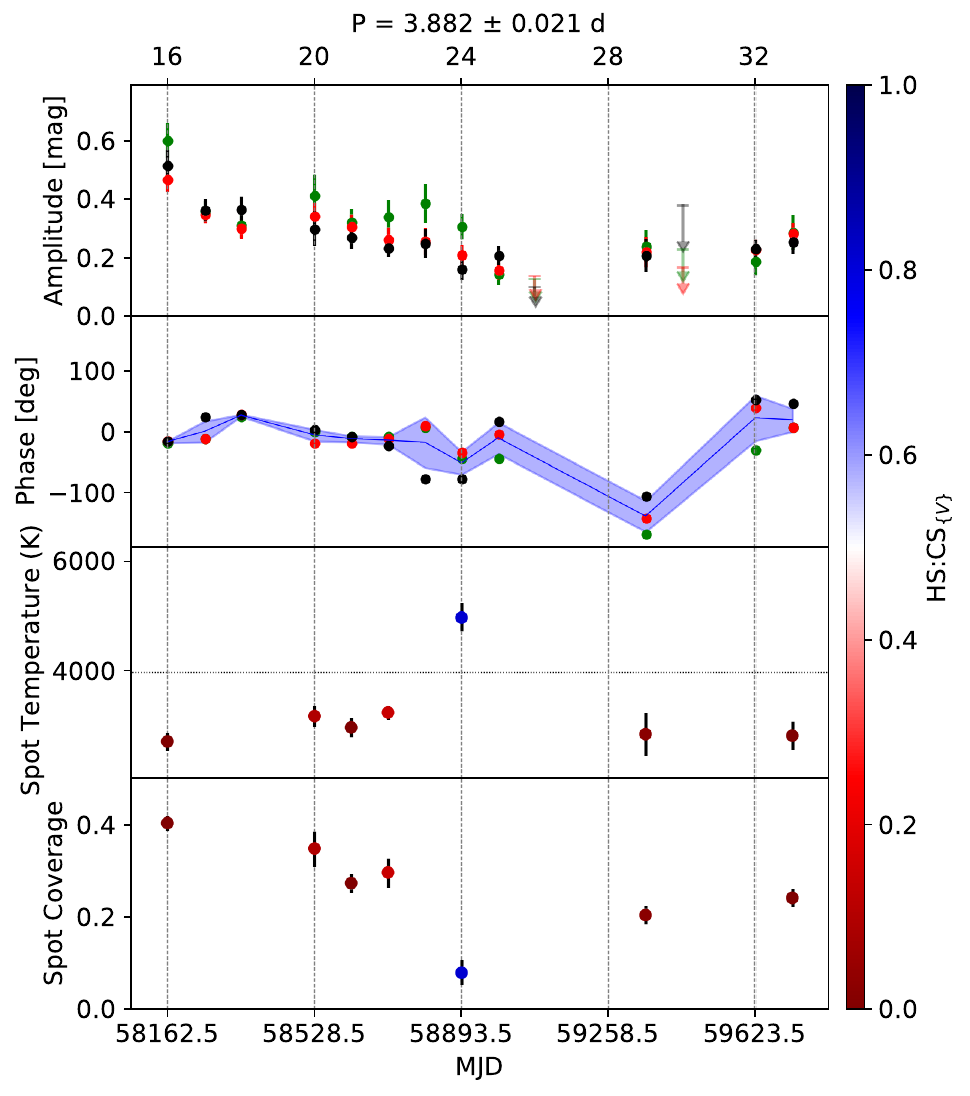}
\caption{As Fig. \ref{fig: ev1} for object 26 \label{fig: ev26}}
\end{figure}

\begin{figure}
\centering
\includegraphics[width=\columnwidth]{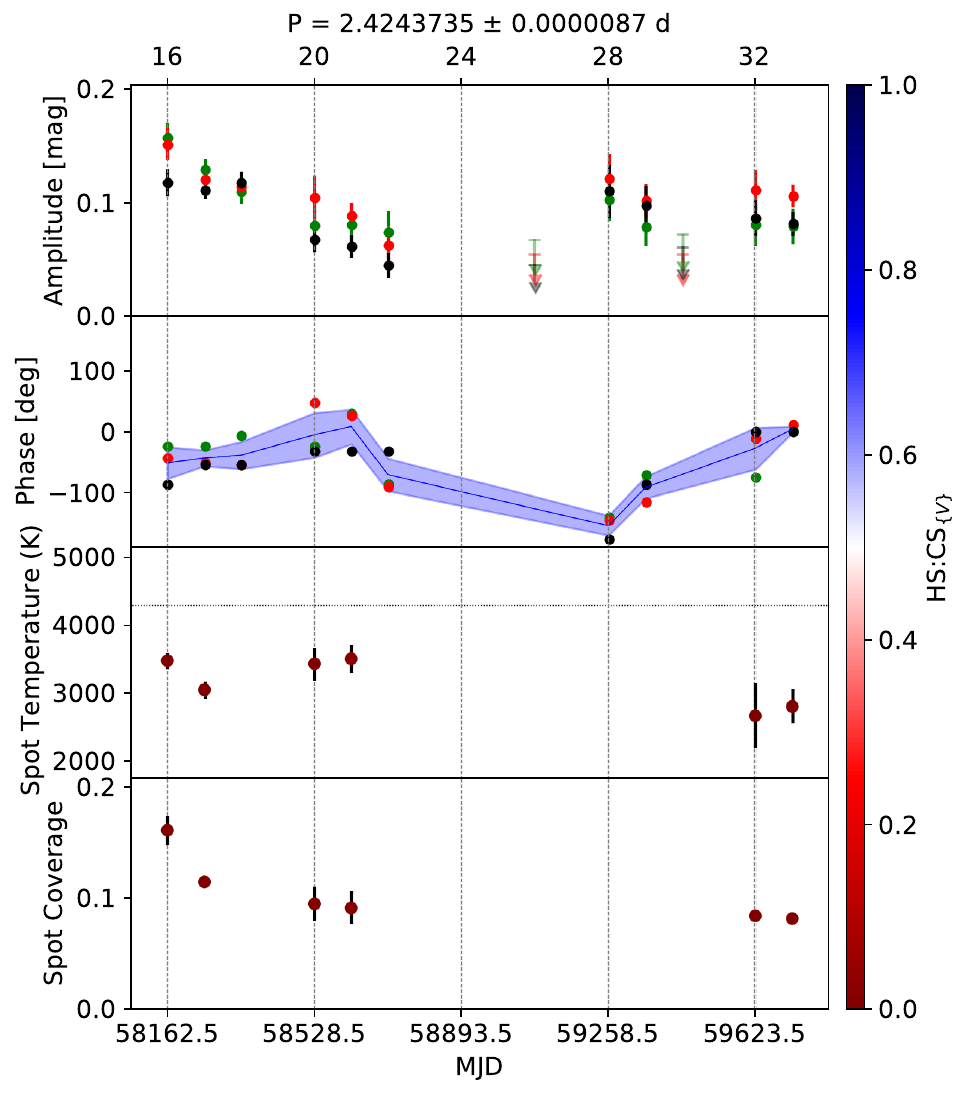}
\caption{As Fig. \ref{fig: ev1} for object 27 \label{fig: ev27}}
\end{figure}

\begin{figure}
\centering
\includegraphics[width=\columnwidth]{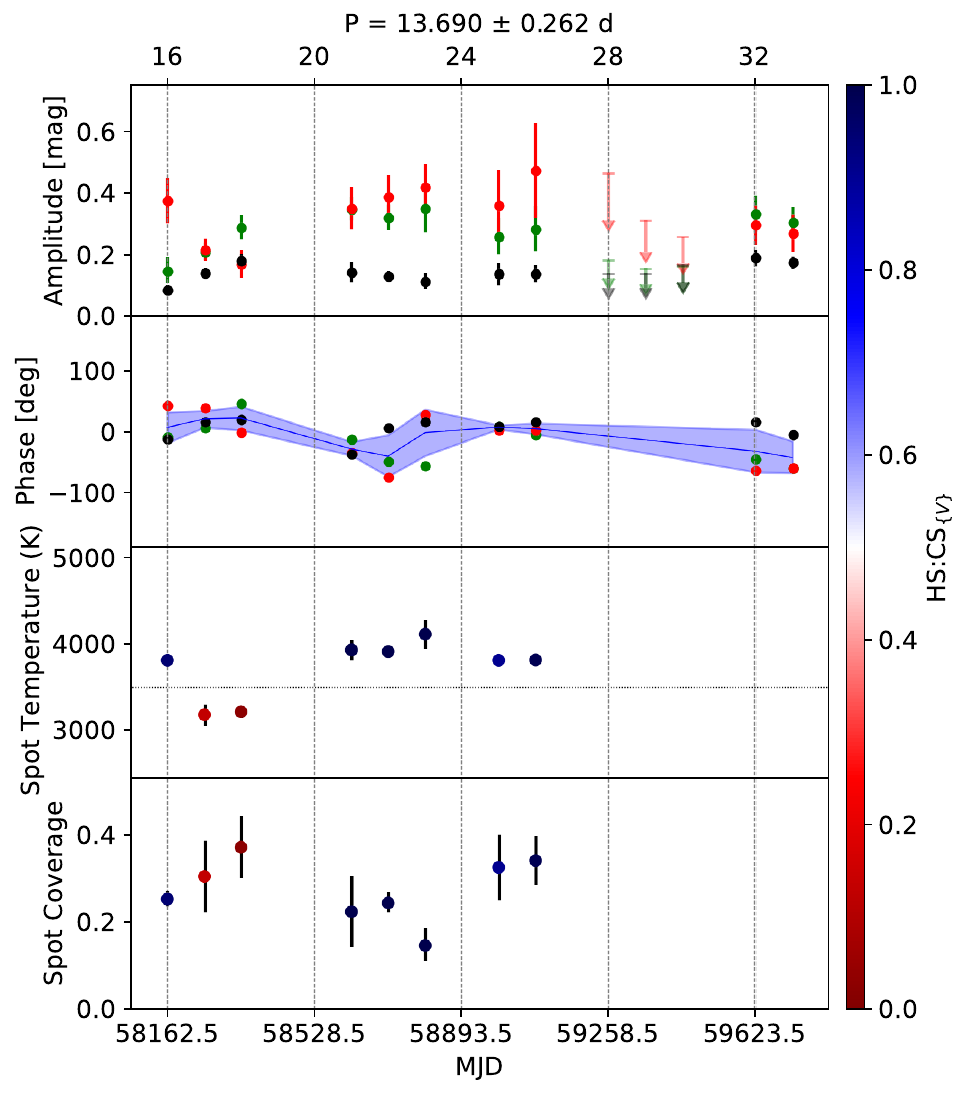}
\caption{As Fig. \ref{fig: ev1} for object 28 \label{fig: ev28}}
\end{figure}

\begin{figure}
\centering
\includegraphics[width=\columnwidth]{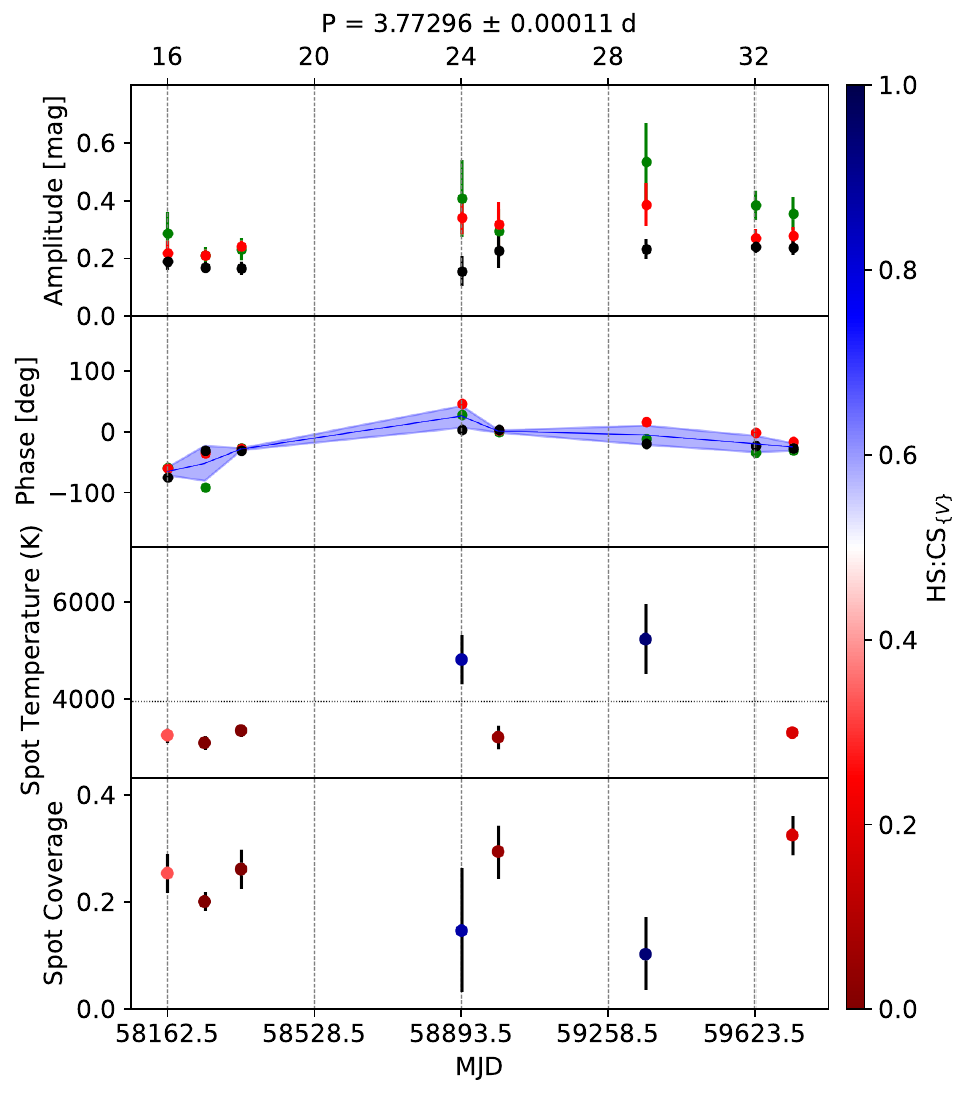}
\caption{As Fig. \ref{fig: ev1} for object 29 \label{fig: ev29}}
\end{figure}

\begin{figure}
\centering
\includegraphics[width=\columnwidth]{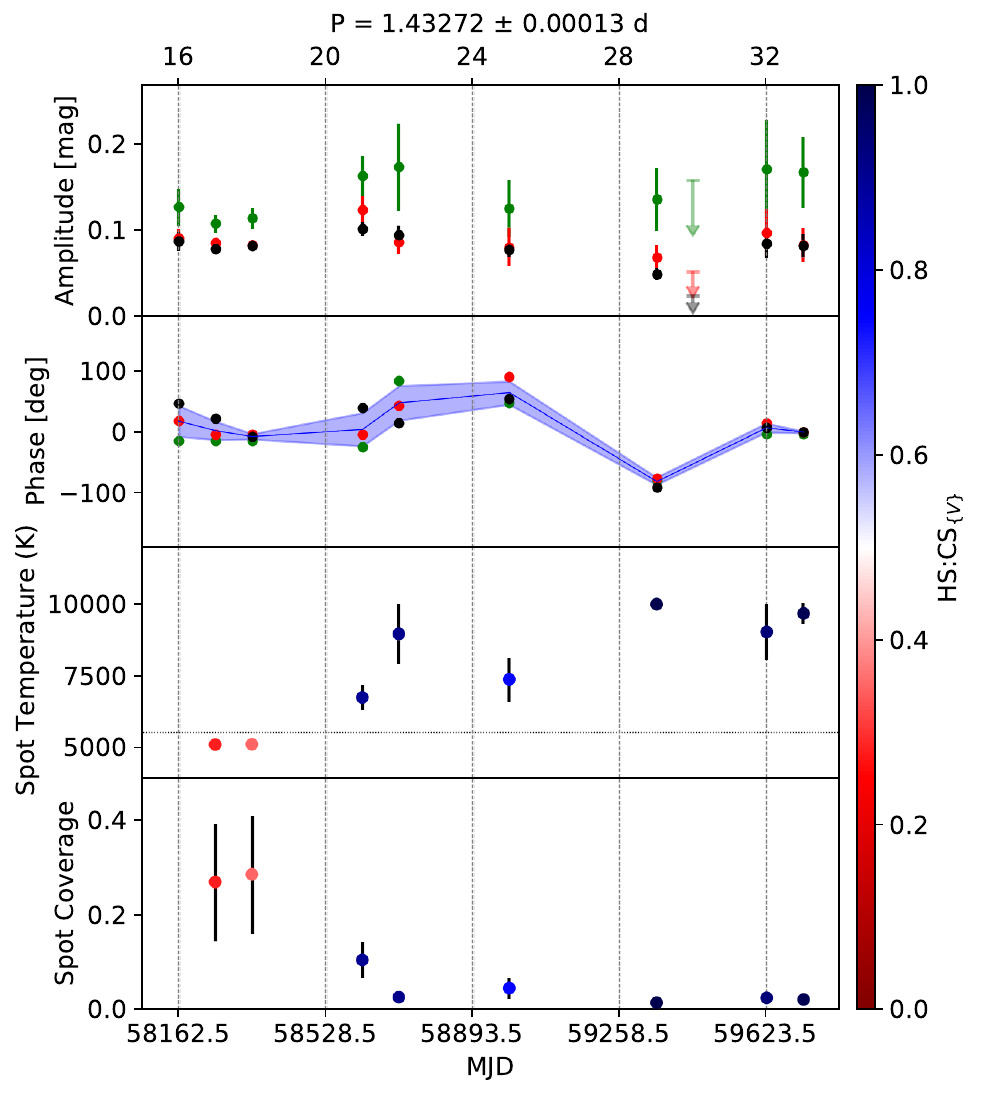}
\caption{As Fig. \ref{fig: ev1} for object 30 \label{fig: ev30}}
\end{figure}

\begin{figure}
\centering
\includegraphics[width=\columnwidth]{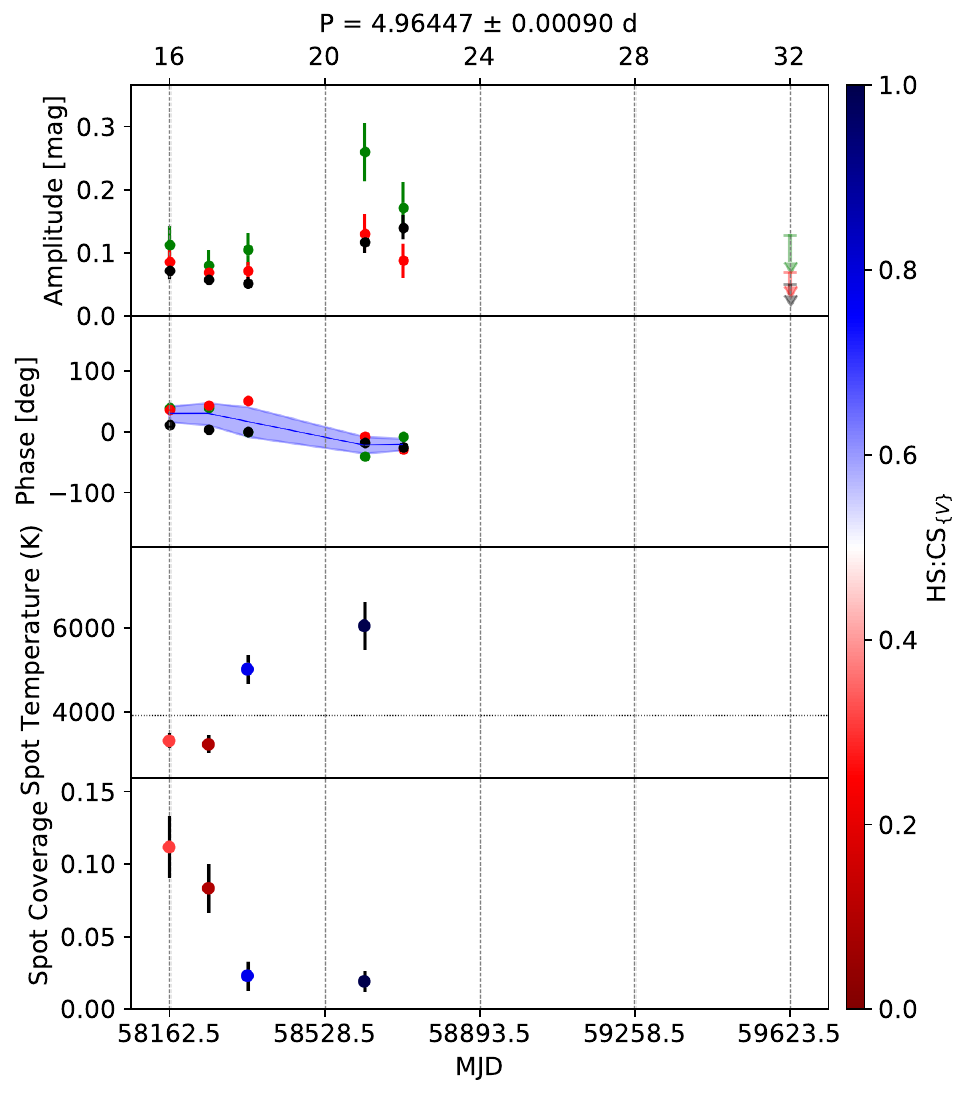}
\caption{As Fig. \ref{fig: ev1} for object 31 \label{fig: ev31}}
\end{figure}

\begin{figure}
\centering
\includegraphics[width=\columnwidth]{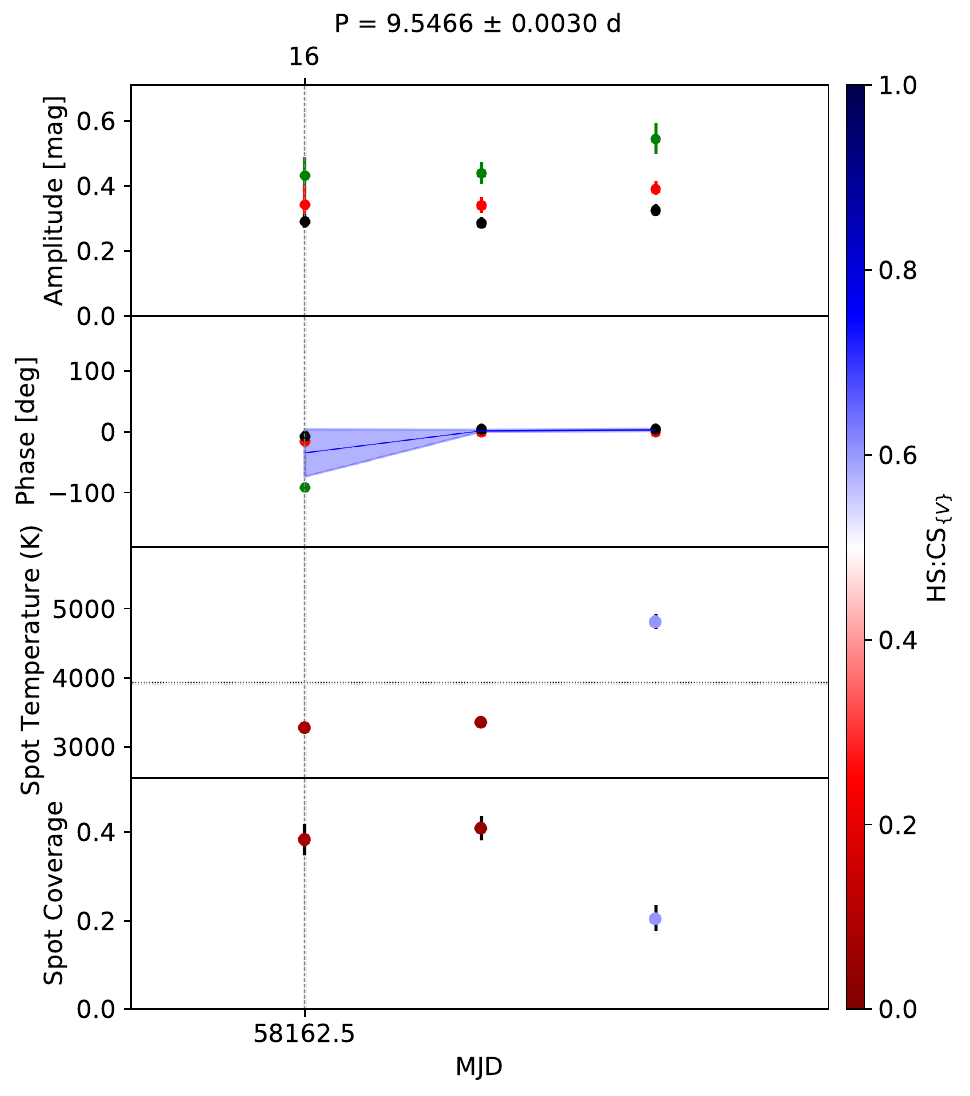}
\caption{As Fig. \ref{fig: ev1} for object 32 \label{fig: ev32}}
\end{figure}

\clearpage
\bsp	
\label{lastpage}
\end{document}